\documentclass[11pt,a4paper]{article}
\pdfoutput=1
\usepackage[pdftex]{graphics}
\usepackage{jheppub}
\usepackage{amsmath,amssymb,amsfonts}
\usepackage{enumitem}
\usepackage{multirow}
\usepackage{array,booktabs}
\usepackage{slashed}

\def\a{\alpha}
\def\b{\beta}
\def\e{\epsilon}
\def\ve{\varepsilon}
\def\m{\mu}
\def\n{\nu}
\def\r{\rho}

\def\s{\sigma}

\def\t{\tau}

\def\L{\Lambda}
\def\iff{\Longleftrightarrow}

\def\det{{\rm det}}

\usepackage{accents}
\usepackage{mathrsfs}
\usepackage{mathtools}

\usepackage[usenames,dvipsnames]{xcolor}
\definecolor{skyblue}{RGB}{34,139,230}
\definecolor{navy}{rgb}{0,0,0.7}
\definecolor{purple}{RGB}{171,1,207}
\definecolor{labelcolor}{RGB}{194, 175, 116}

\usepackage{tikz}
\usetikzlibrary{calc} % to use relative coordinates
\usetikzlibrary{shapes.geometric} % to draw regular polygons
\usetikzlibrary{positioning} % to use right=of 
\usetikzlibrary{fit} % for fit size
\usepackage[a]{esvect} % arrow styling %f
\tikzset{empty/.style = {inner sep = 0pt, outer sep = 0, minimum size = 0}}
\tikzset{b/.style = {inner sep = 2pt, outer sep = 4pt, minimum size = 12pt}}
\tikzset{w/.style = {inner sep = 1pt, outer sep = 2pt, minimum size = 12pt, anchor = west}}
\tikzset{s/.style = {inner sep = 2.5pt, outer sep =2.5pt, minimum size = 1pt, font = \small}}
\usepackage[export]{adjustbox}

\usepackage[final]{showlabels} % use "inline" or "final"

\def\mem{\hspace{0.1em}}
\def\hem{\hspace{0.05em}}
\def\nem{\hspace{-0.1em}}
\def\hnem{\hspace{-0.05em}}
\def\hhem{\hspace{0.025em}}
\def\hhnem{\hspace{-0.025em}}
\def\hhhem{\hspace{0.0125em}}
\def\hhhnem{\hspace{-0.0125em}}

\def\blank{{\,\,\,\,\,}}

\def\qiq{{\quad\implies\quad}}
\def\iq{{{\implies}\quad}}

\def\Re{{\operatorname{Re}}}

\def\be{{\bar{\epsilon}}}

\def\bs{{\bar{\sigma}}}

\def\bDelta{{\bar{\Delta}}}
\def\bomega{{\bar{\omega}}}

\def\bpi{{\bar{\pi}}}
\def\bphi{{\bar{\phi}}}

\def\bpsi{{\smash{\bar{\psi}}\kern0.02em\vphantom{\psi}}}

\def\bchi{{\bar{\chi}}}

\def\bZ{{\bar{Z}}}

\def\bW{{\bar{W}}}
\def\mathe{{\mathrm{e}}}
\def\da{{\dot{\a}}}
\def\db{{\dot{\b}}}
\def\c{{\gamma}}
\def\dc{{\dot{\c}}}
\def\d{{\delta}}
\def\dd{{\dot{\d}}}
\def\rmA{{\mathrm{A}}}
\def\rmB{{\mathrm{B}}}

\def\umi{$\ddot{\smash{\clipbox{0em 0em 0em 0.2em}{\text{i}}}}$}
\def\i{\iota}
\def\bi{{\bar{\iota}}}
\renewcommand{\o}{o} 
\def\bo{{\bar{o}}}
\def\bozo{{\bar{o}zo}}
\def\rambda{{\bar{\lambda}}}
\def\bmu{{\bar{\mu}}}

\def\lsq{{
    \kern-0.037em
    \adjustbox{scale=0.99,valign=c}{$
        {\lfloor \llap{\reflectbox{\rotatebox[origin=c]{180}{$\lfloor$}}}}
    $}
    \kern-0.04em
}}
\def\rsq{{
    \kern-0.04em
    \adjustbox{scale=0.99,valign=c}{$
        {\rlap{\reflectbox{\rotatebox[origin=c]{180}{$\rfloor$}}} \rfloor}
    $}
    \kern-0.037em
}}

\def\one{{\mathsf{1}}}
\def\two{{\mathsf{2}}}
\def\three{{\mathsf{3}}}
\def\four{{\mathsf{4}}}
\def\bone{{\bar{\mathsf{1}}}}
\def\btwo{{\bar{\mathsf{2}}}}
\def\bthree{{\bar{\mathsf{3}}}}
\def\bfour{{\bar{\mathsf{4}}}}

\def\minie{{\textstyle\frac{1}{2}}}
\def\minime{{\textstyle\frac{1}{-2}}}

\def\bz{{\bar{z}}}

\def\ihbar{{i\hbar}}
\newcommand{\Ket}[1]{{\hem\big|\hem{#1}\big\rangle}}
\newcommand{\Bra}[1]{{\big\langle{#1}\hem\big|\hem}}
\newcommand{\BraKet}[2]{{\big\langle{#1}\hem\big|\hem{#2}\big\rangle}}

\newcommand{\wedgez}[2]{{dz^{#1}\hspace{-0.21em}\wedge\hspace{-0.05em}dz^{#2}}}
\newcommand{\wedgebz}[2]{{d\bz^{#1}\hspace{-0.21em}\wedge\hspace{-0.05em}d\bz^{#2}}}

\newcommand{\wedgetwo}[4]{{d{#1}^{#3}\hspace{-0.21em}\wedge\hspace{-0.05em}d{#2}^{#4}}}

\def\swedge{{\mem{\wedge}\,}}

\usepackage{bm}

\newcommand{\dbar}{
    d\kern-.20em\makebox[0pt][l]{$\bar{}$}\kern.20em
}
\newcommand{\deltabar}{
    \delta\kern-.20em\makebox[0pt][l]{$\bar{}$}\kern.20em
}

\def\identity{{\rlap{1} \hskip 1.6pt \adjustbox{scale=1.1}{1}}}

\setlist[itemize]{leftmargin=1.86em,label={\adjustbox{scale=0.8,valign=c}{$\bullet$}}}

\makeatletter
\def\hlinewd#1{
    \noalign{\ifnum0=`}\fi\hrule \@height #1 \futurelet
    \reserved@a\@xhline
}
\makeatother

\def\ellf{{\kern0.08em\ell!\kern0.08em}}

\def\dnab{{d^\nabla\kern-0.2em}}

\def\revA{{\rotatebox[origin=c]{180}{$A$}\kern-0.2em}}
\def\revh{{\rotatebox[origin=c]{180}{$h$}\kern-0.04em}}
\def\mt{{\mathbb{MT}}}
\def\ma{{\mathbb{MA}}}

\def\MPl{{M_\text{Pl}}}
\def\Kerr{{\text{$\sqrt{\text{Kerr\hem}}$}}}

\def\M{{\mathcal{M}}}

\def\doublerightarrow{
    {\hspace{0.2em}\mathrlap{\kern-0.2em\rightarrow}\rightarrow}
}

\newcommand{\strikedgamma}{
    \gamma\kern-.20em\makebox[0pt][l]{\adjustbox{raise=-1.0ex}{$\bar{}$}}\kern.20em
}
\newcommand{\strikedC}{
    C\kern-.60em\makebox[0pt][l]{\adjustbox{raise=-0.6ex}{$\bar{}$}}\kern.60em
}

\title{
    Symplectic Perturbation Theory
    \\[-0.07\baselineskip]
    in Massive Ambitwistor Space:
    \\[0.1\baselineskip]
    A Zig-Zag Theory of Massive Spinning Particles
}

\author[a]{Joon-Hwi Kim}
\author[b,c,d]{Sangmin Lee} 

\affiliation[a]{Department of Physics, California Institute of Technology, Pasadena, CA 91125, U.S.A.}
\affiliation[b]{Department of Physics and Astronomy, Seoul National University, Seoul 08826, Korea}
\affiliation[c]{Center for Theoretical Physics, Seoul National University, Seoul 08826, Korea}
\affiliation[d]{College of Liberal Studies, Seoul National University, Seoul 08826, Korea}

\abstract{
    We develop a 
    theory of massive spinning 
    particles
    interacting with
    background fields
    in four spacetime dimensions
    in which 
    holomorphy and chirality
    play a central role.
    Applying a perturbation theory of symplectic forms
    to the massive twistor space
    as a K\"ahler 
    manifold,
    we find that
    the spin precession behavior of 
    a massive spinning particle
    is directly determined from
    the manner in which
    self-dual and anti-self-dual
    field strengths permeate into ``complex spacetime.''
    Especially,
    the particle shows the minimally coupled precession behavior
    if 
    self-dual field strength continues holomorphically into the complex:
    the Newman-Janis shift.
    In general,
    computing the momentum impulse
    shows that
    the parameters that control
    generic non-holomorphic
    continuations
    are directly related to 
    the coupling constants in
    the massive-massive-massless spinning on-shell amplitude of
    Arkani-Hamed, Huang, and Huang,
    and thus they are interpreted as 
    the single-curvature Wilson coefficients given by Levi and Steinhoff,
    redefined on complex worldlines.
    Finally,
    exact expressions for
    Kerr and \texorpdfstring{{\Kerr}}{√Kerr} actions
    are
    bootstrapped
    in monochromatic self-dual plane-wave backgrounds
    from
    symplectivity and
    a matching between classical scattering and the on-shell amplitude,
    from which we obtain all-order exact impulses of classical observables.
}

\emailAdd{joonhwi@caltech.edu, sangmin@snu.ac.kr}

\begin{document}

\maketitle

\section{Introduction}

Throughout a series of papers,
we take a journey to the physics of
massive spinning particles
% coupled to gauge theory and gravity
in four dimensions.
We investigate their interactions with
gauge theory and gravity
within the context of 
recent developments in the scattering theory of spinning objects \cite{ambikerr0,gmoov,ahh2017,Emond:2020lwi,bah2020kerr,monteiro2014black,arkani2020kerr,guevara2019scattering,chkl2019,aoude2021classical,guevara2019black,Maybee:2019jus,kosower2019amplitudes,guevara2021reconstructing,strominger2021black,Aoude:2022thd,Cangemi:2022bew,Moynihan:2019bor,chung2020kerr,Bern:2020buy,Chung:2019duq,vines2018scattering,vines2016canonical,Levi:2019kgk,levi2020effective,Levi:2015msa,porto2016effective,levi2015leading,marsat2015cubic,blanchet2014gravitational,perrodin2012subleading,Levi:2011eq,levi2010spin-orbit,levi2010spin(1)-spin(2),porto2010next,porto2008spin,goldberger2007effective,goldberger2006effective,porto2006hyperfine,porto2006post}.

It is an interesting fact that
a classic theme of electromagnetism and general relativity,
the equation of motion of a relativistic spinning test particle in an external field,
has lacked a systematic analysis
until recently.
The Thomas-Bargmann-Michel-Telegdi (TBMT) 
% \cite{thomas1926motion,thomas1927kinematics,bargmann1959precession,frenkel1926spinning,frenkel1926elektrodynamik,jackson2008examples,hushwater2014discovery,jackson2014jackson,rafelski2018relativistic}
\cite{thomas1927kinematics,bargmann1959precession,rafelski2018relativistic,frenkel1926spinning,frenkel1926elektrodynamik,thomas1926motion}
and Mathisson-Papapetrou-Tulczyjew-Dixon (MPTD) \cite{mathisson1937neue,papapetrou1951spinning,tulczyjew1959motion,dixon1964covariant,dixon1965classical,Trautman:2002zz,costa2016spacetime,costa2015ssc}
equations
specify the dynamics
up to dipole order.
The latter has been extended up to the quadrupole order as well \cite{Khriplovich:1989ed,yee1993equations,khriplovich1997equations,khriplovich2000equations,pomeranskii2000spinning,deriglazov2016ultrarelativistic,deriglazov2017mathisson}.
The dipole and quadrupole couplings are given by
the gyromagnetic ratio $g$ and the gravimagnetic ratio $\kappa$, respectively.
Meanwhile, a 
particle description of a relativistic spinning object
inherently involves the intricacy of 
the spin constraint \cite{pryce1948mass,costa2015ssc},
due to the arbitrariness in defining the center-of-mass worldline.
Steinhoff \cite{steinhoff2015spin} has formulated the problem in terms of gauge symmetry.
Building upon such a development,
a systematic classification of all higher spin-induced multipole couplings,
in a way independent of the spin gauge redundancies,
was finally given 
by Levi and Steinhoff \cite{Levi:2015msa}
% at the level of action
after the subject gained revived interest 
in the context of 
% effective point-particle description of spinning bodies for 
% the binary inspiral problem
% and effective field theory approach to gravitational physics
the effective field theory approach to gravitational physics
% \cite{vines2018scattering,vines2016canonical,levi2020effective,Levi:2015msa,porto2016effective,levi2015leading,marsat2015cubic,blanchet2014gravitational,perrodin2012subleading,levi2010spin-orbit,levi2010spin(1)-spin(2),porto2010next,porto2008spin,goldberger2007effective,goldberger2006effective,porto2006hyperfine,porto2006post}.
\cite{vines2018scattering,vines2016canonical,Levi:2019kgk,levi2020effective,Levi:2015msa,porto2016effective,levi2015leading,marsat2015cubic,blanchet2014gravitational,perrodin2012subleading,Levi:2011eq,levi2010spin-orbit,levi2010spin(1)-spin(2),porto2010next,porto2008spin,goldberger2007effective,goldberger2006effective,porto2006hyperfine,porto2006post}.
While using the Hanson-Regge spherical top
\cite{Hanson:1974qy,bailey1975lagrangian}
to model the spinning degrees of freedom,
\cite{Levi:2015msa} 
enumerated all
% $(D^{n}(\text{Riemann})){\mem\cdot\mem}(\text{spin})^{n+2}$-type
$(D^{\ell-2}(\text{Riemann})){\mem\cdot\mem}(\text{spin})^{\ell}$-type
terms
as spin gauge invariant
operators in the effective point-particle Lagrangian of an extended spinning body.

Amusingly, 
further insights were gained by
matching Levi and Steinhoff's
spin multipole
Wilson coefficients, $(C_1, C_2, C_3, \cdots)$,
with the coupling constants in Arkani-Hamed, Huang, and Huang \cite{ahh2017}'s 
on-shell massive spinning three-point amplitudes \cite{chkl2019,gmoov,guevara2019scattering,guevara2019black,Chung:2019duq,aoude2021classical,arkani2020kerr}.
In quantum field theory,
one may objectively define minimal coupling
by demanding the amplitudes to have the best high-energy behavior \cite{holstein2006large,ahh2017}.
The matching calculation shows that
the minimal coupling of the massive spinning object to electromagnetism and gravity
is given by 
$C_\ell {\,\mem=\mem\,} 1$ for all $\ell > 0$,
which 
implies gyromagnetic ratio $g = 2C_1 = 2$ 
for the
TBMT equation
and gravimagnetic ratio $\kappa = C_2 = 1$
for the
% quadrupole-order 
extended
MPTD
equation:
the spin precession behavior that the Kerr-Newman black hole is known to exhibit \cite{carter1968global,debney1969solutions,israel1970source,reina1975gyromagnetic,lopez1984extended,burinskii2008dirac,thorne1985laws,khriplovich1997equations,Newman:2002mk,newman2004maxwell,Newman:1973yu,newman1973complex,Newman:1965my-kerrmetric}.
Indeed, it turns out that 
considering the classical spin limit of 
the minimally coupled amplitudes
\cite{chkl2019,arkani2020kerr,guevara2019scattering}
\`a la coherent states 
\cite{aoude2021classical}
provides an amplitudes-level derivation of 
the Newman-Janis shift
% \cite{Newman:1965my-kerrmetric,newman1973complex,Newman:1973yu,Newman:2002mk,newman2004maxwell,newman1974curiosity,Newman:1965tw-janis}
\cite{Newman:1965my-kerrmetric,Newman:1965tw-janis,monteiro2014black}
% \cite{Newman:1965my-kerrmetric,Newman:1965tw-janis,Newman:2002mk,monteiro2014black}
property of 
the Kerr-Newman solution.

% When phrased in a covariant fashion,
% the Newman-Janis shift 
% states that
% the self-dual/anti-self-dual field strength and Weyl curvature of the Kerr-Newman solution 
% follow from those of the Reissner-Nordstr\"om solution 
% by shifting the origin by $\pm iy$,
% where $y$ denotes the spin length vector \cite{adamo2014kerr}.
% % In short, ``spin is imaginary deviation.''
% Newman \cite{newman1973complex,Newman:1973yu,Newman:2002mk,newman2004maxwell,newman1974curiosity}
% understands
% such imaginary displacements
% % $\pm iy$
% as taking place in 
% % the complexified Minkowski space.
% a complex spacetime.
% Then we can think of the Newman-Janis shift as
% analytic continuation of spacetime fields into the complex spacetime
% such that
% (anti-)self-dual part continues (anti-)holomorphically.

However,
there is a discrepancy between
the Lagrangian 
% formulation 
and
the amplitudes-level understanding:
the fact that 
% $g{\,=\,}2$, $\kappa{\,=\,}1$, $\cdots$
% are
% $(g{\,=\,}2, \kappa{\,=\,}1, C_{\ell\hem>\hem2}{\,=\,}1)$
$C_\ell {\,\mem=\mem\,} 1$
describes
the minimal 
spin
precession behavior
is 
quite 
obscure in 
the current
presentation
% formulation
of 
the effective action,
as
it rather treats
% $g{\,=\,}0$ and $\kappa{\,=\,}0$
% $g{\,=\,}0$, $\kappa{\,=\,}0$, $\cdots$
$C_\ell {\,\mem=\mem\,} 0$
as ``minimal''
and implements
% $g{\,=\,}2$ or $\kappa{\,=\,}1$
% $g{\,=\,}2$, $\kappa{\,=\,}1$, $\cdots$ 
$C_\ell {\,\mem\neq\mem\,} 0$
by
additional,
``non-minimal'' interaction Lagrangians.
% thus deviating from the ``origin'' favored by the amplitudes.
% In other words,
% one does not obtain the Kerr black hole 
% by merely covariantizing the Hanson-Regge spherical top model
% but needs to supplement it with an infinite tower of higher multipoles.
Recently, 
% Guevara, Maybee, Ochirov, O'Connell, and Vines \cite{gmoov}
Guevara et al.\,\cite{gmoov} proposed
complex-worldline actions for Kerr and 
% its single copy.
{\Kerr}.
It is tempting to generalize their proposal,
have a fully covariant all-order realization of the actions,
and
develop a 
% (classical)
perturbation theory
that takes
the true minimal coupling
% the Kerr-Newman black hole 
as the ex\-pansion point.
Even before 
considering two-body systems \cite{jakobsen2022conservative} or four-point amplitudes \cite{aoude2022searching},
to our best knowledge,
the test particle equations of motion
that follow from \cite{Levi:2015msa}'s\;effective
% i.e., all-order generalizations of the TBMT and MPTD equations,
action
have not been explicitly spelled out,
at least
in the post-Minkowskian setting that retains
manifest Lorentz invariance.
We aim to
fill these gaps
while providing a new angle on the physics of massive spinning particles
from
% by building upon
a twistorial perspective 
developed
% taken
in \cite{ambikerr0}.

In this first part of our journey, 
we take
% an ``exotic''
% a ``flipped'' 
a ``bottom-up''
approach
to spin precession.
In a sense, we parallel the punchline of the modern S-matrix program.
First,
the effective action is not the starting point.
Rather, we 
work with
Poisson brackets 
% implement interactions with background fields
% as perturbations on the Poisson bracket
and substitute the existence of a local action with
the Jacobi identity (symplectivity),
following the ``symplectic perturbation theory'' approach of \cite{spt}.
Second,
the gauge fields of Yang-Mills or gravity never appear;
we only assume 
% an anonymous 
a
closed two-form 
% $\minie\mem \phi_{\m\n}\hnem(x)\mem \wedgetwo{x}{x}{\m}{\n}$ 
% $\minie\mem \phi_{\smash{\a\da\b\db}\vphantom{\b}}\hhnem(x)\mem \wedgetwo{x}{x}{\da\a}{\db\b}$ 
$\minie\mem \phi_{\smash{\a\da\b\db}\vphantom{\b}}\mem\hhhem \wedgetwo{x}{x}{\da\a}{\db\b}$
on spacetime
as a formal substitute for 
a ``field strength''
(contracted with a ``charge'').
From a set of physical assumptions,
including Lorentz symmetry, little group symmetry, and symplectivity,
we construct
a complete dictionary between 
% equations of motion up to linear order in curvature,
$\mathcal{O}((\text{curvature})^1{\hem\cdot\mem}(\text{spin})^\infty)$ 
spin precession
equations of motion,
% impulses of classical observables,
spin frame and momentum impulses,
and on-shell three-point amplitudes
in Section \ref{sec:Flat}.
Moreover,
in Section \ref{sec:Wave},
we deduce
the exact 
symplectic
structures of Kerr and {\Kerr}
in monochromatic self-dual plane-wave backgrounds
from
a match\-ing between classical scattering
and the on-shell amplitude.
Exact solutions to the resulting all-order equations of motion
derive the impulses 
of classical observables
as exact quantities.
A ``correction'' term that
follows from the symplectivity requirement
induces
precession
of the color charge
for {\Kerr}
and 
an extra spacetime translation
for Kerr.
A top-down derivation of these results
from a realization of the effective action in generic backgrounds
is 
% rather
only
given in 
the follow-up
% the next
paper \cite{ambikerr2}.

A key feature of
% such a ``symplectic perturbation'' \cite{spt} approach to the massive twistor
our formulation of interacting massive spinning particles
is that 
it 
% respects and
fully appreciates
a ``hidden'' complex-geometrical structure
that is inherent in the free theory.
By doing so,
the minimal nature 
of the Kerr-Newman coupling
$C_\ell {\,\mem=\mem\,} 1$
becomes
vividly evident
already
at the classical level.
Let us outline the logic briefly
as an invitation to the rest of the paper.

As a K\"ahler manifold,
the massive twistor space $(Z,\bZ)$
has a ``zig-zag'' Poisson bracket,
meaning that only the brackets between
un-barred (holomorphic, ``zig'')
and barred (anti-holomorphic, ``zag'')
variables are non-zero:
\begin{align}
    \label{eq:int.zebras}
    \{Z,Z\} = 0
    \,,\quad
    \{Z,\bZ\} \neq 0
    \,,\quad
    \{\bZ,\bZ\} = 0
    \,.
\end{align}
In
Section \ref{sec:Flat.eom-min},
we introduce
a graphical notation
that
uses
``white for \textit{$Z\nem$ig}
and black for \textit{$\bZ\nem$ag}.''
The above free theory bracket
is depicted as
white-black 
% ($\{Z,\bZ\} {\mem\neq\mem\hhem\hhhem} 0$)
or
black-white 
% ($\{\bZ,Z\} {\mem\neq\mem\hhem\hhhem} 0$)
junctions:
\begin{align}
    \{Z,\bZ\} \neq 0
    \,,\quad
    \{\bZ,Z\} \neq 0
    \qquad\rightsquigarrow\qquad
    \label{eq:int.v.prop}
    % \omega{}^{-1}
    % \quad=\quad
    \textit{$Z\hnem\hnem$ig}\,
    % \text{zig}
        \,\mem
        \includegraphics[valign=c]{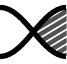}
        \mem\,
    % \text{zag}
    \textit{$\bZ\hnem\hnem$ag}
    \quad\text{or}\quad
    \textit{$\bZ\hnem\hnem$ag}\,
    % \text{zag}
        \,\mem
        \includegraphics[valign=c]{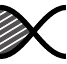}
        \mem\,
    % \text{zig}
    \textit{$Z\hnem\hnem$ig}
    \,.
    % \,,
\end{align}
Then, the perturbed Poisson bracket
of the interacting theory
is given by
summing
over
all possible gluings of
% these ``ribbons'' \eqref{eq:int.v.prop}
these ``ribbons''
while using the ``field strength'' two-form as a ``glue.''
% E.g.,
For instance,
if the ``field strength'' is holomorphic ($dZ \swedge dZ$),
the only allowed gluing is zig-to-zig
so that the
% all-order-exact
% Poisson bracket is just given as
resulting Poisson bracket 
has
% develops
zig-zag, zag-zig, and zag-zag components:
\begin{align}
    \label{eq:int.v.g=2}
    \{\blank,\blank\} 
        {}\,\,\,=\,\,\,{}\hem\hhhem
    &
    \includegraphics[valign=c]{figs/z-bz.pdf}\mem
    {\,\,\mem+\mem\,\,}
    \mem\includegraphics[valign=c]{figs/bz-z.pdf}\mem
    {\,\,\mem-\mem\,\,}
    \includegraphics[valign=c]{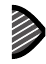}
    \includegraphics[valign=c]{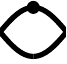}
    \includegraphics[valign=c]{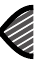}
    \,.
\end{align}
% The candy-shaped diagram shows that 
% holomorphic perturbations induce a ``zag-zag'' bracket.

Remarkably,
this ``zig-zag logic''
provides a derivation of the Newman-Janis shift.
% as elaborated in Section \ref{sec:Flat.o}.
Suppose the particle is minimally coupled.
When the background is self-dual,
the left-handed (zig) massive spinor-helicity variable
\cite{Perjes:1974ra,ahh2017,conde2016spinor,conde2016lorentz},
which is contained in the holomorphic twistor variable $Z$,
should be parallel transported.
For such a spin precession behavior to be implemented as the Hamiltonian equation of motion 
of the massive twistor,
the zig-zag and zag-zag brackets between 
the spinor-helicity variables should vanish
as in \eqref{eq:int.v.g=2}.
This means that
the self-dual ``field strength''
should extend holomorphically 
into the complexified Minkowski space!
The details are elaborated on in Section \ref{sec:Flat.o}.

% the Newman-Janis shift
% leads to a drastic simplification of 
% the all-order perturbation theory.

For a non-minimal coupling
such as
$C_\ell {\,\mem=\mem\,} 0$,
the 
% self-dual
``field strength''
is real and
develops all the
$dZ \swedge dZ$,
$dZ \swedge d\bZ$,
$d\bZ \swedge d\bZ$
components,
so
% so that
the 
% Poisson bracket
diagrammatic expansion
looks like
\begin{align}
\begin{split}
    \label{eq:int.v.g=0}
    \{\blank,\blank\} 
        {}\,\,\,=\,\,\,{}\hem\hhhem
    &
    \includegraphics[valign=c]{figs/z-bz.pdf}\mem
    {\,\,\mem+\mem\,\,}
    \mem\includegraphics[valign=c]{figs/bz-z.pdf}\mem
    % &
    {\,\,\mem-\mem\,\,}
    \includegraphics[valign=c]{figs/l-b.pdf}
    \includegraphics[valign=c]{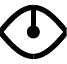}
    \includegraphics[valign=c]{figs/r-b.pdf}
    {\,\,\mem-\mem\,\,}
    \includegraphics[valign=c]{figs/l-b.pdf}
    \includegraphics[valign=c]{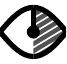}
    \includegraphics[valign=c]{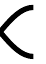}
    {\,\,\mem-\mem\,\,}
    \includegraphics[valign=c]{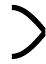}
    \includegraphics[valign=c]{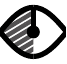}
    \includegraphics[valign=c]{figs/r-b.pdf}
    {\,\,\mem-\mem\,\,}
    \includegraphics[valign=c]{figs/l.pdf}
    \includegraphics[valign=c]{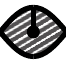}
    \includegraphics[valign=c]{figs/r.pdf}
    \\
    &
    \smash{
    {\,\,\mem+\mem\,\,}
    \includegraphics[valign=c]{figs/l-b.pdf}
    \includegraphics[valign=c]{figs/0+ww.pdf}
    \includegraphics[valign=c]{figs/0+bw.pdf}
    \includegraphics[valign=c]{figs/r-b.pdf}
    {\,\,\mem+\mem\,\,}
    \includegraphics[valign=c]{figs/l-b.pdf}
    \includegraphics[valign=c]{figs/0+ww.pdf}
    \includegraphics[valign=c]{figs/0+bb.pdf}
    \includegraphics[valign=c]{figs/r.pdf}
    {\,\,\mem+\mem\,\,}
    \cdots
    {\,\,\mem+\mem\,\,}
    \includegraphics[valign=c]{figs/l.pdf}
    \includegraphics[valign=c]{figs/0+bb.pdf}
    \includegraphics[valign=c]{figs/0+wb.pdf}
    \includegraphics[valign=c]{figs/r.pdf}
    {\,\,\mem+\mem\,\,}\cdots
    \,.
    }
\end{split}
\end{align}
It never truncates,
in contrast to \eqref{eq:int.v.g=2}.
% We conclude that
We see that
the non-minimal nature of other generic couplings
also becomes 
% obvious
manifest
in our complex-geometrical formulation.

% Such a simplification of perturbation theory provides a
% % rational
% rationale
% for identifying 
% $g{\,=\,}2$/$\kappa{\,=\,}1$ as the true minimal coupling 
% (among three-point couplings)
% already at the classical level.\footnote{
%     The astute reader may have noticed that
%     % A caveat is that 
%     % the truncation of the expansion in self-dual backgrounds
%     % also happen if
%     the simplicity of the expansions
%     is also exhibited by the case when
%     self-dual ``field strength'' extends anti-holomorphically.
%     We call that case ``antipodal minimal coupling.''
%     The resulting equation of motion turns out to be an ``inversion'' of that of the minimal coupling.
% }

These ``zig-zag diagrams'' are 
essentially a Penrose graphical notation \cite{Penrose:1956tensormethods,penrose1971negdim,penrose1984spinors1,Kim:2019eab}.
They depict how 
% (classical)
massive spinning particles
classically
interact with self-dual and anti-self-dual fields.
In this sense,
we adopted the terminology ``zig-zag'' 
as an homage to Penrose's ``zig-zag electron'' \cite{penrose2005road}.
The color scheme and design
of the diagrams
are
also
inspired by 
twistor diagrams 
\cite{arkani2010s-matrix,penrose1973twistor,penrose1975aims,hodges1980twistor,hodges1982twistor,hodges1983moller,hodges1983compton,hodges1985regularization,hodges1985mass,hodges1990feynman,huggett1993cohomology,penrose1998geometric,hodges2005a,hodges2005b}
and \cite{gmoov}'s ``worldsheet'' in the spin-space-time.

The twistor particle program \cite{Perjes:1974ra,Penrose:1974di} is approaching its 50\textsuperscript{th} birthday.
We hope 
our theory of
% coupling the massive twistor particle to background fields,
interacting massive twistor particles,
deeply rooted in holomorphy and chirality,
could 
have the honor of manifesting and realizing
a part of
% Sir 
Roger Penrose's 
geometrical imaginations 
on our universe
and
ambitious proposals
on reformulating our language describing 
% the
nature.

\paragraph{Notations and conventions}
The metric signature is 
% mostly-plus.
$(-,+,+,+)$.
We employ $m,n,r,s,$ $\cdots$ for flat Lorentz indices
while reserving $\m,\n,\r,\s\cdots$ for generic spacetime indices.
It is the former that is converted to
% The former is converted to
% The Lorentz indices are converted to
the left-handed and right-handed  $\mathrm{SL}(2,\mathbb{C})$ spinor indices,
$\a,\b,\c,\d\cdots$ and 
$\da,\db,\dc,\dd\cdots$.
Our spinor conventions are that of \cite{ambikerr0}.
For the $\mathrm{SU}(2)$ massive little group, we use 
$I,J,K,L\cdots$ for the spinors and
$a,b,c,d,\cdots$ for the adjoint.

We use 
the index-free notation
$\pi_\a {\mem\rightsquigarrow\mem} |\pi\rangle$,
$\pi^\a {\mem\rightsquigarrow\hem} \langle\pi|$,
$\bpi^\da {\mem\rightsquigarrow\mem} |\bpi\rsq$,
$\bpi_\da {\mem\rightsquigarrow\hem} \lsq\bpi|$.
% for the index-free notation of $\mathrm{SL}(2,\mathbb{C})$ spinors.
Sometimes we simply make contracted indices implicit as
$\bo_\da\hhem z^{\da\a} \o_\a = \bozo$
or $(\lambda\hem\s_a)_\a{}^I = \lambda_\a{}^J \hhnem (\s_a)_J{}^I$
to avoid clutter.
This does not create any ambiguity because
we never raise/lower spinor indices
except for massless spinors:
no $\lambda^{\a}{}_I$, $\rambda{}^{I\da}$, $z_{\a\da}$, etc.,
and
$\e^{\a\b} \o_\b = \o^\a$
but $\e^{\a\b} \lambda_\b{}^I = (\hhem\e\hem \lambda\hhem)^{\a I}$.

Lorentz indices are converted to spinor indices
as $p_{\a\da} := p_m (\s^m)_{\a\da}$
but with a single exception:
we find it favorable to 
define $x^{\da\a} := \minime (\bs^m)^{\da\a} x^m$
for position-type quantities
to minimize the occurrence of the normalization factor $(-2)$
from $(\s^m)_{\a\da} (\bs^n)^{\da\a} = -2 \eta^{mn}$.
This property 
% inherits to
is passed down to
tetrad \smash{$e^{\da\a} := \minime (\bs^m)^{\da\a}\mem e^m{}_\m\mem dx^\m$},
% tetrad perturbation 
% \smash{$h^{\da\a}
% := \minime (\bs^m)^{\da\a}\hem h^m{}_n\hem dx^n
% =: \minime\mem h^{\da\a}{}_{\smash{\b\db}\vphantom{\da}}\mem dx^{\db\b}$},
four-velocity 
$u^{\da\a} = \dot{x}^{\da\a}$,
% $u^{\da\a} = {dx^{\da\a}\hnem/d\t}$,
% $u^{\da\a} := \minime (\bs^m)^{\da\a} u^m$,
etc.
% and so on.
The conversion rule for two-forms is
$F_{\smash{mn}\vphantom{\b}}\hhnem\hhhnem (\s^m)_{\smash{\a\da}\vphantom{\b}}\hhhnem (\s^n)_{\smash{\b\db}\vphantom{\b}} =: 2\hem \e_{\a\b}\hhnem F_{\smash{\da\db}\vphantom{\b}} + 2 F_{\a\b}\hhem \be_{\smash{\da\db}\vphantom{\b}}$.

Lastly, we employ the notation
$\deltabar(\xi) := 2\pi \delta(\xi)$
for the normalized delta function.

% 
%%[0]%%%%%%%%%%%%%%%%%%%%%%%%%%%%%%%%%%%%%%%%%%%%

% \section{Free theory}
\section{Massive Ambitwistor Space}
\label{sec:MA}

The ``twistor particle program,''
initiated by Penrose \cite{Penrose:1974di},
aims to describe 
particles in four dimensions
% with a set of $n$ twistors.
as a system of twistors.
Massive particles are
regarded as ``composite systems'' of two or more twistors,
as a massive momentum decomposes into two or more null momenta.
Initially, the $\mathrm{SU}(n)$ internal symmetries of such $n$-twistor systems
were associated with 
the zoo of elementary particles:
e.g.,
$n{\,=\,}2$ for the weak isospin doublet of leptons
and
$n{\,=\,}3$ for the flavor symmetry of hadrons
\cite{Penrose:1974di,Perjes:1974ra,Perjes:1976sy,perjes1976evidence,hughston1976twistor,penrose1977twistor,perjes1979unitary,hughston1979twistors,hughston1980programme,perjes1982introduction,perjes1982internal}.
However, it rather turns out that
a massive particle can only be consistently described
with the bi-twistor system \cite{routh2015twistor,okano2017no-go}.
Hence we 
% suggest to
interpret the $\mathrm{SU}(2)$ internal symmetry of a bi-twistor
as a symmetry of a ``kinematic'' origin:
the massive little group.
Then the 
spinor dyads decomposing the massive momentum
\cite{newman1962approachspin,Perjes:1974ra}
are precisely what are nowadays
% conventionally
called the massive spinor-helicity variables
\cite{ahh2017,conde2016spinor,conde2016lorentz}.

At the level of free theory,
the bi-twistor model has been successfully formulated
as a Lagrangian or 
Hamiltonian system
\cite{tod1975dissertation,tod1976two,tod1977some,bette1984pointlike,Bette:1989zt,bette1997extended,fedoruk2003bitwistor,bette2000twistor,bette2004massive,bette2004massive04,bette2004massive05,bette2005twistors-0402150,bette2005two-0503134,fedoruk2007massive,lukierski2014noncommutative,fedoruk2014massive,de2015two,deguchi2016gauged,deguchi2018twistor,albonico2022twistor}.
In \cite{ambikerr0},
it was generalized to incorporate the Regge trajectory
and shown to be equivalent to
the Hanson-Regge spherical top \cite{Hanson:1974qy,bailey1975lagrangian},
which is
another spinning particle model 
that has been developed independently from the twistor side
while being the standard approach
in the study of effective theory for spinning gravitational objects
% \cite{gmoov,chkl2019,aoude2021classical,Chung:2019duq,vines2018scattering,vines2016canonical,levi2020effective,Levi:2015msa,porto2016effective,levi2015leading,marsat2015cubic,blanchet2014gravitational,perrodin2012subleading,levi2010spin-orbit,levi2010spin(1)-spin(2),porto2010next,porto2008spin,goldberger2007effective,goldberger2006effective,porto2006hyperfine,porto2006post}.
\cite{gmoov,chkl2019,aoude2021classical,Chung:2019duq,vines2018scattering,vines2016canonical,Levi:2019kgk,levi2020effective,Levi:2015msa,porto2016effective,levi2015leading,marsat2015cubic,blanchet2014gravitational,perrodin2012subleading,Levi:2011eq,levi2010spin-orbit,levi2010spin(1)-spin(2),porto2010next,porto2008spin,goldberger2007effective,goldberger2006effective,porto2006hyperfine,porto2006post}.
Therefore, a bi-twistor
rather describes a \textit{classical} spinning particle, carrying a spin angular momentum of magnitude a large multiple of $\hbar$.
In the first-quantized theory,
the bi-twistor particle 
becomes a ``universal spin machine''
that can prepare massive states of arbitrary spin,
producing the spectrum of 
\cite{ahh2017}'s massive spinning amplitudes
and 
deriving the mode expansion of massive higher-spin fields given in \cite{ochirov2022chiral}
\cite{ambikerr0,fedoruk2014massive,de2015two,deguchi2016gauged,deguchi2018twistor}.

It is the goal of our journey
% It is our goal in these series of papers
to establish an interacting theory of the massive twistor.
We start with reviewing 
the
details of the free theory.
% Below,
% we review
% essential details of the massive twistor model
% and introduce the notion of
% (projective) massive ambitwistor space.

\subsection{Free theory}
\label{sec:MA.free}

\paragraph{Massive twistor and dual twistor spaces}
The twistor space $\mathbb{T}$ is the carrier space of the fundamental representation of $\mathrm{SU}(2,2)$,
the conformal group of the $(1,3)$-signature flat space.
The massive twistor space $\mt$ is the product $\mathbb{T}\times\mathbb{T}$ endowed with an $\mathrm{SU}(2)$ ``little group'' symmetry that acts from the right as $Z_\rmA{}^I \mapsto Z_\rmA{}^J U_J{}^I$, where $Z_\rmA{}^I\in\mt$ and $U_I{}^J\in\mathrm{SU}(2)$.
The Hermitian metric of $\mathrm{SU}(2,2)$ identifies the conjugate and dual spaces of $\mt$ as $[Z_\rmB{}^I]^* A^{\bar{\rmB}\rmA} =: \bZ_I{}^\rmA \in\mt^*$, while the $\mathrm{SU}(2)$ representation is pseudo-real.

\paragraph{K\"ahler geometry}
$\mathbb{MT}$ is a complex 8-manifold equipped with a K\"ahler structure respecting its symmetry 
% ${\mathrm{Spin}^\mathrm{c}(2,4)\times\mathrm{SU}(2)} 
% \mem\cong\mem 
% (({\mathrm{SU}(2,2){\mem\times\mem}\mathrm{U}(1))/\mathbb{Z}_2)\times\mathrm{SU}(2)}$:
$
{\mathrm{Spin}(2,4)\times\mathrm{U}(2)} 
\mem\cong\mem 
{\mathrm{SU}(2,2){\mem\times\mem}\mathrm{U}(1)\times\mathrm{SU}(2)}$:
% The symplectic and complex structures are given as
\begin{align}
    \label{eq:free-symp}
    \theta^\circ\nem
    &=\hem {\textstyle\frac{i}{2}} \big(\bZ_I{}^\rmA dZ_\rmA{}^I - d\bZ_I{}^\rmA Z_\rmA{}^I\big)
    \qiq
    \omega^\circ\nem = d\theta^\circ\nem
    = i\mem d\bZ_I{}^\rmA \nem\wedge dZ_\rmA{}^I
    \,,\\
    \label{eq:free-complexstr}
    J \mem\hem
    &=\hem i\mem dZ_\rmA{}^I \otimes \frac{\partial}{\partial Z_\rmA{}^I} 
    - i\mem d\bZ_I{}^\rmA \otimes \frac{\partial}{\partial \bZ_I{}^\rmA}
    \,,\quad
    \bar{\partial} \alpha
    = 
    \frac{\partial\alpha_{\cdots}}{\partial \bZ_I{}^\rmA}
    \mem d\bZ_I{}^\rmA \swedge \cdots
    % \,,\\
    % \label{eq:free-metric}
    % &
    % \omega^\circ(J^\top(\blank),\blank)
    % = d\bZ_I{}^\rmA \nem\otimes dZ_\rmA{}^I + dZ_\rmA{}^I \nem\otimes d\bZ_I{}^\rmA
    % \,,
    \,.
\end{align}
% respectively. 
% With \eqref{eq:free-complexstr},
% $\mt$ and $\mt^*$ are the holomorphic and anti-holomorphic subspaces of $\ma$.
The Dolbeault operator in \eqref{eq:free-complexstr} introduces the notion of holomorphy of differential forms as $\bar{\partial}=0$.
The Poisson bracket $\{f,g\}^\circ := \omega^\circ{}^{-1}(df,dg)$
is given as
\begin{align}
    \label{eq:free-PB}
    \{Z_\rmA{}^I,Z_\rmB{}^J\}^\circ = 0
    \,,\quad
    \{Z_\rmA{}^I,\bZ_J{}^\rmB\}^\circ = -i\mem \delta_\rmA{}^\rmB\mem \delta_J{}^I
    \,,\quad
    \{\bZ_I{}^\rmA,\bZ_J{}^\rmB\}^\circ = 0
    \,.
\end{align}
The Noether charges of the $\mathrm{U}(1)$ and $\mathrm{SU}(2)$ symmetries are
\begin{align}
    \label{eq:Noether-U2}
    W_0 := -\minie\hem \bZ_I{}^\rmA Z_\rmA{}^I %= - p_m\mem {y}^m
    \,,\quad
    W_a := \minie\hem Z_\rmA{}^I (\s_a)_I{}^J \bZ_J{}^\rmA %= - |\Delta|\mem \L\hnem^a{}_m\mem {y}^m
    \,.
\end{align}

\paragraph{Infinity and chirality}
The two-component content of a twistor and a dual twistor is
\begin{align}
    \label{eq:Weylcontent}
    Z_{\rmA}{}^I = 
    \begin{pmatrix}
        \lambda_{\a}{}^I
        \\
        i\mu^{\da I}
    \end{pmatrix}
    \,,\quad
    \bar{Z}_I{}^{\rmA} =
    \begin{pmatrix}
        -i\bmu_I{}^{\a}
        &
        \rambda_{I\da}
    \end{pmatrix}
    \,,\quad
    A^{\bar{\rmB}\rmA} =
    % \begin{pmatrix}
    %     0 & {\delta^\db{}_\da\kern-0.16em}
    %     \\
    %     {\delta_\b{}^\a} & 0
    % \end{pmatrix}
    {\renewcommand{\arraycolsep}{0.2em}
    \begin{pmatrix}
        \hspace{0.81em}\mathclap{0}\hspace{0.7em}\mathclap{} &
        \hspace{0.7em}\mathclap{\delta^\db{}_\da\kern-0.16em}\hspace{0.7em}\mathclap{}
        \\
        \hspace{0.81em}\mathclap{\delta_\b{}^\a}\hspace{0.7em}\mathclap{} &
        \hspace{0.7em}\mathclap{0}\hspace{0.7em}\mathclap{}
    \end{pmatrix}}
    \,.
    % \,,
\end{align}
In this block basis,
the infinity twistors and the highest-rank Clifford algebra element $\gamma_\ast := \frac{1}{4!}\hem\ve_{mnrs}\gamma^m\gamma^n\gamma^s\gamma^r$ representing the Hodge star are given as
\begin{align}
    \label{eq:infinity-chirality}
    I^{\rmA\rmB} = 
    \begin{pmatrix}
        \e^{\a\b} & 0\hspace{0.1em}
        \\
        0 & 0\hspace{0.1em}
    \end{pmatrix}
    \,,\quad
    \bar{I}_{\rmA\rmB} = 
    \begin{pmatrix}
        \hspace{0.1em}0 & 0
        \\
        \hspace{0.1em}0 & \hem{-\be^{\da\db}}
    \end{pmatrix}
    \,,\quad
    (\gamma_\ast)_\rmA{}^\rmB =
    {\renewcommand{\arraycolsep}{0.02em}
    \begin{pmatrix}
        -i\delta_\a{}^\b & 0
        \\
        0 & +i\delta^\da{}_\db
    \end{pmatrix}}
    \,,
\end{align}
where $\ve_{0123} \doteq 1$.
These are invariant under infinity-fixing (i.e., Poincar\'e) and origin-fixing conformal transformations, respectively.
We introduce shorthand notations
\begin{align}
    \nonumber
    \label{eq:Delta-def}
    \textstyle
    \Delta := -\frac{1}{2}\mem I^{\rmA\rmB} Z_\rmA{}^I Z_\rmB{}^J \e_{IJ} = \det(\lambda)
    &\textstyle
    \,,\quad
    \bDelta := -\frac{1}{2}\mem \e^{IJ} \bZ_I{}^\rmA \bZ_J{}^\rmB \bar{I}_{\rmA\rmB} = \det(\rambda)
    \,,\\
    (\lambda^{-1})_I{}^\a := - \Delta^{-1} \e^{\a\b} \e_{IJ} \lambda_\b{}^J
    &
    \,,\quad
    (\rambda^{-1})^{\da I} := \bDelta^{-1} \be^{\da\db} \e^{IJ} \rambda_{\smash{J\db}\vphantom{I}}
    \,,\\
    \nonumber
    \textstyle
    \chi 
    % := \frac{i}{4}\mem \bZ_I{}^\rmA (1-i\gamma_\ast)_\rmA{}^\rmB Z_\rmB{}^I
    := \frac{i}{2}\mem \bZ_I{}^\rmA (P_+)_\rmA{}^\rmB Z_\rmB{}^I
    = -\frac{1}{2}\mem\rambda_{I\da}\mem \mu^{\da I}
    &
    \textstyle
    \,,\quad
    \bchi 
    % := -\frac{i}{4}\mem \bZ_I{}^\rmA (1+i\gamma_\ast)_\rmA{}^\rmB Z_\rmB{}^I
    := -\frac{i}{2}\mem \bZ_I{}^\rmA (P_-)_\rmA{}^\rmB Z_\rmB{}^I
    = -\frac{1}{2}\mem\bmu_I{}^\a \lambda_\a{}^I
    % \nonumber
    % \textstyle
    % \chi := \frac{1}{4}\mem \bZ_I{}^\rmA (\gamma_\ast{+\mem}i1)_\rmA{}^\rmB Z_\rmB{}^I
    % = -\frac{1}{2}\mem\rambda_{I\da}\mem \mu^{\da I}
    % &
    % \textstyle
    % \,,\quad
    % \bchi := \frac{1}{4}\mem \bZ_I{}^\rmA (\gamma_\ast{-\mem}i1)_\rmA{}^\rmB Z_\rmB{}^I
    % = -\frac{1}{2}\mem\bmu_I{}^\a \lambda_\a{}^I
    \,,
\end{align}
where $(P_\pm)_\rmA{}^\rmB := \minie (1 \mp i\gamma_*)_\rmA{}^\rmB$
are the right-hand (self-dual) and left-hand (anti-self-dual) projectors.
$\chi$ and $\bchi$ are not holomorphic/anti-holomorphic,
but
the only non-vanishing Poisson brackets between $\Delta$, $\bDelta$, $\chi$, $\bchi$ are still
the ``skew-diagonal'' ones:
% $\{\chi,\bDelta\}^\circ = \bDelta$ and $\{\bchi,\Delta\}^\circ = \Delta$.
\begin{align}
    \{\chi,\bDelta\}^\circ = \bDelta
    \,,\quad
    \{\bchi,\Delta\}^\circ = \Delta
    \,.
\end{align}
% \begin{align}
%     \label{eq:freePB-chiphi}
%     \{\chi,\Delta\}^\circ = 0
%     \,,\quad
%     \{\chi,\bDelta\}^\circ = \bDelta
%     \,,\quad
%     \{\bchi,\Delta\}^\circ = \Delta
%     \,,\quad
%     \{\bchi,\bDelta\}^\circ = 0
%     \,;\quad
%     \{\chi,\bchi\}^\circ = 0
%     \,,
% \end{align}

\paragraph{Fibration and ``hybrid'' description}
The complexified Minkowski space 
\cite{penrose1973twistor,penrose1967twistoralgebra,newman1973complex,Newman:1973yu,Newman:2002mk,newman2004maxwell,newman1974curiosity}
$\mathbb{CE}^{1,3}$ is the vector space $\mathbb{C}^4$ equipped with the 
metric $\eta_{mn} \doteq \mathrm{diag}(-1,+1,+1,+1)$
that has signature $(1,3)$ on the real section.
The massive twistor space $\mt$ fibers over $\mathbb{CE}^{1,3}$
as the trivial spin-frame bundle $\mathbb{CE}^{1,3}\times\mathbb{C}^4$.
The massive incidence relation can be thought of as describing such a fibration
and is invertible
in the sub-bundle
$\mathbb{CE}^{1,3}\times\mathrm{GL}(2,\mathbb{C})$
where the spin frames are restricted to be non-degenerate.
As a result, we are equipped with two coordinate systems,
``twistor'' $(Z,\bZ)$ for $\mt$
and ``hybrid'' \cite{fedoruk2007massive} $(\lambda,\rambda,z,\bz)$
for the sub-bundle:
% descriptions:
\begin{align}
\begin{split}
    \label{eq:incidence}
    \textstyle
    z^{\da\a} := (\mu\lambda^{-1})^{\da\a}
    \,,\quad
    \bz^{\da\a} := (\rambda^{-1}\hnem\bmu)^{\da\a}
    \quad\Longleftrightarrow\quad
    \mu^{\da I} 
    = 
    {z}^{\da\a} \lambda_\a{}^I
    \,,\quad
    \bmu_I{}^\a 
    = 
    \rambda_{I\da}\mem {\bz}^{\da\a}
    \,.
\end{split}
\end{align}
The symplectic potential, form, and the Poisson bivector appear in the hybrid basis as
\begin{align}
    \label{eq:free-symp-hybrid}
    \theta^\circ
    &= - \rambda_{I\da}\mem d{x}^{\da\a} \lambda_\a{}^I - 
    {i\big(
        \rambda_{I\da}\mem {y}^{\da\a} d\lambda_\a{}^I 
        - d\rambda_{I\da}\mem {y}^{\da\a} \lambda_\a{}^I
    \big)}
    \,,\\
    \nonumber
    \omega^\circ
    &= \rambda_{I\da}\mem {d{\bz}^{\da\a}\nem\nem \wedge d\lambda_\a{}^I} 
    - {d\rambda_{I\da} \wedge d{z}^{\da\a}} \lambda_\a{}^I 
    % - {d\rambda_{I\da} \wedge d{z}^{\da\a}} \lambda_\a{}^I 
    % + \rambda_{I\da}\mem {d{\bz}^{\da\a}\nem\nem \wedge d\lambda_\a{}^I} 
    - 2i\mem d\rambda_{I\da} \wedge {y}^{\da\a} d\lambda_\a{}^I
    \,,\\
    \nonumber
    \omega^\circ{}^{-1}
    &= (\rambda^{-1})^{\da I} \frac{\partial}{\partial \lambda_\a{}^I} \nem\wedge\nem \frac{\partial}{\partial {\bz}^{\da\a}}
    - (\lambda^{-1})_I{}^\a \frac{\partial}{\partial {z}^{\da\a}\vphantom{{\bz}^{\da}}} \nem\wedge\nem \frac{\partial}{\partial \rambda_{I\da}}
    - 2i{y}^{\da\b} (\rambda^{-1}\nem\lambda^{-1})^{\db\a}
    \frac{\partial}{\partial {z}^{\da\a}\vphantom{{\bz}^{\da}}} 
    \nem\wedge\nem \frac{\partial}{\partial {\bz}^{\db\b}}
    \,.
\end{align}
By the vector fields $\partial/\partial\lambda$, $\partial/\partial\rambda$, $\partial/\partial{z}$, $\partial/\partial{\bz}$, we always refer to the $(\lambda,\rambda,{z},{\bz})$ basis.
The real structure that $\mathbb{CE}^{1,3}$ inherits from $\mt$ defines real coordinates $x^m := \frac{1}{2}(z^m {\mem+\mem} \bz^m)$, $y^m$ $:= \frac{1}{2i}(z^m {\mem-\mem} \bz^m)$.
The non-vanishing $\{z^{\da\a},\bz^{\da\a}\}^\circ = -2iy^{\da\b} (\rambda^{-1}\nem\lambda^{-1})^{\db\a}$ then boils down to
\begin{align}
\begin{split}
    \label{eq:xy-nc}
    \textstyle
    \{{x}^m,{x}^n\}^\circ = \{{y}^m,{y}^n\}^\circ = S^{mn}\hnem/\Delta\bDelta
    \,,\quad
    \{{x}^m,{y}^n\}^\circ = 
    2({y}^{(m}p^{n)} - \minie\hem \eta^{mn} y {\mem\cdot\mem} p)
    /\Delta\bDelta
    \,,
\end{split}
\end{align}
where $S^{mn}$ and $p_m$ are defined in \eqref{eq:solve-constraints}.
Therefore, the real spacetime $\mathbb{E}^{1,3}$ (the real section $y^m=0$ of $\mathbb{CE}^{1,3}$) is noncommutative upon canonical quantization.
This motivates us to regard holomorphic
% or anti-holomorphic
geometric objects as physically more fundamental than the real ones.

\paragraph{Spin-space-time}

At this point,
let us
introduce a new terminology
``\textit{spin-space-time}.''
In general, 
% the ``spin-space-time'' of a particle is
it refers to
a complex manifold
in which
spacetime is embedded as a real section and
deviations from the real section describe the spin length.
Such a notion of a complex manifold was first introduced by
Newman
\cite{newman1974curiosity,newman1974collection,Newman:2002mk,newman2004maxwell,Newman:1973yu,newman1973complex,Newman:1965my-kerrmetric,Newman:1965tw-janis}.
He called it ``complex spacetime,''
but
% we would like to
let us 
introduce a new term 
% ``spin-space-time''
to distinguish it from a mere complexification of spacetime
while emphasizing its physical semantics that the imaginary directions describe spin.
As we have just described in \eqref{eq:xy-nc},
the spin-space-time
of a free spinning particle in special relativity
is the complexified Minkowski space
of twistor theory \cite{newman1974curiosity}.

\paragraph{Spherical top interpretation}
The twistor space $\mathbb{T}$
can be viewed as a spinorial representation of a massless spinning
particle's constrained phase space in which the mass-shell constraint is explicitly solved in terms of spinor-helicity variables \cite{penrose1973twistor,shirafuji1983lagrangian}.
Similarly, the massive twistor space $\mt$ provides a spinorial rephrasing of a massive spinning particle's degrees of freedom \cite{Penrose:1974di,Perjes:1974ra,fedoruk2014massive},
yet not only solving the mass-shell constraint
with 
the spinorial frame variables
but also the spin supplementary condition \cite{ambikerr0}.

The equivalence between
the massive twistor model and the Hanson-Regge spherical top
is best illustrated by
rewriting the symplectic potential \eqref{eq:free-symp-hybrid}.
Up to a term, it becomes the symplectic potential
of 
the covariantly gauge-fixed Hanson-Regge spherical top:
\begin{align}
    \label{eq:solve-constraints}
    &
    % \bigg\{\,
    {\renewcommand{\arraystretch}{1.1}
    \renewcommand{\arraycolsep}{0em}
    \begin{array}{rl}
        p_{\a\da} := - \lambda_\a{}^I \rambda_{I\da}
        &\,,\quad
        % (\L\hnem^{-1})^a{}_{\a\da} := |\Delta|^{-1} \lambda_\a{}^I (\s^a)_I{}^J \rambda_{J\da}
        \L\hnem^{\da\a}{}_a := -|\Delta|\mem (\rambda^{-1})^{\da I} (\s_a)_I{}^J\hhnem (\lambda^{-1})_J{}^\a
        %%% option 1: \L\hnem^{\da\a}{}_A = -|\Delta|\mem (\rambda^{-1})^{\da I} (\s_A)^J{}_I (\lambda^{-1})_J{}^\a, (\s_0)_I{}^J = -\delta_I{}^J
        %%% option 2: take (\s_0)_I{}^J = \delta_I{}^J.
        \,,\\
        S^{mn} := \ve^{mnrs}\mem {y}_r p_s
        % S^m{}_n := \ve^{ms}{}_{nr}\mem {y}^r p_s
        &\,,\quad
        (\L\hnem^{-1})^a{}_{\a\da} :=
        |\Delta|^{-1}\hem
        \lambda_\a{}^I
        \hhnem (\s^a)_I{}^J \hem
        \rambda_{J\da}
        \,,
    \end{array}}
    \\
    \label{eq:theta-free-HR}
    &
    \iq
    \theta^\circ\nem 
    \,=\,
    % &
    p_m\hem d{x}^m \nem\hnem + \minie\mem S^n{}_m\mem d\L\hnem^m{}_a (\L\hnem^{-1})^a{}_n
    % + W_0 \mem{d\psi}
    + W_0\mem\hhem d\big(\hem{ 
        {\textstyle\frac{i}{2}}\hnem
        \log(\bDelta/\Delta)
    }\hhnem\big)
    % \,,\quad
    % % (\L\hnem^{-1})^a{}_{\a\da} 
    % % := |\Delta|^{-1} \lambda_\a{}^I\hnem (\s_a)_I{}^J\hem \rambda_{J\da}
    % (\L\hnem^{-1})^a{}_m 
    % := \eta_{mn}\mem\hhnem \L\hnem^n{}_b \mem\delta^{ba}
    \,,
    % \,.
\end{align}
where $(\s_a)_I{}^J$, ${a=1,2,3}$, are the $\mathrm{SU}(2)$ sigma matrices.
Note that $(\L\hnem^{-1})^a{}_m 
= \eta_{mn}\mem\hhnem \L\hnem^n{}_b \mem\delta^{ba}$.
% so that
% $(\L\hnem^{-1})^a{}_m\mem \L^m{}_b = \delta^a{}_b$,
% $\L^m{}_a\mem (\L\hnem^{-1})^a{}_n = \delta^m{}_n + p^m p_n / |\Delta|^2$.

We emphasize that,
unlike in \cite{ambikerr0},
$p_m$, $\L\hnem^m{}_a$, $S_{mn}$ are
simply \textit{defined}
% exactly 
as the above in this paper.
They
are the particle's momentum, body frame, and spin angular momentum, whereas $x^m$ describes the particle's Cartesian position
in spacetime.
Quotienting $\mt$ by the $\mathrm{U}(1)$ action of $\{\blank,W_0\}^\circ$,
% (which translates to shifting the angle variable ${\textstyle\frac{i}{2}}\hnem \log(\bDelta/\Delta)$),
one obtains a $15$-dimensional submanifold
coordinatized by $p_m$, $\L\hnem^m{}_a$, $x^m$, $y^m$.
With these variables,
the parameterization \eqref{eq:solve-constraints} of the spherical top
``almost'' solves the covariant spin condition.
Fixing the value of $W_0 = - p_m\mem {y}^m$ then finally eliminates the redundancy ${y}^m \sim {y}^m + k p^m$ inherent in $S_{mn} = \ve_{mnrs}\mem {y}^r p^s$.
% so that the time evolution is uniquely determined by the real mass-shell constraint.
As a result, we are left with a $14$-dimensional constrained phase space, which we denote as $\mt_\diamond$.

\paragraph{Massive ambitwistor space}
So far, $Z_\rmA{}^I$ and $\bZ_I{}^\rmA$ have denoted the holomorphic and anti-holomorphic coordinates of $\mt$.
Meanwhile,
$\mt$ can also be obtained by
first considering
the product space $\ma := \mt\times\mt^* \ni (Z,\bZ)$
where $Z_\rmA{}^I$ and $\bZ_\rmA{}^I$ are \textit{independent complex variables}
and then imposing the Lorentzian-signature reality condition $[Z_\rmB{}^I]^* A^{\bar{\rmB}\rmA} =: \bZ_I{}^\rmA$ later.
Conversely, geometric structures in $\mt$,
such as \eqref{eq:free-symp}-\eqref{eq:free-PB},
can be promoted to $\ma$
by dropping the reality condition.
Then the spin frames $\lambda_\a{}^I$, $\rambda_{I\da}$ in \eqref{eq:Weylcontent} describe complexified massive kinematics,
and the holomorphic and anti-holomorphic subspaces of $\mathbb{CE}^{1,3}$
are promoted to separate complex vector spaces
so that $z^m$ and $\bz^m$ are not related by complex conjugation.
We call this space $\ma$
the massive ambitwistor space.

Further, from the definition of the projective (massless) ambitwistor space \cite{Mason:2005kn,geyer2014ambitwistor,geyer2016ambitwistor},
\begin{align}
    \mathbb{A}_\diamond
    \,:=\, \Big\{\hem{
        (Z,\bZ) {\,\in\,} 
        \mathbb{T}{\mem\times\mem}\mathbb{T}^*
        \,\Big|\,
        \bZ^\rmA Z_\rmA = 0
    }\mem\hem\Big\}
    \Big/
    (
        {Z_\rmA}\mem\partial/\partial{Z_\rmA}
        - {\bZ^\rmA}\partial/\partial{\bZ^\rmA}
    )
    \,,
\end{align}
it is natural to define the ``projective massive ambitwistor space'' as
\begin{align}  
    \label{eq:PMAdef}
    \mathbb{MA}_\diamond
    \,:=\, \Big\{\hem{
        (Z,\bZ) {\,\in\,}
        \mathbb{MT}{\mem\times\mem}\mathbb{MT}^*
        \,\Big|\,
        W_0 = 0
    }\mem\hem\Big\}
    \Big/ \{\blank,W_0\}^\circ
    \,.
\end{align}
Then the constrained phase space $\mt_\diamond$, which we described earlier,
is the real section of $\ma_\diamond$.
We 
% work in
will assume
this complexified setting from now.
% From now, we work in this complexified setting,
% leaving the imposition of the reality condition implicit.
By abuse of terminology, we continue to use 
% the non-complexified 
terms such as
``holomorphic'' and ``anti-holomorphic'' 
as well as
``complex conjugate,'' ``real-valued,'' etc.,
omitting the premise ``when restricted on the real section.''

Some twistor particle models have interpreted a certain $\mathrm{U}(1)$ subgroup of the $\mathrm{U}(2)$ internal symmetry of a bi-twistor as the electromagnetic gauge group
\cite{newman1974curiosity,tod1975dissertation,Bette:1989zt,bette2000twistor,bette2004massive,bette2004massive04,bette2005twistors-0402150,bette2004massive05,bette2005two-0503134,fedoruk2007massive,deguchi2016gauged}.
However, 
in our case,
% we clarify that
the internal $\mathrm{U}(2)$ is purely of a ``kinematic'' origin.
% However, 
% we clarify that
% the internal $\mathrm{U}(2)$ is purely of a ``kinematic'' origin
% in our case.
$W^a$ are the body frame components of the Pauli-Lubanski pseudovector,
associated with the massive little group.
$W_0$ is the redundancy inherent in formulating the spin length as a Lorentz-covariant (pseudo)vector:
\begin{align}
    \label{eq:PauliLubanski}
    W_0
    = - p_m\mem {y}^m
    \,,\quad
    W^a
    = - (\lambda\hem\s^a\rambda)_{\a\da}\hem y^{\da\a} 
    = -|\Delta| (\L\hnem^{-1})^a{}_m\hem y^m
    \,.
\end{align}
The physical spin length pseudovector is given by the ``spatial'' projection
$-(y^m {\mem-\mem\mem} p^m\hem p{\mem\cdot\mem}y/p^2) = \L\hnem^m{}_a\mem W^a/|\Delta| =: a^m$.
% \begin{align}
%     -(y^m {\mem-\mem\mem} p^m\hem p{\mem\cdot\mem}y/p^2)     \mem=\mem
%     \L\hnem^m{}_a\mem W^a/|\Delta| 
%         \mem=:\mem
%     a^m
%     \,.
% \end{align}
Note that this ambiguity of $y^m$ is precisely that of the impact parameter.
As already indicated in \eqref{eq:PMAdef},
we find the gauge slice $W_0 = 0$ particularly preferable.

\paragraph{Free theory time evolution}
In \cite{ambikerr0}, 
the massive twistor 
% spinning particle 
model is generalized
to incorporate the Regge trajectory---a function 
$m(\vec{W}^2)$ of the spin-squared $\vec{W}^2 := W^a W_a$
that controls the mass of the spinning particle \cite{Hanson:1974qy,steinhoff2015spin}.
Taking $(m(\vec{W}^2) - \bDelta)$ and $(m(\vec{W}^2) - \Delta)$
as Hamiltonian constraints
with Lagrange multipliers $\kappa$ and $\bar{\kappa}$,
the equation of motion reads
\begin{align}
    % \frac{d}{d\s}
    d/d\s
    = 
    \kappa \{\blank, m(\vec{W}^2) {\mem-\mem} \bDelta\}^\circ
    + \bar{\kappa} \{\blank, m(\vec{W}^2) {\mem-\mem} \Delta\}^\circ
    \,,\quad
    \bDelta = m(\vec{W}^2) = \Delta
    \,.
\end{align}
%% "d/d\s real" ~ "\s real"
The proposed gauge-fixing
% condition 
$\dot{W}_0 {\mem=\mem} 0$
is invariant under $x^m$-translations,
conforms to the reality requirement 
($f{\mem\hem\in\mem\mem}\mathbb{R} {\mem\implies\mem} df\hnem/d\s\in\mathbb{R}$ after imposing the reality condition),
and puts $\kappa {\mem-\mem} \bar{\kappa} = 0$ on the mass-shell constraint surface.
% $\Delta = m(\vec{W}^2) = \bDelta$.
The remaining one degree of freedom, $(\kappa {\mem+\mem} \bar{\kappa})/2$, accounts for the reparameterization of the worldline.
We use a ``constant einbein'' gauge:
\begin{align}
    \label{eq:constant-einbein}
    &
    (\kappa {\mem+\mem} \bar{\kappa})\mem d\s =: d\tau
    \,,\\
    \label{eq:eom-free}
    &
    \iq
    % \frac{d}{d\t}
    d/d\tau
    = 
    \minime \{\blank,\bDelta{\mem+\mem}\Delta\}^\circ
    +
    % 2m'\hnem(\vec{W}^2) W^a \{\blank,W_a\}^\circ
    \Omega^a \{\blank,W_a\}^\circ
    \,,\quad
    \Omega^a := 2m'\hnem(\vec{W}^2) W^a
    \,.
\end{align}
% \begin{align}
%     \label{eq:eom-free}
%     {-2}((\kappa + \bar{\kappa})/2)\mem d\s \mem=:\mem d\tau
%     \qiq
%     d/d\tau
%     = 
%     \minime \{\blank,\bDelta{\mem+\mem}\Delta\}^\circ
%     +
%     2m'\hnem(\vec{W}^2) W^a \{\blank,W_a\}^\circ
%     \,.
% \end{align}
Unfixing the gauge is easy.
This $\t$ coincides with the proper time for solutions to \eqref{eq:eom-free}.

\paragraph{Color phase space}
\label{par:color-ps}

For describing the {\Kerr} particle,
we will sometimes assume
an extension of the phase space
by color degrees of freedom.
The color phase space can be
modeled as a fermionic phase space $\mathbb{C}^{0|N}$
(or either $\mathbb{C}^{N}$ for a bosonic model).
Let $\psi^i$, $\bpsi_i$ be 
its holomorphic and anti-holomorphic coordinates.
% These essentially are worldline restrictions of
% the fermion fields:
% holomorphic and anti-holomorphic sections of a vector bundle over spacetime whose typical fiber is $\mathbb{C}^{0|N}$.
The free theory is governed by the K\"ahler form $i\mem d\bpsi_i \mem{\wedge}\, d\psi^i$,
and the color phase space 
enjoys
the $\mathrm{U}(N)$ symmetry.

A subgroup $G$ of the $\mathrm{U}(N)$ can be gauged by coupling to spacetime gauge fields
while the remaining color degrees of freedom are suppressed by
% introducing 
Lagrange multipliers
(worldline gauge fields).
This means that the equation of motion \eqref{eq:eom-free}
gets appended by terms from a set of Hamiltonian constraints.
For example,
$(n - \bpsi_i \psi^i) = 0$
reduces $\mathrm{U}(N)$ to
$G = \mathrm{SU}(N)$.
% \footnote{
%     (Semiclassical) quantization further demands $n$
%     to be an integer.
% }

Let $(t_a)^i{}_j$ be anti-Hermitian generators of $G$ such that
$(t_a t_b)^i{}_i = -\frac{1}{2}\mem \delta_{ab}$ and
$([t_a,t_b])^i{}_j = (t_c)^i{}_j\hem f^c{}_{ab}$.
The color charge of the particle 
in the adjoint representation
is given by $q_a := i\mem \bpsi_i (t_a)^i{}_j \psi^j$,
and it follows 
from $\{\psi^i,\bpsi_j\}^\circ = -i\mem \delta^i{}_j$
that
$\{q_a,q_b\}^\circ = q_c\mem f^c{}_{ab}$,
so
% so that 
the Hamilto\-nian flows of $q_a$ 
realize $G$.
The Lagrange multipliers disappear
in the $\dot{q}_a$ equation of motion.

With this understanding,
we derive equations of motion
while gauging the whole group for simplicity:
$G {\,=\,} \mathrm{U}(N)$.
It is straightforward to generalize the results
to $G {\,\neq\,} \mathrm{U}(N)$.

\subsection{Coupling to background fields}
\label{sec:MA.coupling}

Now, we start to move on to the interacting theory.

\paragraph{Symplectic perturbation theory}

Following \cite{spt},
we understand interactions
as \textit{perturbations on the symplectic structure}
while retaining the \textit{same Hamiltonian}.
This
idea traces back to 
Souriau \cite{souriau1970structure,torrence1973gauge,souriau1974modele,sniatycki1974prequantization,sternberg1978classical}
and also to Feynman \cite{dyson1990feynman}.
The phase space
is equipped with two symplectic forms
$\omega^\circ$ and $\omega = \omega^\circ + \omega'$,
the former defining the free theory
and the latter defining the interacting theory.
The key equation is that
the perturbed Poisson bracket 
$\{f,g\} = \omega^{-1}(df,dg)$
is given by a geometric series expansion
\begin{align}
\begin{split}
    \label{eq:pert1-series}
    \omega^{-1} 
    \mem&=\mem
    \omega^\circ{}^{-1}
    - \omega^\circ{}^{-1} {\,\omega'\mem} \omega^\circ{}^{-1}
    +  \omega^\circ{}^{-1} {\,\omega'\mem} \omega^\circ{}^{-1}  {\,\omega'\mem} \omega^\circ{}^{-1}
    - \cdots
    \,,\\
    \{f,g\}
    \mem&=\mem
    \{f,g\}^\circ 
    - \{f,\zeta^i\}^\circ \, \omega'_{ij} \, \{\zeta^j,g\}^\circ
    + \{f,\zeta^i\}^\circ \, \omega'_{ij} \, \{\zeta^j,\zeta^k\}^\circ \, \omega'_{kl} \, \{\zeta^l,g\}^\circ
    - \cdots
    \,,
\end{split}
\end{align}
where $\zeta^i$ denote local coordinates 
such that $\omega' = \minie\mem \omega'_{ij}\mem d\zeta^i \swedge d\zeta^j$.
See \cite{spt} for various examples concerning familiar physical systems.
Here, we directly jump to the massive ambitwistor
without
% covering preliminary examples.
a warm-up.
The massive ambitwistor space
is now granted a new symplectic form given by ``\eqref{eq:free-symp} plus perturbation,''
$\omega = \omega^\circ + \omega'$.

\paragraph{Assumptions on the interactions}

Various interactions can be systematically classified
in the symplectic perturbation language
as if classifying interaction Hamiltonians.
First of all, we restrict our attention to interactions that ``decouple'' from the internal $\mathrm{U}(2)$:
\begin{subequations}
    \label{eq:decouple-W}
\begin{align}
% \begin{split}
    \label{eq:decouple-W0}
    \omega'(\blank,\{\blank,W_0\}^\circ) = 0
    &\quad\iff\quad
    \{f,W_0\} = \{f,W_0\}^\circ
    \,,\\
    \label{eq:decouple-Wa}
    \omega'(\blank,\{\blank,W_a\}^\circ) = 0
    &\quad\iff\quad
    \{f,W_a\} = \{f,W_a\}^\circ
    \,.
% \end{split}
\end{align}
\end{subequations}
Quotienting $\ma$ by the $\mathrm{U}(2)$ group action of $\{\blank,W_0\}^\circ$ and $\{\blank,W_a\}^\circ$,
one finds that the symplectic perturbation $\omega'$ should be composed only of $dp_m$, $dz^m$, $d\bz^m$.
If the phase space gets extended by some additional degree of freedom, say $Q$,
then $\omega'$ can also involve $dQ$,
provided that $Q$ has vanishing free-theory Poisson bracket with 
% $p_m$, $z^m$, $\bz^m$.
$W_0$ and $W_a$.

The first condition \eqref{eq:decouple-W0}
states that $\omega'$ reduces down to 
the projective massive ambitwistor space
$\ma_\diamond$.
% It is compulsory because 
% we identify 
% $W_0$ as a gauge degree of freedom.
It is compulsory, as 
we identify $\{\blank,W_0\}^\circ$ as a gauge direction.
On the other hand,
the second condition \eqref{eq:decouple-Wa}
is an optional assumption that simplifies our discussion
and can be violated by having a $d\L\hnem^m{}_a$ component in $\omega'$.
It implies that $W^a$ continues to be conserved in the interacting theory as in the free theory
so that the precession of $\L\hnem^m{}_a$ and $y^m$ are synchronized.\footnote{
    For example,
    $\omega' = d(\minie\mem q\ell^2 F_{mn}\hnem(z,\bz)\hem \L\hnem^m{}_a\hem d\L\hnem^n{}_b\mem \delta^{ab})$
    leads to a torque 
    $\smash{q\ell^2 (-\dot{\vec{B}})} = q\ell^2(\vec{\partial}{\mem\times\mem}\vec{E})$
    % $q\ell^2(\vec{\partial}{\mem\times\mem}\vec{E})$
    % \smash{$q\ell^2 (-\dot{\vec{B}})$}
    in the momentarily co-moving frame of the particle:
    torque from the induced electric field,
    due to a finite extension $\ell$ of the body.
    In a regime where the body is effectively ``rigid,'' 
    a sensible candidate
    for the length scale $\ell$
    will be
    % $\ell = \sqrt{\hem \pi \hnem\big/ 2m(\smash{\vec{W}^2})\hem m'(\smash{\vec{W}^2}) \hem}$,
    $\ell \sim (\hem \pi \hnem/\hem 2m(\smash{\vec{W}^2})\hem m'\hnem(\smash{\vec{W}^2}) ){\kern-0.07em}^{1/2}$,
    given a Regge trajectory $m(\vec{W}^2)$.
}
% Then the particle is no longer ``point-like''
% in the sense that an additional characteristic length scale other than $\hbar/|p|$ and $|\vec{W}|/|p|$ is introduced.\footnote{
%     In this sense, black holes are point-like (despite their spin-induced multipoles), neutron stars are not.
%% (a BH really resides in complex ST?)
% }
It
% \eqref{eq:decouple-Wa}
leads to a weaker condition,
\begin{align}
    \label{eq:decouple-WW}
    \omega'(\blank,\{\blank,\vec{W}^2\}^\circ) = 0
    &\quad\iff\quad
    \{f,\vec{W}^2\} = \{f,\vec{W}^2\}^\circ
    \,,
\end{align}
which implies that the torque that external fields exert on the body
does not alter the rotational kinetic energy.
In the language of three-point amplitudes,
this means that
% one is
we are
restricting to the equal-mass sector.

% To sum up,
% \eqref{eq:decouple-W0}, \eqref{eq:decouple-Wa}, and \eqref{eq:decouple-WW} respectively implies that
% $W_0$, $W^a$, and $\vec{W}^2$ are constants of motion.

\paragraph{Interacting theory time evolution}
By the very idea of symplectic perturbation theory,
the Hamiltonian equation of motion 
% reads
is given by the same Hamiltonian constraints
$(m(\vec{W}^2) - \bDelta)$
and 
$(m(\vec{W}^2) {\mem-\mem} \Delta)$,
but the bracket is different.
Provided the requirement \eqref{eq:decouple-W0},
our reparameterization gauge fixing in the free theory equally applies to the interacting theory:
\begin{align}
    \label{eq:eom-int}
    d/d\t
    = \minime \{\blank,\bDelta{\mem+\mem}\Delta\}
    + \Omega^a\{\blank,W_a\}^\circ
    \,.
\end{align}
In virtue of \eqref{eq:decouple-WW},
one simply needs to add the Regge evolution term \textit{of the free theory} 
to the constant-mass equation of motion.
Hence, for simplicity,
we often assume constant mass when we derive equations of motion.
The Regge term
% , of course,
generates internal $\mathrm{SU}(2)$ rotation:
\begin{align}
    \label{eq:free-internalmotion}
    \{\blank,W_a\}^\circ
    &=
    % &=
    % \frac{1}{2i} \nem \left(\nem
    %     (Z \s_a)_\rmA{}^I \mem \frac{\partial}{\partial Z_\rmA{}^I}
    %     - (\s_a \bZ)_I{}^\rmA \mem \frac{\partial}{\partial \bZ_I{}^\rmA}
    % \right)
    \frac{1}{2i} \nem \left(\hhhnem{
        (\lambda \hem \s_a)_\a{}^I \hhem \frac{\partial}{\partial \lambda_\a{}^I}
        - (\s_a \rambda)_{I\da} \mem \frac{\partial}{\partial \rambda_{I\da}}
    }\hem\right)
    \,.
\end{align}
One can easily check that
$W_0$, $W^a$, $\Omega^a$, $m(\vec{W}^2)$, etc.
are constants of motion under \eqref{eq:eom-int}.

\subsection{A derivation of the Newman-Janis shift}
% \subsection{Deriving the Newman-Janis shift from the massive ambitwistor space}
\label{sec:Flat.o}

Having described general aspects of
% coupling the massive ambitwistor particle to background fields,
the interacting theory,
let us give a more detailed look on 
the implications of 
applying
symplectic perturbation theory 
to the massive ambitwistor space
and provide an overview of the next section.

% \paragraph{The ``zig-zag logic''}

% Various phase spaces were considered in the work \cite{spt}, 
% but perhaps 
Symplectic perturbation theory 
perhaps
becomes 
the most interesting in K\"ahler manfiolds.
The symplectic structure of a K\"ahler manifold is $(1,1)$.
(We 
% gave
have given
a nickname ``zig-zag'' to such a $(1,1)$ property.)
This implies that
holomorphic (anti-holomorphic) coordinates 
remain 
% to be
Poisson-commutative
under holomorphic (anti-holomorphic)
perturbations on the symplectic form.
In particular,
for K\"ahler vector spaces,
% such as the massive twistor space,
holomorphic (anti-holomorphic) perturbations on the symplectic form leave the holomorphic (anti-holomorphic) subspace as a Lagrangian submanifold.\footnote{
    Note that
    symplectic perturbations can also be regarded as deformations of the complex structure if one retains the K\"ahler metric of the free theory.
}
% A ``zig-zig'' perturbation induces ``zag-zag'' bracket;
% a ``zag-zag'' perturbation induces ``zig-zig'' bracket.
This ``zig-zag logic'' applies to the massive twistor space,
as it is a K\"ahler vector space.

Meanwhile,
holomorphy and chirality are inherently linked
% in the twistor space:
in twistor theory:
the left-handed and right-handed
% spin frames,
spinor-helicity variables,
$\lambda$ and $\rambda$,
% $\lambda_\a{}^I$ and $\rambda_{I\da}$,
are contained respectively in the holomorphic and anti-holomorphic twistor variables,
% $Z_\rmA{}^I$ and $\bZ_I{}^\rmA$.
$Z$ and $\bZ$.
When combined with the ``zig-zag logic,''
this feature of the twistor space 
has a remarkable physical implication.

Suppose
a massive ambitwistor particle
under a constant Regge trajectory
is
put in a self-dual background.
If the coupling is minimal,
the left-handed spin frame should be parallel-transported.
Think of gravity:
if the left-handed spinor bundle is flat,
$\dot{\lambda}_\a{}^I$ should be
nothing but zero
in the gauge where the connection coefficients vanish
because ``\textit{gravity is geometry}.''\footnote{
    For electromagnetism or Yang-Mills,
    one can 
    % we
    appeal to the ``double copy'' relationship
    between the Thomas-Bargmann-Michel-Telegdi (TBMT) equation and the Mathisson-Papapetrou-Tulczyjew-Dixon (MPTD) equation,
    described in Appendix \ref{app:BMT-MPD}.
    In particular, see \eqref{eq:prec-eq.frames} and
    \eqref{eq:frames-BMT}-\eqref{eq:frames-MPD}.
}
% \footnote{
%     A premise is that we gauge-fix the bundle 
%     such that the connection coefficients vanish.
%     % Employing the ``covariant'' symplectic perturbation scheme
%     % we introduce in our follow-up paper \cite{ambikerr2}
%     % avoids this caveat.
%     This caveat, the gauge fixing,
%     is avoided in
%     the ``covariant'' symplectic perturbation scheme
%     we introduce in 
%     % our follow-up paper 
%     \cite{ambikerr2}.
%     Note the two versions of ``double copy'' 
%     of spin precession equations of motion,
%     % that can be drawn between the TBMT and MPTD equations,
%     explained in Appendix \ref{app:BMT-MPD}.
% }
For $\dot{\lambda}_\a{}^I = 0$ to be implemented
as the Hamiltonian equation of motion
% (Hamiltonian flow generated by $\minime\mem (\Delta {\mem+\mem} \bDelta)$),
(i.e., $\{\lambda_\a{}^I,
    \bDelta {\,+\,} \Delta
\} = 0$),
the brackets $\{\lambda_\a{}^I,\lambda_\b{}^J\}$ and $\{\lambda_\a{}^I,\rambda_{\smash{J\db}\vphantom{\b}}\}$ should vanish.
The zig-zag structure
% of the massive twistor space
then asserts that
the spin-space-time part of the symplectic perturbation
should have $(2,0)$ components only.
Therefore, the Newman-Janis shift 
is reborn 
% as a geometric prescription of minimal coupling: 
from the massive ambitwistor space
as a geometric prescription of minimal coupling:
self-dual field strength extends holomorphically to the spin-space-time
\cite{Newman:1965my-kerrmetric,Newman:1965tw-janis,Newman:2002mk,monteiro2014black,adamo2014kerr}.

In short, 
the Newman-Janis shift
% \cite{Newman:1965my-kerrmetric,Newman:1965tw-janis,Newman:2002mk,monteiro2014black}
can be 
% thought of
regarded as
% as a derived fact
% that traces back to 
a fact that derives from
the zig-zag nature of the massive ambitwistor space,
provided that we 
% identify $\dot{\lambda}_\a{}^I = 0$
% as the minimal spin precession behavior
% of a massive spinning particle.
take the fact that
$\dot{\lambda}_\a{}^I = 0$
specifies the minimal spin precession behavior
of a massive spinning particle
as an input.
% This could be argued from the equivalence principle,

Let us elaborate on how 
a 
``zig-zig'' perturbation $dz \swedge dz$ induces
the ``zag-zag'' bracket $\{\rambda,\rambda\}$.
Recall first
how the 
% Poisson
brackets of a 
charged
scalar particle
% coupled to electromagnetism
follow from
the expansion
\eqref{eq:pert1-series}.
Through the free theory's 
% bracket
$\{x^m,p_n\}^\circ {\mem\,=\,\mem} \delta^m{}_n$,
coupling 
to
$\omega' = \minie\mem \phi_{mn}\hnem(x)\mem dx^m \swedge dx^n$
leads to
the bracket
% $\{p_{\a\da},p_{\smash{\b\db}\vphantom{\da}}\} =
% \phi_{\smash{\a\da\b\db}}\hhnem(x)
% $
$\{p_m,p_n\} = \phi_{mn}\hnem(x)$
between ki\-netic momenta (generators of gauge-covariant translations)
in the interacting theory:
$\{p_m,p_n\} = - \{p_m,x^r\}^\circ \,\phi_{rs}\hnem(x)\mem \{x^s,p_n\}^\circ$.
% 
% For the massive ambitwistor,
In the same way,
for the spinning particle,
coupling to
a holomorphic perturbation
% $\omega' = \minie\mem 
% \phi_{\smash{\a\da\b\db}\vphantom{\b}}\hhnem(z)\mem
% % qF_{\smash{\a\da\b\db}\vphantom{\b}}\hnem(z)\mem
% \wedgetwo{z}{z}{\da\a}{\db\b}$
$\omega' = \minie\mem \phi_{mn}\hnem(z)\mem dz^m \swedge dz^n$
leads to
$\{\lambda,\lambda\} = 0$,
$\{\lambda,\rambda\} = 0$,
$\{\rambda,\rambda\} \sim \phi$
through the non-vanishing 
free theory
bracket $\{\rambda,z\}^\circ$:
$\{\rambda_{I\da},\rambda_{\smash{J\db}\vphantom{I\da}}\} = - \{\rambda_{I\da},z^r\}^\circ \,\phi_{rs}\hnem(z)\mem \{z^s,\rambda_{\smash{J\db}\vphantom{I\da}}\}^\circ$.\footnote{
    Note that
    % These brackets
    the brackets 
    $\{\lambda,\lambda\}$,
    $\{\lambda,\rambda\}$,
    $\{\rambda,\rambda\}$
    are
    % indeed
    ``square roots'' of $\{p_{\a\da},p_{\smash{\b\db}\vphantom{\da}}\}$,
    as
    $p_{\a\da} = -\lambda_\a{}^I \rambda_{I\da}$.
    % Hence,
    % Accordingly, 
    The spinor-helicity variables decompose
    the kinetic momentum,
    not the canonical momentum.
}

Going further,
we can expect that 
the particle will 
get 
non-minimally coupled
% exhibit non-minimal spin precession behaviors
if the 
spin-space-time part of the
self-dual field strength involves
$dz \swedge d\bz$ and $d\bz \swedge d\bz$ components as well:
the left-handed spin frame 
is not parallel transported even if the background is left-flat.
It remains to question
how big the 
% set
space
of 
% coupling constants 
Wilson coefficients
that such a non-holomorphic analytic continuation of the self-dual field strength
% can describe.
can cover
is.

In the next section, 
we verify these 
% prescriptions for minimal and non-minimal couplings
claims and expectations
by computing equations of motion 
and further amplitudes.
Before that, let us introduce 
% some
a few
terminologies.

\paragraph{Spin-space-time part of the symplectic perturbation}

The criteria \eqref{eq:decouple-W}
allows having $dp_m$, $d\psi^i$, $d\bpsi_i$
as well as spin-space-time components $dz^m$, $d\bz^m$
in 
% the symplectic perturbation.
$\omega'$.
Meanwhile, the non-spin-space-time components
can be ignored 
for obtaining spin frame equations of motion
and computing amplitudes.
Hence, in the next section,
we consider $\omega'$ of the form
\begin{align}
    \label{eq:earthomega'}
    \omega'
    {}&{}=\,
    {^{(2,0)}\omega'_{mn}}\mem
        (\minie\mem \wedgetwo{z}{z}{m}{n})
    + 
    {^{(1,1)}\omega'_{mn}}\mem
        \wedgetwo{z}{\bz}{m}{n}
    +
    {^{(0,2)}\omega'_{mn}}\mem
        (\minie\mem \wedgetwo{\bz}{\bz}{m}{n})
    \,,
\end{align}
regardless of its closure.

We assume that a spacetime two-form
$\phi = \minie\mem \phi_{mn}\hnem(x)\mem dx^m \swedge dx^n$
is given as a background field,
and 
% the symplectic perturbation $\omega'$ on the massive ambitwistor space
\eqref{eq:earthomega'}
arises from it.
We give the nickname ``field strength'' to $\phi$.
While ${^{(2,0)}\omega'_{mn}}$ and ${^{(0,2)}\omega'_{mn}}$
are antisymmetric by construction,
${^{(1,1)}\omega'_{mn}}$
can have both antisymmetric and symmetric components.
However, the symmetric component vanishes on the real section
as $dx^{(m} \swedge dx^{n)} = 0$
and cannot be 
% simply
generated 
from the antisymmetric ``field strength'' $\phi$
at linear order.
% and cannot be interpreted as a field strength 
% extended from real spacetime.
Therefore,
up to linear order in $\phi$ 
we can say that
all of the components
in \eqref{eq:earthomega'}
are antisymmetric
and thus can be split into self-dual and anti-self-dual parts.
% Hence, it is possible to split them
% into self-dual and anti-self-dual parts.
As a result, we rewrite \eqref{eq:earthomega'} as
$\omega' = \omega'^+ + \omega'^- + \mathcal{O}(\phi^2)$,
where
\begin{subequations}
    \label{eq:arbomega'}
\begin{align}
    \label{eq:arbomega'+}
    \omega'^+
    {}&{}=\,
    {^0\omega'^+_{mn}}\mem
        (\minie\mem \wedgetwo{z}{z}{m}{n})
    + 
    {^1\omega'^+_{mn}}\mem
        \wedgetwo{z}{\bz}{m}{n}
    +
    {^2\omega'^+_{mn}}\mem
        (\minie\mem \wedgetwo{\bz}{\bz}{m}{n})
    \,,\\
    \label{eq:arbomega'-}
    \omega'^-
    {}&{}=\,
    {^0\omega'^-_{mn}}\mem
        (\minie\mem \wedgetwo{\bz}{\bz}{m}{n})
    + 
    {^1\omega'^-_{mn}}\mem
        \wedgetwo{z}{\bz}{m}{n}
    +
    {^2\omega'^-_{mn}}\mem
        (\minie\mem \wedgetwo{z}{z}{m}{n})
    % \,,
\end{align}
\end{subequations}
% where 
% all of the components
% ${^0\omega'^\pm_{mn}}$,
% ${^1\omega'^\pm_{mn}}$,
% ${^2\omega'^\pm_{mn}}$
% are antisymmetric,
% ${^0\omega'^+_{mn}}$,
% ${^1\omega'^+_{mn}}$,
% ${^2\omega'^+_{mn}}$
% are self-dual,
% and 
% ${^0\omega'^-_{mn}}$,
% ${^1\omega'^-_{mn}}$,
% ${^2\omega'^-_{mn}}$
% are anti-self-dual.
such that
% where
${^0\omega'^+_{mn}}$,
${^1\omega'^+_{mn}}$,
${^2\omega'^+_{mn}}$
are self-dual
and 
${^0\omega'^-_{mn}}$,
${^1\omega'^-_{mn}}$,
${^2\omega'^-_{mn}}$
are anti-self-dual.
It is reasonable to assume that
$\omega'^\pm = \mathcal{O}(\phi^\pm{}^1)$,
where $\phi^\pm$ are the self-dual and anti-self-dual parts of $\phi$.
The $\mathcal{O}(\phi^2)$ part in 
$\omega' = \omega'^+ + \omega'^- + \mathcal{O}(\phi^2)$
contains terms like $\phi^+\phi^-$.

\paragraph{Heavenly\,vs.\,Earthly}

For $\omega'$ to be 
a real two-form,
$\omega'^+$ and $\omega'^-$ should be complex conjugate
to each other.
% Ignoring the $\mathcal{O}(\phi^2)$ term,
\begin{align}
    \label{eq:earth}
    \omega'
    \mem=\mem
    \omega'^+ + \omega'^-
    + \mathcal{O}(\phi^2)
    \,,\quad
    \omega'^-\nem = [\omega'^+]^*
    \quad\text{(earthly perturbation)}
    \,.
\end{align}
While the fields we experience in our ``real'' macroscopic world are described by 
symplectic structures
like \eqref{eq:earth}
that reduce to a real two-form upon imposing the $(1,3)$-signature reality condition,
it is also physically meaningful to consider purely self-dual or anti-self-dual configurations
\cite{shaviv1975general,newman1976heaven,penrose1976nonlinear,penrose1976curvedtwistor,hansen1978metric,ko1981theory,penrose1992h,plebanski1975some,Plebanski:1977zz,belavin1975pseudoparticle,hawking1977gravitational}.
To this end, we allow complex symplectic perturbations,
which should not come as a surprise because we have already complexified $\omega^\circ$.
Following Newman \cite{shaviv1975general,newman1976heaven} and Pleba\'nski \cite{plebanski1975some,Plebanski:1977zz},
we call purely self-dual or anti-self-dual complexified cases ``heavenly'' and the real case ``earthly'':
\begin{align}
\begin{split}
    \label{eq:heaven}
        \omega'
        &{}={}\mem
        \omega'^+
        \quad\text{(heavenly perturbation, self-dual)}
        \,,\\
        \omega' 
        &{}={}\mem
        \omega'^-
        \quad\text{(heavenly perturbation, anti-self-dual)}
        % \,,
        \,.
\end{split}
% \\
%     \label{eq:earth}
%     \omega'
%     &{}={}\mem
%     \omega'^+\nem + \omega'^-
%     \,,\quad
%     \omega'^-\nem = [\omega'^+]^*
%     \quad\text{(earthly perturbation)}
%     \,.
\end{align}
We 
% will
present heavenly and earthly 
% symplectic 
perturbations at once
by describing only the self-dual part $\omega'^+$.
Our main focus is on the heavenly case
for both conceptual and practical reasons:
a) the geometry of a single (``nonlinear'') field quantum
is given as a heavenly configuration
\cite{penrose1976nonlinear,penrose1976curvedtwistor},
% so it is natural to study particle scattering in heavenly backgrounds, 
and
b) earthly equations easily follow from
the corresponding heavenly equations
by taking the ``$\hem2\hem\Re$'' value
if one is only interested in linear perturbative order in
% $\omega'$.
$\phi$.

%
%%[1A]%%%%%%%%%%%%%%%%%%%%%%%%%%%%%%%%%%%%%%%%%%%%

\section{From Symplectic Perturbations to Amplitudes}
\label{sec:Flat}

% % In this section,
% We study
% universal implications of relating chirality and holomorphy
% in the complexified Minkowski space
% % in $\mathbb{CE}^{1,3}$
% on the linear-order perturbative physics
% of the massive twistor particle.

\subsection{Minimal equation of motion}
\label{sec:Flat.eom-min}

\paragraph{Linking self-duality with holomorphy}

First, we consider the case where
the self-dual symplectic perturbation is given by a holomorphic two-form in
the spin-space-time:
% the complexified Minkowski space:
% $\mathbb{CE}^{1,3}$:
\begin{align}
    \label{eq:GMOOVform}
    \omega'^+
    &={}\mem
    \phi^+_{mn}\hnem(z)\mem (\minie\mem dz^m \swedge dz^n)
    \mem=\mem
    \phi_{\smash{\da\db}\vphantom{\b}}(z)\mem 
    \e_{\a\b}(\wedgez{\da\a}{\db\b})
    \quad\text{(minimal coupling)}
    \,.
    % \,,
\end{align}
% according to the argument given in the previous section.
In terms of \eqref{eq:arbomega'},
this has vanishing
${^1\omega'^\pm_{mn}}$
and
${^2\omega'^\pm_{mn}}$.
The argument given in the previous section
claims that \eqref{eq:GMOOVform} describes the minimal coupling of the particle to the background.\footnote{
    For another motivation for relating chirality and holomorphy,
    consider
    the Ward correspondence 
    \cite{ward1977selfdual,atiyah1977instantons,witten2004perturbative,huggett1994introduction,ward1991twistor}
    in twistor theory.
    It
    % which
    derives self-dual gauge fields on spacetime from holomorphic vector bundles over the projective twistor space.
    Note that it is natural to identify the complexified Minkowski space of massless and massive theories.
    % The gist of the Ward correspondence is that translations along
    % a self-dual
    % null plane are holonomy-free for self-dual field configurations.
    % A similar argument can be made in the massive setting by computing the $\{c^I\rambda_{I\da},c^J\rambda_{\smash{J\db}\vphantom{\da}}\}$ bracket with a constant spinor $c^I$ under a symplectic perturbation of the form \eqref{eq:GMOOVform}.
    The gist of the Ward correspondence is that translations along
    an anti-self-dual
    null plane are holonomy-free for self-dual field configurations.
    The same argument can be made in 
    % the 
    % massive
    % % massive particle
    % % current
    our
    setting by computing the bracket 
    $\{\eta^\a p_{\a\da}, \eta^\b p_{\smash{\b\db}\vphantom{\da}}\}$ 
    with a constant spinor $\eta^\a$ under a symplectic perturbation of the form \eqref{eq:GMOOVform}.
}

\paragraph{Earthly Poisson bracket}

Let us consider the earthly case first.
Applying \eqref{eq:pert1-series}, the perturbed Poisson bracket reads
(we ignore the potentially existent $\mathcal{O}(\phi^2)$ part of $\omega'$)
\begin{align}
    \label{eq:pert1-zigzag}
{\renewcommand{\arraystretch}{1.1}
\renewcommand{\arraycolsep}{0em}
\begin{array}{rl}
    \{f,g\} 
    \mem=\mem \{f,g\}^\circ
    \mem
    &{}- \{f,{z}^r\}^\circ {\,\phi^+_{rs}({z})\mem} \{{z}^s,g\}^\circ
    + \{f,{\bz}^m\}^\circ {\,\phi^- _{mn}\hnem({\bz})\mem}
    \{{\bz}^n,{z}^r\}^\circ {\,\phi^+_{rs}({z})\mem} \{{z}^s,g\}^\circ
    \\
    &{}- \{f,{\bz}^r\}^\circ {\,\phi^- _{rs}({\bz})\mem} \{{\bz}^s,g\}^\circ
    + \{f,{z}^m\}^\circ {\,\phi^+_{mn}\hnem({z})\mem}
    \{{z}^n,{\bz}^r\}^\circ {\,\phi^- _{rs}({\bz})\mem} \{{\bz}^s,g\}^\circ
    \\
    &{}+ \cdots
    \,,
\end{array}
}
\end{align}
provided \eqref{eq:earth} and \eqref{eq:GMOOVform}.
% assuming \eqref{eq:earth} and \eqref{eq:GMOOVform}.
% Observe that the skew-diagonalization is lost from the first perturbative order 
% so that
% $\mathbb{MT}$ and $\mathbb{MT}^*$ are no longer Lagrangian submanifolds.
The holomorphic/anti-holomorphic skew-diagonalization is lost, and $\mathbb{MT}$ and $\mathbb{MT}^*$ are no longer Lagrangian submanifolds.

\paragraph{Zig-zag diagrams}

To reduce
bulkiness in writing these equations down,
we devise a handy graphical notation motivated by the alternating 
% ``\textit{$z$ig-$\bz$ag}'' 
``$zig\text{-}\bz ag$''
pattern of $\phi^+\hnem(z)$ and $\phi^- \hnem(\bz)$ in \eqref{eq:pert1-zigzag}.
For example, the Hamiltonian vector field generated by $\bDelta$ reads
\begin{align}
\begin{split}
    \label{eq:pert1-zigzag-Delta}
    \{\blank,\bDelta\}
    \mem=\mem
    \{\blank,\bDelta\}^\circ
    &- \{\blank,{z}^r\}^\circ {\,\phi^+_{rs}({z})\mem} \{{z}^s,\bDelta\}^\circ
    \\
    &
    + \{\blank,{\bz}^m\}^\circ {\,\phi^- _{mn}\hnem({\bz})\mem}
    \{{\bz}^n,{z}^r\}^\circ {\,\phi^+_{rs}({z})\mem} \{{z}^s,\bDelta\}^\circ
    + \cdots
    \,.
\end{split}
\end{align}
In the graphical notation, \eqref{eq:pert1-zigzag-Delta} appears as \eqref{eq:X-bDelta},
\begin{subequations}
\begin{align}
    \label{eq:X-bDelta}
    \{\blank,\bDelta\}
    &\,\,=\,\mem
    \includegraphics[valign=c]{figs/l.pdf}
    \includegraphics[valign=c]{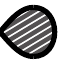}
    -
    \includegraphics[valign=c]{figs/l-b.pdf}
    \includegraphics[valign=c]{figs/v+.pdf}
    \includegraphics[valign=c]{figs/xD-b.pdf}
    +
    \includegraphics[valign=c]{figs/l.pdf}
    \includegraphics[valign=c]{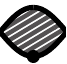}
    \includegraphics[valign=c]{figs/v+.pdf}
    \includegraphics[valign=c]{figs/xD-b.pdf}
    -
    \includegraphics[valign=c]{figs/l-b.pdf}
    \includegraphics[valign=c]{figs/v+.pdf}
    \includegraphics[valign=c]{figs/v-.pdf}
    \includegraphics[valign=c]{figs/v+.pdf}
    \includegraphics[valign=c]{figs/xD-b.pdf}
    +
    \cdots
    \,,\\
    \label{eq:X-Delta}
    \{\blank,\Delta\}
    &\,\,=\,\mem
    \includegraphics[valign=c]{figs/l-b.pdf}
    \includegraphics[valign=c]{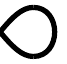}
    -
    \includegraphics[valign=c]{figs/l.pdf}
    \includegraphics[valign=c]{figs/v-.pdf}
    \includegraphics[valign=c]{figs/xD.pdf}
    +
    \includegraphics[valign=c]{figs/l-b.pdf}
    \includegraphics[valign=c]{figs/v+.pdf}
    \includegraphics[valign=c]{figs/v-.pdf}
    \includegraphics[valign=c]{figs/xD.pdf}
    -
    \includegraphics[valign=c]{figs/l.pdf}
    \includegraphics[valign=c]{figs/v-.pdf}
    \includegraphics[valign=c]{figs/v+.pdf}
    \includegraphics[valign=c]{figs/v-.pdf}
    \includegraphics[valign=c]{figs/xD.pdf}
    +
    \cdots
    \,,
\end{align}
\end{subequations}
while complex conjugation ``flips'' \eqref{eq:X-bDelta} to \eqref{eq:X-Delta}.
The rule is that each graphical element corresponds to 
a differential-geometric object.
The ``external legs'' and ``vertices'' are
\begin{align}
\begin{split}
    \label{eq:elemz-vert}
    \omega^\circ{}^{-1}(\blank,d\Delta)
    \,=\,  
    \includegraphics[valign=c]{figs/l-b.pdf}
    \includegraphics[valign=c]{figs/xD.pdf}
    &\,,\quad
    \minie\mem \phi^+_{mn}\hnem(z)\mem \wedgetwo{z}{z}{m}{n}
    \,=\,  
    \includegraphics[valign=c]{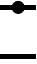}\hem
    \,,
    \\
    \omega^\circ{}^{-1}(\blank,d\bDelta)
    \,=\,  
    \includegraphics[valign=c]{figs/l.pdf}
    \includegraphics[valign=c]{figs/xD-b.pdf}
    &\,,\quad
    \minie\mem \phi^- _{mn}\hnem(\bz)\mem \wedgetwo{\bz}{\bz}{m}{n}
    \,=\,  
    \includegraphics[valign=c]{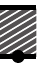}\hem
    \,,
\end{split}
\end{align}
while color-flipping ``propagators'' represent the skew-diagonalized free Poisson bivector:
\begin{align}
    \label{eq:elemz-prop}
    \omega^\circ{}^{-1}(\blank,\blank)
    \quad = \quad  
    \hem
    \includegraphics[valign=c]{figs/z-bz.pdf}
    % \quad\text{or}\quad
    \,\,+\,\,
    \includegraphics[valign=c]{figs/bz-z.pdf}
    \,\,.
\end{align}
Juxtaposing these elements
\eqref{eq:elemz-vert}-\eqref{eq:elemz-prop}
in a line 
by 
attaching
``zigs'' (holomorphic, white) 
to zigs
and ``zags'' (anti-holomorphic, black)
to zags
% while
% alternating between the ``zigs'' (holomorphic, white) and the ``zags'' (anti-holomorphic, black)
gives a valid ``zig-zag diagram'' (inspired by Penrose's idea ``zig-zag electron'' \cite{penrose2005road}).
One can imagine a diagram as a ribbon and say that it should always twist once before ``interacting'' with an ``operator insertion'' $\omega'^\pm$.
One side of the ribbon is white;
the other side is black.
By 
% enumerating all the zig-and-zag contractions of the elements \eqref{eq:elemz-vert}-\eqref{eq:elemz-prop},
% i.e., 
``summing over zig-zags,''
we can intuitively figure out 
how the time-evolution generators $\bDelta$ and $\Delta$ ``propagate'' to the holomorphic and anti-holomorphic sectors.

In this notation,
the Poisson bivector
$\omega^{-1}$
is given by the sum of the components
of the following block matrix:\footnote{
% of a block matrix\footnote{
    Our initial motivation for borrowing the term ``zig-zag'' 
    from the zig-zag electron
    was that this block matrix resembles Fig.\,25.2 of \cite{penrose2005road}.
}
\begin{align}
    \nem\nem
    \left(\nem
    {\renewcommand{\arraycolsep}{0.5em}
    \renewcommand{\arraystretch}{1.5}
    \begin{array}{l|l}
            -
            \includegraphics[valign=c,scale=1.0]{figs/l.pdf}
            \includegraphics[valign=c,scale=1.0]{figs/v-.pdf}
            \includegraphics[valign=c,scale=1.0]{figs/r.pdf}
            +
            \includegraphics[valign=c,scale=1.0]{figs/l.pdf}
            \includegraphics[valign=c,scale=1.0]{figs/v-.pdf}
            \includegraphics[valign=c,scale=1.0]{figs/v+.pdf}
            \includegraphics[valign=c,scale=1.0]{figs/v-.pdf}
            \includegraphics[valign=c,scale=1.0]{figs/r.pdf}
            - \cdots
        &
            \includegraphics[valign=c,scale=1.0]{figs/z-bz.pdf}
            -
            \includegraphics[valign=c,scale=1.0]{figs/l.pdf}
            \includegraphics[valign=c,scale=1.0]{figs/v-.pdf}
            \includegraphics[valign=c,scale=1.0]{figs/v+.pdf}
            \includegraphics[valign=c,scale=1.0]{figs/r-b.pdf}
            + \cdots
            \phantom{\displaystyle\frac{}{0_{0_0}}}
        \\\hline
            \includegraphics[valign=c,scale=1.0]{figs/bz-z.pdf}
            -
            \includegraphics[valign=c,scale=1.0]{figs/l-b.pdf}
            \includegraphics[valign=c,scale=1.0]{figs/v+.pdf}
            \includegraphics[valign=c,scale=1.0]{figs/v-.pdf}
            \includegraphics[valign=c,scale=1.0]{figs/r.pdf}
            + \cdots
        &
            -
            \includegraphics[valign=c,scale=1.0]{figs/l-b.pdf}
            \includegraphics[valign=c,scale=1.0]{figs/v+.pdf}
            \includegraphics[valign=c,scale=1.0]{figs/r-b.pdf}
            +
            \includegraphics[valign=c,scale=1.0]{figs/l-b.pdf}
            \includegraphics[valign=c,scale=1.0]{figs/v+.pdf}
            \includegraphics[valign=c,scale=1.0]{figs/v-.pdf}
            \includegraphics[valign=c,scale=1.0]{figs/v+.pdf}
            \includegraphics[valign=c,scale=1.0]{figs/r-b.pdf}
            - \cdots
    \end{array}
    }
    \nem\right)
    .
    \nem
\end{align}
\phantom{.}\\[-1.5\baselineskip]

\paragraph{Heavenly Poisson bracket}
Next, consider the heavenly case.
The
% zig-zag
expansions truncate at linear order
if we have the zig vertex only.
The Poisson bracket is \textit{exactly} given as
\begin{align}
    \label{eq:omegainv-heavenly}
    \omega^{-1}
    =
    \includegraphics[valign=c]{figs/l.pdf}
    \includegraphics[valign=c]{figs/r-b.pdf}
    +
    \includegraphics[valign=c]{figs/l-b.pdf}
    \includegraphics[valign=c]{figs/r.pdf}
    -
    \includegraphics[valign=c]{figs/l-b.pdf}
    \includegraphics[valign=c]{figs/v+.pdf}
    \includegraphics[valign=c]{figs/r-b.pdf}
    \,.
\end{align}
The holomorphic massive twistor space $\mt$ remains to be a Lagrangian submanifold,
\begin{align}
    \label{eq:PB-heavenly-unaltered}
    \{Z_\rmA{}^I,Z_\rmB{}^J\} = 0
    \,,\quad
    \{Z_\rmA{}^I,\bZ_J{}^\rmB\} = -i\mem \delta_J{}^I \delta_\rmA{}^\rmB
    \,,
\end{align}
while the anti-holomorphic $\mt^*$ becomes Poisson-noncommutative as
% \begin{align}
%     \label{eq:PB-heavenly-candy}
%     \{\bZ_I{}^\rmA,\bZ_J{}^\rmB\} 
%     &=
%     % -
%     % d\bZ_I{}^\rmA
%     % \includegraphics[valign=c]{figs/l-b.pdf}
%     % \includegraphics[valign=c]{figs/v+.pdf}
%     % \includegraphics[valign=c]{figs/r-b.pdf}
%     % d\bZ_J{}^\rmB
%     % =
%     \e_{IJ}\,
%     \phi^{\rmA\rmB}(Z)
%     \,,\quad
%     \phi^{\rmA\rmB}(Z)
%     :=
%     (2/\Delta)\mem
%     % \frac{2}{\Delta}\mem
%     \phi_{\smash{\dc\dd}\vphantom{\b}}(z)\mem
%     \rho^{\dc\rmA} \rho^{\dd\rmB}
%     \,,
% \end{align}
\begin{align}
    \label{eq:PB-heavenly-candy}
    \{\bZ_I{}^\rmA,\bZ_J{}^\rmB\} 
    &=
    \frac{2}{\Delta}\mem
    \e_{IJ}\mem
    \phi_{\smash{\dc\dd}\vphantom{\b}}(z)\mem
    \rho^{\dc\rmA} \rho^{\dd\rmB}
    \,,
\end{align}
where $\rho^{\dc\rmA}$ is the ``converter'' 
that follows from
pulling $dz^{\da\a}$ back from $\mathbb{CE}^{1,3}$ to $\mt$:
\begin{align}
\begin{split}
    \label{eq:hingedef}
    \rho^{\dc\rmA}
    :=
    \begin{pmatrix}
        -iz^{\dc\a} & \delta^\dc{}_\da
    \end{pmatrix}
    % \,,\quad
    % \lho_\rmA{}^\c
    % :=
    % \begin{pmatrix}
    %     \delta_\a{}^\c
    %     \\
    %     i\bz^{\da\c}
    % \end{pmatrix}
    \qiq
    % \quad\implies
    % \\
    &
    dz^{\da\a} = -i (\rho\mem dZ\lambda^{-1})^{\da\a}
    % \,,\quad
    % d\bz^{\da\a} = i (\rambda^{-1} d\bZ\lho)^{\da\a}
    \,,\quad
    \rho^{\dc\rmA} Z_\rmA{}^I=0
    % \,,\quad
    % \bZ_I{}^\rmA \lho_\rmA{}^\c =0
    %%%%%%
    % \,,\\
    % &
    % \omega'^+\hnem(z)
    % = 
    % % -\frac{1}{\Delta}\mem
    % \omega'_{\smash{\dc\dd}\vphantom{\b}}(z)\mem
    % \big({
    %     - \Delta^{-1}
    %     \rho^{\dc\rmA}\rho^{\dd\rmB}
    %     dZ_\rmA{}^I\nem\nem\wedge dZ_\rmB{}^J
    %     \e_{IJ}
    % }\big)
    % % \,,\quad
    % % \bS^{\c\d} 
    % % =
    % % +\frac{1}{\bDelta}\mem
    % % \e^{IJ}
    % % d\bZ_I{}^\rmA\nem\nem\wedge d\bZ_J{}^\rmB
    % % \lho_\rmA{}^\c\lho_\rmB{}^\d
    %%%%%%
    \,.
\end{split}
\end{align}
For reference, we unpack \eqref{eq:PB-heavenly-candy} in the hybrid basis:
\begin{align}
\begin{split} 
    \label{eq:PB-heavenly-rambda-bz}
    \{\rambda_{I\da},\rambda_{\smash{J\db}\vphantom{\da}}\}
    &= (2/\Delta)\mem \e_{IJ} 
    \mem\phi_{\smash{\da\db}\vphantom{\b}}(z)
    \,,\\
    \{\rambda_{I\da},\bz^{\db\b}\}
    &= (4i/\Delta)\mem 
    (\rambda^{-1})^{\db J}\mem
    \e_{IJ}\mem
    \phi_{\smash{\da\dd}\vphantom{\b}}(z)\mem y^{\dd\b}
    \,,\\
    \{\bz^{\da\a},\bz^{\db\b}\} 
    &= (8/\Delta\bDelta)\mem \be^{\da\db} \phi_{\smash{\dc\dd}}(z)\mem y^{\dc\a} y^{\dd\b}
    \,.
\end{split}
\end{align}
From the candy-shaped diagram
$\includegraphics[valign=c]{figs/l-b.pdf}
\includegraphics[valign=c]{figs/v+.pdf}
\includegraphics[valign=c]{figs/r-b.pdf}$,
one intuitively sees that
the ``zig'' symplectic perturbation induces Poisson non-commutativity on the ``zag'' massive twistor space.
For the particle to be minimally coupled,
the self-dual field strength should only appear in $\{\rambda_{I\da},\rambda_{\smash{J\db}\vphantom{I\da}}\}$ among the spinorial frame brackets,
and hence we find \eqref{eq:GMOOVform}.
% that \eqref{eq:GMOOVform} describes the minimal coupling.

\paragraph{Heavenly equation of motion}
Further, let us check the minimal spin precession behav\-ior by deriving the equation of motion with \eqref{eq:GMOOVform}.
For the constant-mass case,
\eqref{eq:eom-int} gives
\begin{align}
    \label{eq:eom-heavenly}
    \mem \frac{d}{d\t}
    &= \frac{1}{-2}\hnem \left(
        \includegraphics[valign=c]{figs/l.pdf}
        \includegraphics[valign=c]{figs/xD-b.pdf}
        -
        \includegraphics[valign=c]{figs/l-b.pdf}
        \includegraphics[valign=c]{figs/v+.pdf}
        \includegraphics[valign=c]{figs/xD-b.pdf}
    \right)
    +
    \frac{1}{-2}\hnem \left(
        \includegraphics[valign=c]{figs/l-b.pdf}
        \includegraphics[valign=c]{figs/xD.pdf}
    \right)
    \,.
\end{align}
$\bDelta$ and $\Delta$ generate holomorphic and anti-holomorphic translations in the free theory:
\begin{align}
\begin{split}
    \label{eq:holomorhictranslations}
    &
    \{\blank,\bDelta\}^\circ
    =
    \includegraphics[valign=c]{figs/l.pdf}
    \includegraphics[valign=c]{figs/xD-b.pdf}
    =
    % = 
    % -i \e^{IJ} \bZ_J{}^\rmB \bar{I}_{\rmA\rmB} \frac{\partial}{\partial Z_\rmA{}^I}
    \bDelta (p^{-1})^{\da\a} \frac{\partial}{\partial{z}^{\da\a}}
    \,,\quad
    \{\blank,\Delta\}^\circ
    =
    \includegraphics[valign=c]{figs/l-b.pdf}
    \includegraphics[valign=c]{figs/xD.pdf}
    =
    % =
    % -i I^{\rmA\rmB} Z_\rmB{}^J \e_{IJ} \frac{\partial}{\partial \bZ_I{}^\rmA}
    \Delta (p^{-1})^{\da\a} \frac{\partial}{\partial{\bz}^{\da\a}}
    \,.
\end{split}
\end{align}
The interaction with 
% the self-dual background
$\omega'^+$
performs a
local
chiral (complexified) Lorentz transformation:
\begin{align}
    \label{eq:heavenly-3ptX}
    -
    \includegraphics[valign=c]{figs/l-b.pdf}
    \includegraphics[valign=c]{figs/v+.pdf}
    \includegraphics[valign=c]{figs/xD-b.pdf}
    &= \frac{-2}{\Delta}\hnem
    \left(
        \rambda_{\smash{I\db}\vphantom{\b}}\mem \phi^\db{}_\da(z)\hem
        \frac{\partial}{\partial\rambda_{I\da}}
        +
        2i\mem \phi^\da{}_\db(z)\mem y^{\db\a}
        \frac{\partial}{\partial\bz^{\da\a}}
    \right)
    \,.
    % \kappa \bDelta(p^{-1})^{\da\a} \frac{\partial}{\partial z^{\da\a}}
    % + \bar{\kappa} \Delta(p^{-1})^{\da\a} \frac{\partial}{\partial \bz^{\da\a}}
\end{align}
Note that $2i\mem\hem \partial/\partial \bz^{\da\a}|_z = -\partial/\partial y^{\da\a}|_z$.
As $\mt$ is Poisson-commutative, the zig sector of the particle gets decoupled from the interaction:
\begin{align}
\begin{split}
    \label{eq:eom-heavenly-hol}
    \dot{Z}_\rmA{}^I
    &=
    \minime
    \left\langle
        dZ_\rmA{}^I
        ,
        \includegraphics[valign=c]{figs/l.pdf}
        \includegraphics[valign=c]{figs/xD-b.pdf}
    \right\rangle
    \,,\\
    \dot{\bZ}_I{}^\rmA
    &= 
    \minime
    \left\langle
        d\bZ_I{}^\rmA
        ,
        \includegraphics[valign=c]{figs/l-b.pdf}
        \includegraphics[valign=c]{figs/xD.pdf}
    \right\rangle
    +
    \minime
    \left\langle
        d\bZ_I{}^\rmA
        ,
        - \includegraphics[valign=c]{figs/l-b.pdf}
        \includegraphics[valign=c]{figs/v+.pdf}
        \includegraphics[valign=c]{figs/xD-b.pdf}
    \right\rangle
    \,.
\end{split}
\end{align}
The angle bracket here represents the contraction between one-forms and vectors.
The linear coupling in the $\dot{\bZ}_I{}^\rmA$ equation of motion encodes all changes in the dynamics due to $\omega'$,
while $\omega'$ need not be ``small.''
Such linearization or exact solvability is indeed the characteristic of heavenly physics.

\paragraph{Minimal spin precession}
From \eqref{eq:holomorhictranslations}-\eqref{eq:eom-heavenly-hol},
we find
%% \delta\kappa and \delta\bar{\kappa} eom in the ps action has put \Delta = m_0 and \bDelta = m_0.
\begin{align}
\begin{split}
    \label{eq:eom-heavenly-components}
    \dot{\lambda}_\a{}^I = 0
    &\,,\quad
    \textstyle
    \dot{\rambda}_{I\da} = \frac{1}{m_0}\mem \rambda_{\smash{I\db}\vphantom{I}} \, \phi^\db{}_\da(z)
    \,,\\
    \dot{z}^{\da\a} = \minime m_0(p^{-1})^{\da\a}
    &\,,\quad
    \textstyle
    \dot{\bz}^{\da\a}
    = \minime m_0(p^{-1})^{\da\a}
    + \frac{1}{m_0}\mem
    2i\mem\phi^\da{}_\db(z)\mem y^{\db\a}
    \,.
\end{split}
\end{align}
In turn, it follows that the 
% (``external'')
precession of the right-handed spin frame $\rambda_{I\da}$ and the precession of spin length pseudovector $y^{\da\a}$ are \textit{exactly} synchronized:
% assuming constant Regge trajectory:
\begin{align}
    \label{eq:eom-heavenly-xy}
    i\Big({\hem
        \dot{x}^{\da\a} - \minime m_0(p^{-1})^{\da\a} 
    }\Big)
    \,=\, \dot{y}^{\da\a}
    \mem=\, 
    \textstyle
    -\frac{1}{m_0}\mem 
    \phi{}^\da{}_\db(z)\mem y^{\db\a}
    \,.
\end{align}
Thus, we conclude that the prescription \eqref{eq:GMOOVform} ``universally'' leads to the minimal spin precession behavior,
regardless of what specific field $\phi$ describes!

\begin{figure}[t]
    \centering
    \includegraphics[scale=1.25, valign=c]{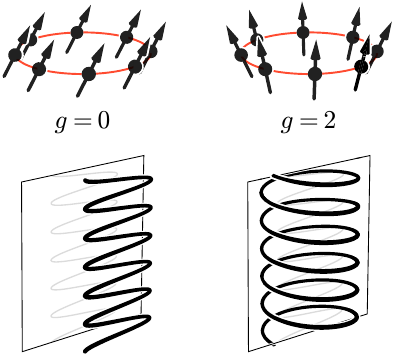}
    \caption{
        Geometric characterization of the gyromagnetic ratio $g$.
        A charged spinning particle exhibits both orbital and spin precession in an external ``magnetic'' field.
        If $g{\mem=\mem}2$, the two precessions are locally synchronized;
        such a motion appears as an isotropic spiral trajectory as seen in the particle's spin-space-time.
        In contrast, $g{\mem=\mem}0$ translates to a ``flattened'' sinusoidal curve.
        % Spin-space-time offers a new angle on spin precession.
        % The visual reasoning suggests that the field strength should somehow extend into the complex.
        Notice how spin-space-time offers a new angle on spin precession.
        The visual reasoning suggests that the field strength should somehow extend into the complex.
        The isotropy of $g{\mem=\mem}2$ points to the equal blending of all $dx$-$dy$ components in $dz \swedge dz$;
        the ``flatness'' of $g{\mem=\mem}0$ points to the reality of $dx \swedge dx$.
    }
    \label{fig:g=0-g=2}
\end{figure}

\paragraph{Earthly equation of motion and \texorpdfstring{$\bm{C_\ell{\,\mem=\mem\,}1}$}{C\_ℓ=1}}

Finally, we spell out the constant-mass earthly equation of motion at $\mathcal{O}(\omega'^1)$
in the vectorial language by taking ``{\hhem$2\hem\Re$}'' to \eqref{eq:eom-heavenly-components}:
\begin{align}
\begin{split}
    \label{eq:g=2eom}
    m_0\mem \dot{p}_m
    \simeq -\mem p_n\mem 
        % {\phi}{}^n{}_m\hnem(z,\bz)
        (\hem \phi^+\hnem(z) {\,+\,} \phi^-\hnem(\bz) \hhhem)^n{}_m
    &\,,\quad
    m_0\mem \dot{\L}^m{}_a
    \simeq \mem
        % {\phi}{}^m{}_n\hnem(z,\bz)
        (\hem \phi^+\hnem(z) {\,+\,} \phi^-\hnem(\bz) \hhhem)^m{}_n
    \mem \L\hnem^n{}_a
    \,,\\
    m_0 \dot{x}^m
    \simeq p^m - \mem
        % {\ast\phi}{}^m{}_n\hnem(z,\bz)
        *(\hem \phi^+\hnem(z) {\,+\,} \phi^-\hnem(\bz) \hhhem)^m{}_n
    \mem y^n
    &\,,\quad
    m_0\mem \dot{y}^m
    \simeq \mem
        % {\phi}{}^m{}_n\hnem(z,\bz)
        (\hem \phi^+\hnem(z) {\,+\,} \phi^-\hnem(\bz) \hhhem)^m{}_n
    \mem y^n
    \,.
\end{split}
\end{align}
The orbital ($p_m$) and spin ($\L\hnem^m{}_a$, $y^m$)
precessions 
are synchronized at 
% linear order in $\omega'$
$\mathcal{O}(\phi^1)$
(Figure \ref{fig:g=0-g=2}).
Taylor-expanding
the fields
around the real section,
we recognize
$C_\ell{\,\mem=\mem\,}1$
(see \eqref{eq:prec-eq.vec}),
in particular
% the gyromagnetic ratio 
$g{\,=\,}2$
at $\mathcal{O}(y^1)$
and 
% the gravimagnetic ratio
$\kappa{\,=\,}1$
at $\mathcal{O}(y^2)$
by referring to TBMT and MPTD
equations.

\paragraph{Spin as a chiral Wick rotation of deviation}

Meanwhile, note that
the evolution of $x^m$ 
deviates from this synchronized precession ($p^m$)
by a zitterbewegung that is ``dual'' to the spin precession ($-{\ast\phi}{}^m{}_n\mem y^n$).\footnote{
    Note that the covariant spin gauge eliminates zitterbewegung in the free theory.
}
This zitterbewegung reflects the ``spin-induced spacetime noncommutativity'' \eqref{eq:xy-nc} and its resolution in the complex combinations:
% % % it is a vestige of the covariant spin gauge
% % % and the resulting complex spacetime representation,
% % % where \textit{deviation in real spacetime} and \textit{spin} are put on equal footing.
% % the covariant spin gauge puts \textit{deviation in real spacetime} and \textit{spin} on equal footing.
% % % This term vanishes if, for example, $\omega'_{mn}$ has only magnetic components and $y^m$ is spatial.
% \textit{deviation} in real spacetime and \textit{spin} are put on equal footing.
The covariant spin gauge fixing has put
\textit{deviation} in real spacetime and \textit{spin} 
on an equal footing.
We give a closer look at this 
% ``chiral Wick rotation''
% ``Wick rotation''
dual
relationship between deviation and spin
in \cite{ambikerr2}.

\subsection{Non-minimal equation of motion}
\label{sec:Flat.eom-nonmin}

We now return to the generic
class of
spin-space-time symplectic perturbations \eqref{eq:arbomega'}.

\paragraph{Maximally non-minimal coupling}

A typical example of a non-holomorphic 
spin-space-time
symplectic perturbation
is a closed two-form that localizes on the spacetime:
% 
% Let us start with a typical example of a non-holomorphic symplectic perturbation:
% a closed two-form that localizes on the real section,
% As a typical example of a non-holomorphic symplectic perturbation,
% consider a closed two-form that localizes on the real section:
\begin{align}
\begin{split}
    \label{eq:realform}
    \omega'^+
    &=
    {\textstyle\frac{1}{2}\,}
    \phi^+_{mn}\hnem(x)\mem\wedgetwo{x}{x}{m}{n}
    \,,\\
    &=
    {\textstyle\frac{1}{4}\,}
    \phi^+_{mn}\hnem(x)\mem
    \big(
         \minie\mem \wedgetwo{z}{z}{m}{n}
         + dz^m {\hem\otimes\mem\hem} d\bz^n
         + d\bz^m {\hem\otimes\mem\hem} dz^n
         + \minie\mem \wedgetwo{\bz}{\bz}{m}{n}
    \big)
    \,,\\
    &=:
    \frac{1}{4}\nem\nem
    \left(\,
        \includegraphics[valign=c]{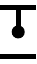}
        +
        \includegraphics[valign=c]{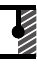}
        +
        \includegraphics[valign=c]{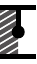}
        +
        \includegraphics[valign=c]{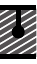}
    \,\right)
    \,.
\end{split}
\end{align}
In terms of \eqref{eq:arbomega'}, we have
% ${^0\omega'^+_{mn}} = {^1\omega'^+_{mn}} = {^2\omega'^+_{mn}} = \frac{1}{4}\mem \phi^+_{mn}\hnem(x)$.
\begin{align}
    {^0\omega'^+_{mn}} 
    \mem=\mem
    {^1\omega'^+_{mn}} 
    \mem=\mem
    {^2\omega'^+_{mn}} 
    \mem=\mem
    {\textstyle\frac{1}{4}}\mem \phi^+_{mn}\hnem(x)
    \,.
\end{align}
At first sight, 
the symplectic perturbation
\eqref{eq:realform}
may seem like a ``minimal'' coupling prescription
because it simply 
% couples the particle to
brings
the ``field strength'' two-form
without any post-processing,
such as the Newman-Janis shift.
However, in the holomorphic/anti-holomorphic basis,
it looks rather complicated.
In fact, the zig-zag representation discloses its \textit{maximally non-minimal} nature.
We find three new types of zig-zag vertices,
which
cause a proliferation of zig-zag interaction scenarios:
\begin{align}
    \{\blank,\bDelta\}
    =\,&
    \includegraphics[valign=c]{figs/l.pdf}
    \includegraphics[valign=c]{figs/xD-b.pdf}
    -
    \includegraphics[valign=c]{figs/l-b.pdf}
    \includegraphics[valign=c]{figs/0+ww.pdf}
    \includegraphics[valign=c]{figs/xD-b.pdf}
    -
    \includegraphics[valign=c]{figs/l.pdf}
    \includegraphics[valign=c]{figs/0+bw.pdf}
    \includegraphics[valign=c]{figs/xD-b.pdf}
    \\
    \nonumber
    &
    +\nem
    \includegraphics[valign=c]{figs/l-b.pdf}
    \includegraphics[valign=c]{figs/0+wb.pdf}
    \includegraphics[valign=c]{figs/0+ww.pdf}
    \includegraphics[valign=c]{figs/xD-b.pdf}
    \nem+\nem
    \includegraphics[valign=c]{figs/l.pdf}
    \includegraphics[valign=c]{figs/0+bb.pdf}
    \includegraphics[valign=c]{figs/0+ww.pdf}
    \includegraphics[valign=c]{figs/xD-b.pdf}
    \nem+\nem
    \includegraphics[valign=c]{figs/l-b.pdf}
    \includegraphics[valign=c]{figs/0+ww.pdf}
    \includegraphics[valign=c]{figs/0+bw.pdf}
    \includegraphics[valign=c]{figs/xD-b.pdf}
    \nem+\nem
    \includegraphics[valign=c]{figs/l.pdf}
    \includegraphics[valign=c]{figs/0+bw.pdf}
    \includegraphics[valign=c]{figs/0+bw.pdf}
    \includegraphics[valign=c]{figs/xD-b.pdf}
    \nem+
    \cdots
    \vphantom{\adjustbox{raise=-2.0ex}{$0$}}
    \,,\\
    \{\blank,\Delta\}
    =\,&
    \includegraphics[valign=c]{figs/l-b.pdf}
    \includegraphics[valign=c]{figs/xD.pdf}
    -
    \includegraphics[valign=c]{figs/l-b.pdf}
    \includegraphics[valign=c]{figs/0+wb.pdf}
    \includegraphics[valign=c]{figs/xD.pdf}
    -
    \includegraphics[valign=c]{figs/l.pdf}
    \includegraphics[valign=c]{figs/0+bb.pdf}
    \includegraphics[valign=c]{figs/xD.pdf}
    \\
    \nonumber
    &
    +\nem
    \includegraphics[valign=c]{figs/l-b.pdf}
    \includegraphics[valign=c]{figs/0+wb.pdf}
    \includegraphics[valign=c]{figs/0+wb.pdf}
    \includegraphics[valign=c]{figs/xD.pdf}
    \nem+\nem
    \includegraphics[valign=c]{figs/l.pdf}
    \includegraphics[valign=c]{figs/0+bb.pdf}
    \includegraphics[valign=c]{figs/0+wb.pdf}
    \includegraphics[valign=c]{figs/xD.pdf}
    \nem+\nem
    \includegraphics[valign=c]{figs/l-b.pdf}
    \includegraphics[valign=c]{figs/0+ww.pdf}
    \includegraphics[valign=c]{figs/0+bb.pdf}
    \includegraphics[valign=c]{figs/xD.pdf}
    \nem+\nem
    \includegraphics[valign=c]{figs/l.pdf}
    \includegraphics[valign=c]{figs/0+bw.pdf}
    \includegraphics[valign=c]{figs/0+bb.pdf}
    \includegraphics[valign=c]{figs/xD.pdf}
    \nem+
    \cdots
    \,.
\end{align}
The expansions never truncate,
regardless of whether one assumes pure self-duality or not.
Besides, one can argue that 
$\phi^+_{mn}\hnem(x)$ actually contains an infinite number of derivative couplings as $\phi^+_{mn}\hnem(z) - iy^r \phi^+_{mn,r}(z) + \cdots$
% if seen from the holomorphic spacetime.
if one takes the holomorphic 
% spacetime
Minkowski space
as more fundamental than the real spacetime.
In any respect, this is by no means ``minimal.''

\paragraph{Maximally non-minimal spin precession: \texorpdfstring{$\bm{C_{\ell\hem>\hem0}{\mem\,=\,\mem}0}$}{C\_\{ℓ>0\}=0}}
% \paragraph{Maximally non-minimal spin precession}
% \paragraph{\texorpdfstring{\textit{g}{\kern0.1em}={\kern0.1em}0 spin precession}{g=0 spin precession}}

Let us compute the $\mathcal{O}(\omega'^1)$ equation of motion
that follows from \eqref{eq:realform}.
% It turns out that \eqref{eq:realform} describes the gyromagnetic ratio $g=0$.
First of all, we find
\begin{subequations}
    \label{eq:all3ptX}
\begin{align}
    \label{eq:all3ptX-0}
    -
    % \rlap{\hspace{16.5pt}\adjustbox{scale=0.65,valign=c,raise=12pt}{$\bm{(+)}$}}
    \includegraphics[valign=c]{figs/l-b.pdf}
    \includegraphics[valign=c]{figs/0+ww.pdf}
    \includegraphics[valign=c]{figs/xD-b.pdf}
    &=\,\, 
    \frac{(-2)}{\Delta}\,
    \omega'^\db{}_{\da}(x)
    \nem
    \left(
        \rambda_{I\db}\mem \hem
        \frac{\partial}{\partial\rambda_{I\da}}
        +
        2iy^{\da\c}
        \frac{\partial}{\partial\bz^{\db\c}}
    \right)
    \,,\\
    \label{eq:all3ptX-1ramb}
    -
    % \rlap{\hspace{16.5pt}\adjustbox{scale=0.65,valign=c,raise=12pt}{$\bm{(+)}$}}
    \includegraphics[valign=c]{figs/l-b.pdf}
    \includegraphics[valign=c]{figs/0+wb.pdf}
    \includegraphics[valign=c]{figs/xD.pdf}
    &=\,\, 
    \frac{(-2)}{\bDelta}\,
    \omega'^\db{}_{\da}(x)
    \nem
    \left(
        \rambda_{I\db}\mem \hem
        \frac{\partial}{\partial\rambda_{I\da}}
        +
        2iy^{\da\c}
        \frac{\partial}{\partial\bz^{\db\c}}
    \right)
    \,,\\
    \label{eq:all3ptX-1lamb}
    -
    % \rlap{\hspace{16.5pt}\adjustbox{scale=0.65,valign=c,raise=12pt}{$\bm{(+)}$}}
    \includegraphics[valign=c]{figs/l.pdf}
    \includegraphics[valign=c]{figs/0+bw.pdf}
    \includegraphics[valign=c]{figs/xD-b.pdf}
    &=\,\, 
    \frac{(-2)}{\Delta}\,
        p_{\a\da}\mem
        \omega'^\da{}_\db\hnem(x)\hem
        (p^{-1})^{\db\b}
        \nem
        \left(
            \lambda_\b{}^I\hem
            \frac{\partial}{\partial\lambda_\a{}^I}
            - 
            2iy^{\dc\a} \hem
            \frac{\partial}{\partial z^{\dc\b}}
        \right)
    \,,\\
    \label{eq:all3ptX-2}
    -
    % \rlap{\hspace{16.5pt}\adjustbox{scale=0.65,valign=c,raise=12pt}{$\bm{(+)}$}}
    \includegraphics[valign=c]{figs/l.pdf}
    \includegraphics[valign=c]{figs/0+bb.pdf}
    \includegraphics[valign=c]{figs/xD.pdf}
    &=\,\, 
    \frac{(-2)}{\bDelta}\,
        p_{\a\da}\mem
        \omega'^\da{}_\db\hnem(x)\hem
        (p^{-1})^{\db\b}
        \nem
        \left(
            \lambda_\b{}^I\hem
            \frac{\partial}{\partial\lambda_\a{}^I}
            - 
            2iy^{\dc\a} \hem
            \frac{\partial}{\partial z^{\dc\b}}
        \right)
    \,.
\end{align}
\end{subequations}
The first two lines are equal to \eqref{eq:heavenly-3ptX} on the mass-shell constraint surface.
The last two lines involve an interesting combination \smash{$p_{\a\da}\mem \omega'^\da{}_\db\hem (p^{-1})^{\db\b}$}.
% This particular sandwiching ``inverts''
% a self-dual bivector with respect to the momentum direction:
This particular sandwiching has a geometrical interpretation as a reflection 
across the momentum direction:
% with respect to the plane orthogonal to the momentum direction:
% \begin{align}
% \begin{split}
%     \label{eq:sigma-inversion}
%     p_{\a\dc}\hem (\bs^m)^{\dc\c} p_{\c\da}
%     = p^2\mem I^m{}_n (\s^n)_{\a\da}
%     &\,,\quad
%     (p^{-1})^{\da\c} (\s^m)_{\c\dc} (p^{-1})^{\dc\a}
%     = p^{-2}\mem I^m{}_n (\bs^n)^{\da\a}
%     \,,\\
%     p_{\a\dc} (\bs^{mn})^\dc{}_\dd (p^{-1})^{\dd\b}
%     = -I^m{}_r I^n{}_s (\s^{rs})_\a{}^\b
%     &\,,\quad
% \end{split}
% \end{align}
\begin{align}
\begin{split}
    \label{eq:sigma-inversion}
    (p\mem \bs^m  p)_{\a\da}
    = p^2\mem I^m{}_n (\s^n)_{\a\da}
    &\,,\quad
    (p^{-1} \s^m\mem p^{-1})^{\da\a}
    = p^{-2}\mem I^m{}_n (\bs^n)^{\da\a}
    \,,\\
    (p\mem \bs^{mn} p^{-1})_\a{}^\b
    = -I^m{}_r I^n{}_s (\s^{rs})_\a{}^\b
    &\,,\quad
    (p^{-1} \s^{mn}\mem p)^\da{}_\db
    = -I^m{}_r I^n{}_s (\bs^{rs})^\da{}_\db
    \,,\\
    I^m{}_n := \delta^m{}_n - 2p^mp_n/p^2
    &\,,\quad
    \ve_{pqkl}\hem I^p{}_m I^q{}_n I^k{}_r I^l{}_s
    = -\ve_{mnrs}
    \,.
\end{split}
\end{align}
These identities are easily understood if
one recalls the inversive geometry of the 
% conformal
Clifford/conformal
algebra
\cite{lasenby2004convariant}
and that the inversion map $p^m \mapsto p^m\hnem/p^2$ has the Jacobian $p^{-2}(\delta^m{}_n - 2p^mp_n/p^2)$.
For later use, we define 
% the following notation 
a notation ``${_{\mathcal{I}\nem}F_{mn}}$''
for 
% an arbitrary two-form
two-forms
$F_{mn}$:
\begin{align}
\begin{split}
    \label{eq:IF-def}
    % \accentset{\adjustbox{raise=1.75ex}{$\smash{ 
    %     \scalebox{1}[-1]{\smash{$\widehat{}$}}
    % }$}}{\omega'}_{mn}
    % \mathrlap{\kern0.1em\adjustbox{raise=0.25ex,scale=0.625}{$\aoverbrace[L1R]{\phantom{\scalebox{1.6}{$F$}}}$}}F_{mn}
    % \mathrlap{\scalebox{1.5}[1]{\adjustbox{raise=1.25ex,scale=0.6}{$\aoverbrace[L1R]{\phantom{F}}$}}}F_{mn}
    {_{\mathcal{I}\nem}F_{mn}}
    := F_{rs}\mem I^r{}_m I^s{}_n
    \qiq 
    \left\{
    {\renewcommand{\arraycolsep}{0em}
    \renewcommand{\arraystretch}{1.1}
    \begin{array}{rl}
        p_{\a\da}\mem F^\da{}_\db\mem (p^{-1})^{\db\b}
        &= -
            % \mathrlap{\scalebox{1.5}[1]{\adjustbox{raise=1.25ex,scale=0.6}{$\aoverbrace[L1R]{\phantom{F}}$}}}F
            (_{\mathcal{I}\nem}F)
        _\a{}^\b
        \,,\\
        (p^{-1})^{\da\a}\mem F_\a{}^\b\mem p_{\b\db}
        &= -
            % \mathrlap{\scalebox{1.5}[1]{\adjustbox{raise=1.25ex,scale=0.6}{$\aoverbrace[L1R]{\phantom{F}}$}}}F
            (_{\mathcal{I}\nem}F)
        ^\da{}_\db
        \,.
    \end{array}}
    \right.
\end{split}
\end{align}
The heavenly equation of motion 
follows from collecting
\eqref{eq:holomorhictranslations}, \eqref{eq:all3ptX}, and \eqref{eq:IF-def}.
% From \eqref{eq:holomorhictranslations}, \eqref{eq:all3ptX}, and \eqref{eq:IF-def},
% one obtains the $\mathcal{O}(\omega'^1)$ constant-mass equation of motion.
Then the earthly equation of motion is then obtained by simply replacing $\phi^+_{mn}\hnem(x)$ with $\phi_{mn}\hnem(x)$,
as the real perturbation $dx \swedge dx$ acts ``symmetrically'' on the zig and zag sectors: 
% \begin{align}
% \begin{split}
%     \label{eq:g=0eom}
%     &m_0\frac{dp_m}{d\tau}
%     \simeq - p_n\mem \phi{}^n{}_m\hnem(x)
%     \,,\quad
%     m_0\frac{d\L\hnem^m{}_a}{d\tau}
%     \simeq
%     -\frac{1}{{m_0}^2}\mem
%     p^m p^r\mem\phi_{rs}\hnem(x)\mem \L\hnem^s{}_a
%     \,,
%     \\
%     &m_0\frac{dx^m}{d\tau}
%     \simeq p^m 
%     - \tilde{\eta}^{mr}\mem{\ast\phi_{rs}}\hnem(x)\mem \tilde{\eta}^{sn} y_n
%     \,,\quad
%     \tilde{\eta}^{mn} := \eta_{mn} - p^m p^n /p^2
%     \,,
%     \\
%     &m_0\frac{dy^m}{d\tau}
%     \simeq 
%     % - \frac{1}{{m_0}^2}\mem p^mp^r\mem\phi_{rs}\hnem(x)\mem y^s
%     % + \frac{W_0}{{m_0}^2}\,{\phi}{}^m{}_n(x)\mem p^n
%     % 
%     % -\frac{1}{{m_0}^2}
%     % \Big({
%     %     p^mp^r \phi_{rs}\hnem(x)\mem \eta^{sn}
%     %     + \eta^{mr}\mem \phi_{rs}\hnem(x)\mem p^sp^n
%     % }\Big) y_n
%     % 
%     -\frac{1}{{m_0}^2}\mem
%     p^mp^r \phi_{rs}\hnem(x)\mem \eta^{sn} y_n
%     \,.
% \end{split}
% \end{align}
\begin{align}
\begin{split}
    \label{eq:g=0eom}
    m_0 \mem\dot{p}_m
    \simeq - p_n\mem \phi{}^n{}_m\hnem(x)
    \,,\quad
    &\textstyle
    m_0 \mem\dot{\L}^m{}_a
    \simeq
    -\frac{1}{{m_0}^2}\mem
    p^m p^r \phi_{rs}\hnem(x)\mem \L\hnem^s{}_a
    \,,
    \\
    m_0\dot{x}^m
    \simeq p^m 
    % - \tilde{\eta}^{mr}\mem{\ast\phi_{rs}}\hnem(x)\mem \tilde{\eta}^{sn} y_n
    - (\eta^{mr} {-\mem} p^mp^r\nem\hnem/p^2)\mem{\ast\phi_{rs}}\hnem(x)\mem y^s
    % - \tilde{\eta}^{mr}\mem{\ast\phi_{rs}}\hnem(x)\mem y^s
    \,,\quad
    &\textstyle
    m_0 \mem\dot{y}^m
    \simeq 
    -\frac{1}{{m_0}^2}\mem
    p^mp^r \phi_{rs}\hnem(x)\mem y^s
    % \,,
    \,.
\end{split}
\end{align}
% where $\tilde{\eta}^{mn} := \eta^{mr} {-\mem} p^mp^r\nem\hnem/p^2$.
\eqref{eq:g=0eom} agrees with \eqref{eq:prec-eq.vec}
if $C_{\ell\hem>\hem0}{\mem\,=\,\mem}0$:
$g{\,=\,}0$ for TBMT 
and $\kappa{\,=\,}0$ for MPTD.
Evidently,
the precession behaviors of $p_m$ and $y^m$ are not synchronized.

To sum up, 
our formalism is the simplest with the holomorphic symplectic perturbation \eqref{eq:GMOOVform}
and gets ``maximally'' complicated with \eqref{eq:realform}:
all the $(2,0)$-, $(1,1)$-, and $(0,2)$-parts appear in \eqref{eq:realform}
with an ``equal'' magnitude.
This contrasts sharply
with the traditional approach,
which treats the real-supported symplectic structure \eqref{eq:realform} as the default
and implements the minimal gyromagnetic ratio $g{\,=\,}2$
by an additional ``non-minimal'' spin interaction Hamiltonian $H' = q\vec{B}{\mem\cdot\mem}\vec{S}$
% (and similarly for the gravimagnetic ratio).
(and similarly for $\kappa$).
While the traditional approach 
creates an unfortunate discrepancy between
the classical formalism and
the amplitudes-level understanding of \cite{holstein2006large,ahh2017},
our perturbation theory
is literally minimal at the true minimal coupling
$C_{\ell}{\mem\,=\,\mem}1$
($g{\,=\,}2$, $\kappa{\,=\,}1$, $\cdots$).
Therefore, 
we argue that
our
``zig-zag''
formulation 
provides a rationale for identifying 
$C_{\ell}{\mem\,=\,\mem}1$
% $g{\,=\,}2$, $\kappa{\,=\,}1$, $\cdots$
as the true minimal coupling
and 
$C_{\ell\hem>\hem0}{\mem\,=\,\mem}0$
($g{\,=\,}2$, $\kappa{\,=\,}0$, $\cdots$) 
as the maximally non-minimal coupling
already at the classical level.
The massive twistor representation of massive spinning particles has guided us to do so,
as the zig-zag structure---already inherent in the \textit{free theory} as the skew-diagonalization of the Poisson bracket---asserts that one should understand everything in the complex basis.
% asserts that the canonical basis to comprehend a symplectic perturbation is the complex basis.

\paragraph{Generic heavenly equation of motion}
% \paragraph{Non-minimal heavenly equation of motion}

Finally, we spell out the $\mathcal{O}(\omega'^1)$ heavenly equation of motion that follows from the generic perturbation \eqref{eq:arbomega'+}.
It is easily obtainable by recycling the calculations of \eqref{eq:all3ptX}:
\begin{align}
\begin{split}
    \label{eq:eom-nonmin}
    m_0\hem
    \frac{d}{d\tau}
    \,\simeq\,\mem
    \frac{p^{\da\a}}{-2}
    \hem\frac{\partial}{\partial x^{\da\a}}
    % \\
    &{}
    + 
        (\hem{^{0+1}\hnem\omega'})^\db{}_\da
        \hnem\left(
            \rambda_{I\db}\mem  \frac{\partial}{\partial\rambda_{I\da}}
            + 
            2iy^{\da\c}
            \frac{\partial}{\partial\bz^{\db\c}}
        \right)
    \\
    &{}
    + 
        % (
        % \mathrlap{\kern0.1em\scalebox{4.2}[1]{\adjustbox{raise=2.3ex,scale=0.6}{$\aoverbrace[L1R]{\phantom{F}}$}}}
        % \hem{^{1+2}\omega'})_\a{}^\b
        ({{}^{1+2}\mathllap{\adjustbox{raise=-0.2ex}{$_\mathcal{I}\hem$}}\omega'})_\a{}^\b
        \hnem\left(
            - %
            \lambda_\b{}^I\hem
            \frac{\partial}{\partial\lambda_\a{}^I}
            % - 
            +
            2iy^{\dc\a} \hem
            \frac{\partial}{\partial z^{\dc\b}}
        \right)
    % \,.
    \,,
\end{split}
\end{align}
where 
% we have denoted
% \begin{align}
%     {^{0+1}\omega'^+_{mn}}
%     := {^0\omega'^+_{mn}}
%     +  {^1\omega'^+_{mn}}
%     \,,\quad
%     {^{1+2}\omega'^+_{mn}}
%     := {^1\omega'^+_{mn}}
%     +  {^2\omega'^+_{mn}}
%     \,.
% \end{align}
$
    {^{0+1}\omega'^+_{mn}}
    := {^0\omega'^+_{mn}}
    +  {^1\omega'^+_{mn}}
$,
$
    {^{1+2}\omega'^+_{mn}}
    := {^1\omega'^+_{mn}}
    +  {^2\omega'^+_{mn}}
$.
% \eqref{eq:eom-nonmin} boils down to
In other words,
\begin{align}
\begin{split}
    \label{eq:eom-nonmin.hybrid}
    m_0\hem \dot{\lambda}_\a{}^I
    \simeq -
    % ({{}^{1+2}\mathllap{\adjustbox{raise=-0.2ex}{$_\mathcal{I}\hem$}}\omega'(z,\bz)})_\a{}^\b
    (^{1+2}\mathllap{\adjustbox{raise=-0.2ex}{$_\mathcal{I}\hem$}}\omega')_\a{}^\b
    \hem\lambda_\b{}^I
    &\,,\quad
    m_0\hem \dot{z}^{\da\a}
    \simeq 
    \minime{p^{\da\a}}
    + 2i\hem
    y^{\da\b}\hem
    (^{1+2}\mathllap{\adjustbox{raise=-0.2ex}{$_\mathcal{I}\hem$}}\omega')_\b{}^\a
    \,,\\
    m_0\hem \dot{\rambda}_{I\da}
    \simeq \rambda_{I\db}\hem
    (\hem^{0+1}\hnem\omega')^\db{}_\da
    &\,,\quad
    m_0\hem \dot{\bz}^{\da\a}
    \simeq 
    \minime{p^{\da\a}}
    + 2i\mem
    (\hem^{0+1}\hnem\omega')^\da{}_\db
    \mem y^{\db\a}
    \,.
\end{split}
\end{align}
In the vectorial language,
we find
\begin{align}
\begin{split}
    \label{eq:eom-nonmin.vector}
    m_0\mem \dot{p}_m
    &\simeq
    \big(\hem
        {{^{0+1}\hnem\omega'^+}}
        - {{{}^{1+2}\mathllap{\adjustbox{raise=-0.2ex}{$_\mathcal{I}\hem$}}\omega'^+}}
    \big)\hnem{}_m{}^n\mem p_n
    \simeq -p_n\mem \omega'^+{}^n{}_m
    \,,\\
    m_0\mem \dot{\Lambda}^m{}_a
    &\simeq
    \big(\hem
        {{^{0+1}\hnem\omega'^+}}
        - {{{}^{1+2}\mathllap{\adjustbox{raise=-0.2ex}{$_\mathcal{I}\hem$}}\omega'^+}}
    \big)\hnem{}^m{}_n\mem \L\hnem^n{}_a
    \,,\\
    m_0\mem \dot{y}^m
    &\simeq 
    \big(\hem
        {{^{0+1}\hnem\omega'^+}}
        - {{{}^{1+2}\mathllap{\adjustbox{raise=-0.2ex}{$_\mathcal{I}\hem$}}\omega'^+}}
    \big)\hnem{}^m{}_n\mem y^n
    \,,\\
    m_0\dot{x}^m
    &\simeq 
    p^m
    -
    {\ast}\big(\hem
        {{^{0+1}\hnem\omega'^+}}
        - {{{}^{1+2}\mathllap{\adjustbox{raise=-0.2ex}{$_\mathcal{I}\hem$}}\omega'^+}}
    \big)\hnem{}^m{}_n\mem y^n
    \,.
\end{split}
\end{align}
% Obtaining the earthly equation of motion is straightforward.
Note that ${}_\mathcal{I}$ and $\ast$ anti-commute.

\paragraph{Arbitrary \textit{g} value}

Before moving on to the Wilson coefficients,
let us discuss two more examples for further concreteness.
First, 
it is instructive to think of
a certain interpolation between the two extremes \eqref{eq:GMOOVform} and \eqref{eq:realform}:
% that reproduces the arbitrary-$g$ $\mathcal{O}(\omega'^1)$ equation of motion in the small spin-derivative limit $(y{\hem\cdot\hem}\partial)\hem \omega' \ll \omega'$:
% \begin{subequations}
\begin{align}
    \label{eq:omega'-arbitraryg}
\begin{split}
    % \label{eq:omega'-arbitraryg+}
    \omega'^+
    &{}=
    \phi_{\smash{\da\db}\vphantom{\b}}({x{\mem+\mem}igy/2})\,\hem
    \e_{\a\b}\mem d(x{\mem+\mem}igy/2)^{\da\a} \swedge d(x{\mem+\mem}igy/2)^{\db\b}
    \,,\\
    &{}=:
    {
    \frac{1}{4}
    \bigg(\hnem\smash{
        (1+g/2)^2\,
        \mem
        % \rlap{\adjustbox{scale=0.5375,valign=c,raise=11.9pt}{$\bm{(+)}$}}
        \includegraphics[valign=c]{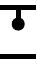}\mem
        +
        (1-(g/2)^2)\,
        \mem
        % \rlap{\adjustbox{scale=0.5375,valign=c,raise=11.9pt}{$\bm{(+)}$}}
        \includegraphics[valign=c]{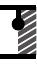}\mem
        +
        (1-(g/2)^2)\,
        \mem
        % \rlap{\adjustbox{scale=0.5375,valign=c,raise=11.9pt}{$\bm{(+)}$}}
        \includegraphics[valign=c]{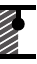}\mem
        +
        (1-g/2)^2\,
        \mem
        % \rlap{\adjustbox{scale=0.5375,valign=c,raise=11.9pt}{$\bm{(+)}$}}
        \includegraphics[valign=c]{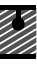}\mem
        \,
    }\bigg)
    }
    \,,
% \end{split}
    \\
% \begin{split}
    % \label{eq:omega'-arbitraryg-}
    \omega'^-
    &{}=
    \phi_{\smash{\da\db}\vphantom{\b}}({x{\mem-\mem}igy/2})\,\hem
    \be_{\smash{\da\db}\vphantom{\b}}\mem d(x{\mem-\mem}igy/2)^{\da\a} \swedge d(x{\mem-\mem}igy/2)^{\db\b}
    \,,\\
    &{}=:
    {
    \frac{1}{4}
    \bigg(\hnem
        (1+g/2)^2\,
        \mem
        % \rlap{\adjustbox{scale=0.5375,valign=c,raise=-11.9pt}{$\bm{(-)}$}}
        \adjustbox{valign=c}{{\includegraphics[rotate=180]{figs/insertg+bb.pdf}}}\mem
        +
        (1-(g/2)^2)\,
        \mem
        % \rlap{\adjustbox{scale=0.5375,valign=c,raise=-11.9pt}{$\bm{(-)}$}}
        \adjustbox{valign=c}{{\includegraphics[rotate=180]{figs/insertg+wb.pdf}}}\mem
        +
        (1-(g/2)^2)\,
        \mem
        % \rlap{\adjustbox{scale=0.5375,valign=c,raise=-11.9pt}{$\bm{(-)}$}}
        \adjustbox{valign=c}{{\includegraphics[rotate=180]{figs/insertg+bw.pdf}}}\mem
        +
        (1-g/2)^2\,
        \mem
        % \rlap{\adjustbox{scale=0.5375,valign=c,raise=-11.9pt}{$\bm{(-)}$}}
        \adjustbox{valign=c}{{\includegraphics[rotate=180]{figs/insertg+ww.pdf}}}\mem
        \,
    \bigg)
    }
    \,.
\end{split}
\end{align}
% \end{subequations}
% This symplectic perturbation
% reproduces the arbitrary-$g$ equation of motion at \smash{$\mathcal{O}(\omega'^1)$} and the
% % small spin-derivative 
% limit $(y{\hem\cdot\hem}\partial)\hem \omega' \ll \omega'$.
% The resulting self-dual heavenly impulse conforms to the kinematics \eqref{eq:3kin-real} with $\theta = g$,
% and the amplitude is partially Newman-Janis shifted as $\mathe^{(g/2)\hem\three{\mem\cdot\mem}a}$.
% We have spelled out the anti-self-dual symplectic perturbation also for reference.
For reference,
we have spelled out the anti-self-dual counterpart as well.
As indicated in the diagrammatic format,
each part of $\omega'^+$ is given as
% $
%     {^0\omega'^+_{mn}}
%     = ((1{\mem+\mem}g/2)^2\hnem/4)\mem \phi^+_{mn}\hnem(x{\mem+\mem}igy/2)
% $,
% $
%     {^1\omega'^+_{mn}}
%     = ((1 {\mem-\mem} (g/2)^2)/4)\mem \phi^+_{mn}\hnem(x{\mem+\mem}igy/2)
% $,
% and
% $
%     {^2\omega'^+_{mn}}
%     = ((1{\mem-\mem}g/2)^2\hnem/4)\mem \phi^+_{mn}\hnem(x{\mem+\mem}igy/2)
% $.
\begin{align}
\begin{split}
    {^0\omega'^+_{mn}}
    &\mem=\mem\hhem 
        {\textstyle\frac{1}{4}}\hem
        (1 + g/2)^2\, \phi^+_{mn}\hnem(x{\mem+\mem}igy/2)
    \,,\\
    {^1\omega'^+_{mn}}
    &\mem=\mem\hhem
        {\textstyle\frac{1}{4}}\hem
        (1 - (g/2)^2)\, \phi^+_{mn}\hnem(x{\mem+\mem}igy/2)
    \,,\\
    {^2\omega'^+_{mn}}
    &\mem=\mem\hhem
    {\textstyle\frac{1}{4}}\hem
    (1 - g/2)^2\, \phi^+_{mn}\hnem(x{\mem+\mem}igy/2)
    \,.
\end{split}
\end{align}
The heavenly equation of motion \eqref{eq:eom-nonmin.vector} boils down to
% \begin{align}
% \begin{split}
%     m_0\hem \dot{\lambda}_\a{}^I
%     = -
%     \frac{1{\mem-\mem}g/2}{2}\mem
%     (^{\phantom{\mathcal{I}}}\mathllap{\adjustbox{raise=-0.2ex}{$_\mathcal{I}\hem$}}\phi)_\a{}^\b
%     \hem\lambda_\b{}^I
%     &\,,\quad
%     m_0\hem \dot{z}^{\da\a}
%     = 
%     \minime{p^{\da\a}}
%     + 
%     \frac{1{\mem-\mem}g/2}{2}\mem
%     2i\hem y^{\da\b}\hem
%     (^{\phantom{\mathcal{I}}}\mathllap{\adjustbox{raise=-0.2ex}{$_\mathcal{I}\hem$}}\phi)_\b{}^\a
%     \,,\\
%     m_0\hem \dot{\rambda}_{I\da}
%     = 
%     \frac{1{\mem+\mem}g/2}{2}\mem
%     \rambda_{I\db}\hem
%     \phi^\db{}_\da
%     &\,,\quad
%     m_0\hem \dot{\bz}^{\da\a}
%     = 
%     \minime{p^{\da\a}}
%     +
%     \frac{1{\mem+\mem}g/2}{2}\mem
%     \phi^\da{}_\db
%     \mem 2i\hem y^{\db\a}
%     \,.
% \end{split}
% \end{align}
\begin{align}
    \label{eq:neom-arbg}
\begin{split}
    m_0\mem \dot{p}_m
    &=
    -p_n\mem \phi^+{}^n{}_m\hnem(x{\mem+\mem}igy/2)
    \,,\\
    m_0\hem\hhem \dot{\Lambda}^m{}_a
    &=
    \bigg(\mem{
        \frac{g}{2}\, \phi^+{}^m{}_n\hnem(x{\mem+\mem}igy/2)
        - \frac{2{\mem-\mem}g}{2}\, \frac{\hhem{p^m p_l}\hhem}{m^2}\, \phi^+{}^l{}_n\hnem(x{\mem+\mem}igy/2)
    }\bigg)
    \mem \L^n{}_a
    \,,\\
    m_0\mem \dot{y}^m
    &= 
    \bigg(\mem{
        \frac{g}{2}\, \phi^+{}^m{}_n\hnem(x{\mem+\mem}igy/2)
        - \frac{2{\mem-\mem}g}{2}\, \frac{\hhem{p^m p_l}\hhem}{m^2}\, \phi^+{}^l{}_n\hnem(x{\mem+\mem}igy/2)
    }\bigg)
    \mem y^n
    \,,\\
    m_0\dot{x}^m
    &= 
    p^m
    +
    \bigg(\mem{
        - {\ast}\phi^+{}^m{}_n\hnem(x{\mem+\mem}igy/2)
        - \frac{2{\mem-\mem}g}{2}\, \frac{\hhem{p^m p_l}\hhem}{m^2}\, {\ast}\phi^+{}^l{}_n\hnem(x{\mem+\mem}igy/2)
    }\bigg)
    \mem y^n
    \,.
\end{split}
\end{align}
Replacing $\phi^+{}^m{}_n\hnem(x{\mem+\mem}igy/2)$
with $\phi^+{}^m{}_n\hnem(x)+ (g/2)\mem {\ast}((y{\mem\cdot\mem}\partial) \phi^+\hnem(x))^m{}_n + \cdots$,
one finds that \eqref{eq:neom-arbg} reproduces the
arbitrary-$g$ TBMT equation \eqref{eq:BMT}
in the small spin-derivative limit $(y{\hem\cdot\hem}\partial)\hem \phi \ll \phi$.
It is interesting that
the projector structure of the TBMT equation
derives from the reflection matrix $I^m{}_n$,
which traces back to the inversive geometry
of the Clifford algebra
(cf.\,\cite{doran1996spacetime}).

\paragraph{Antipodal minimal coupling}

When we were introducing the ansatz \eqref{eq:GMOOVform},
the reader may have questioned the possibility of
extending the self-dual part anti-holomorphically:
\begin{align}
\begin{split}
    \label{eq:omega'-antipodal}
    \omega'^+
    &\hem=\,\mem
    \bphi_{\smash{\a\b}\vphantom{\b}}(\bz)\mem 
    \be_{\smash{\da\db}\vphantom{\b}}(\wedgebz{\da\a}{\db\b})
    \,\,=\,\,\hem
    \includegraphics[valign=c,scale=1.0]{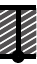}\hem
    \,,\\
    \omega'^-
    &\hem=\,\mem
    \phi_{\smash{\da\db}\vphantom{\b}}(z)\mem 
    \e_{\smash{\a\b}\vphantom{\b}}(\wedgez{\da\a}{\db\b})
    \,\,=\,\,\hem
    \includegraphics[valign=c,scale=1.0]{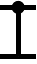}\hem
    \,.
\end{split}
\end{align}
In terms of \eqref{eq:omega'-arbitraryg},
this corresponds to $g {\,=\,} -2$.
The combinatorics of the zig-zag expansions
is the same as that of the minimal coupling:
\begin{align}
\begin{split}
    &
    \omega^{-1}
    =
    \includegraphics[valign=c]{figs/l.pdf}
    \includegraphics[valign=c]{figs/r-b.pdf}
    +
    \includegraphics[valign=c]{figs/l-b.pdf}
    \includegraphics[valign=c]{figs/r.pdf}
    -
    \includegraphics[valign=c]{figs/l.pdf}
    \includegraphics[valign=c]{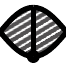}
    \includegraphics[valign=c]{figs/r.pdf}
    \quad
    \text{(heavenly)}
    \,,\\
    &
    {\renewcommand{\arraystretch}{1.5}
    \renewcommand{\arraycolsep}{0em}
    \begin{array}{rl}
        \omega^{-1}
        =
        \includegraphics[valign=c]{figs/l.pdf}
        \includegraphics[valign=c]{figs/r-b.pdf}
        +
        \includegraphics[valign=c]{figs/l-b.pdf}
        \includegraphics[valign=c]{figs/r.pdf}
        {}&{}-
        \includegraphics[valign=c]{figs/l.pdf}
        \includegraphics[valign=c]{figs/A+.pdf}
        \includegraphics[valign=c]{figs/r.pdf}
        -
        \includegraphics[valign=c]{figs/l-b.pdf}
        \includegraphics[valign=c]{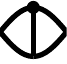}
        \includegraphics[valign=c]{figs/r-b.pdf}
        \\
        {}&{}+
        \includegraphics[valign=c]{figs/l-b.pdf}
        \includegraphics[valign=c]{figs/A-.pdf}
        \includegraphics[valign=c]{figs/A+.pdf}
        \includegraphics[valign=c]{figs/r.pdf}
        +
        \includegraphics[valign=c]{figs/l.pdf}
        \includegraphics[valign=c]{figs/A+.pdf}
        \includegraphics[valign=c]{figs/A-.pdf}
        \includegraphics[valign=c]{figs/r-b.pdf}
        + \cdots
        \quad
        \text{(earthly)}
        \,.
    \end{array}}
\end{split}
\end{align}
We call this case ``antipodal minimal coupling,''
in the sense that its equation of motion is a certain ``inversion'' of that of the minimal coupling.
The self-dual heavenly equation of motion reads
\begin{align}
\begin{split}
    \label{eq:neom-antipodal}
    m_0\hem \dot{\lambda}_\a{}^I
    = -
    (^{\phantom{\hhem\mathcal{I}}}\mathllap{\adjustbox{raise=-0.2ex}{$_\mathcal{I}\hem$}}\bphi(\bz))_\a{}^\b
    \hem\lambda_\b{}^I
    &\,,\quad
    m_0\hem \dot{z}^{\da\a}
    = 
    \minime{p^{\da\a}}
    + 2i\hem
    y^{\da\b}\hem
    (^{\phantom{\hhem\mathcal{I}}}\mathllap{\adjustbox{raise=-0.2ex}{$_\mathcal{I}\hem$}}\bphi(\bz))_\b{}^\a
    \,,\\
    m_0\hem \dot{\rambda}_{I\da}
    = 0
    &\,,\quad
    m_0\hem \dot{\bz}^{\da\a}
    = 
    \minime{p^{\da\a}}
    \,.
    % \,,
\end{split}
\end{align}
One finds the same equations 
by
taking complex conjugate to \eqref{eq:eom-heavenly-components}
and then replacing
$\bphi_\a{}^\b\hnem(\bz)$ with
$(^{\phantom{\hhem\mathcal{I}}}\mathllap{\adjustbox{raise=-0.2ex}{$_\mathcal{I}\hem$}}\bphi(\bz))_\a{}^\b$.

\paragraph{Wilson coefficients on real worldlines}
% \paragraph{\texorpdfstring{Single-$\bm{\omega'}$ Wilson coefficients}{Single-ω' wilson coefficients}}

From the examples \eqref{eq:realform} and \eqref{eq:omega'-arbitraryg},
and further \eqref{eq:omega'-antipodal},
one can conjecture that \eqref{eq:arbomega'+} has sufficient degrees of freedom to implement all types of
% three-point
non-minimal
% interactions
couplings
linear in the ``field strength'':
Levi and Steinhoff \cite{Levi:2015msa}'s Wilson coefficients.

A scalar field $\phi(x)$ on the real section of $\mathbb{CE}^{1,3}$
holomorphically extends into the complex as $\Phi(x,y)$
by the Cauchy-Riemann flow
$\Phi(x,0) = \phi(x)$,
$\partial\Phi/\partial y^m = i\mem \partial\Phi/\partial x^m$
$\implies$
$\Phi(x,y) = \exp(iy^m\partial/\partial x^m)\mem \phi(x)
= \phi(x{\mem+\mem}iy)$.\footnote{
    To avoid clutter,
    we drop the complexification symbol
    % we drop the ``$^\mathbb{C}$'' symbol
    % indicating complexification
    and denote $\phi^\mathbb{C}\hnem(x{\mem+\mem}iy)$ just as $\phi(x{\mem+\mem}iy)$.
}
If the Cauchy-Riemann (holomorphy)
condition ${\partial\Phi\hhhnem/\partial y^m} = i\mem {\partial\Phi\hhhnem/\partial x^m}$ is dropped,
the most general ``analytic'' continuation is then given by
replacing the exponential with a suitably convergent power series
as
$\Phi(x,y) = \big(\mem {1 + \frac{1}{1!}\mem C_1 (iy^m\partial/\partial x^m) + \frac{1}{2!}\mem C_2 (iy^m\partial/\partial x^m)^2 + \cdots\hem} \mem\big)\mem \phi(x)$.
This ``non-holomorphic analytic continuation'' equally applies to differential forms 
by replacing 
$y^m\partial/\partial x^m$ with $\pounds_y$,
where $y := y^m\hem \partial/\partial x^m|_y$ denotes the spin length as a vector field in $\mathbb{CE}^{1,3}$.
% (and in $\mathbb{MA}$ as well).
Then the constants $C_1,C_2,\cdots$ are essentially the Wilson coefficients of \cite{Levi:2015msa}.

Hence,
we consider generic 
% non-minimal
symplectic perturbations of the form
\begin{align}
    \label{eq:Wilson.Cdef}
    \omega'^\pm =
    C^\pm\hnem(\pm i\pounds_y)\, 
    [\hem\minie\mem \phi^\pm_{mn}\hnem(x)\mem \wedgetwo{x}{x}{m}{n}\hem]
    % \omega'^+\hnem(z,\bz) =
    % C^+\nem(i\pounds_y)\, \omega'^+\hnem(x)
    % \,,\quad
    % \omega'^-\hnem(z,\bz) =
    % C^-\nem(- i\pounds_y)\, \omega'^-\hnem(x)
    \,,\quad
    C^\pm\nem(\xi)
    =\textstyle \sum_{\ell=0}^\infty 
    {\textstyle\frac{1}{\ell!}\mem}
    C^\pm_\ell\mem \xi^\ell
    \,,
\end{align}
with $C^\pm\nem(\xi)$ being complex-analytic functions such that
$[C^+\nem(\xi)]^* = C^-\nem(\bar{\xi})$ and
$|C^\pm\nem(0)| = 1$.
The closure condition is guaranteed by $[d,\pounds_y] = 0$,
provided that $d(\minie\mem \phi_{mn}\hnem(x)\mem \wedgetwo{x}{x}{m}{n}) = 0$.
% 
% $C^\pm\nem(0)$ are not necessarily equal to $1$.
Note that we 
leave
% have left
a possibility for
$C^\pm\nem(0)$
to carry a phase:
$C^\pm\nem(0) = \mathe^{\mp ih\varphi}$.
As argued 
% % around
% % \eqref{eq:dualrotation}-\eqref{eq:starfruit},
in \cite{spt},
$
    \omega'^\pm
        \mapsto
    \mathe^{\mp ih\varphi}\mem \omega'^\pm
$
describes the electric-magnetic duality rotation
for helicity-$h$ fields
if certain conditions are met.

\paragraph{Wilson coefficients on complex worldlines}

The minimal coupling \eqref{eq:GMOOVform} is described by
the usual analytic continuation,
$C^+\nem(\xi) = C^-\nem(\xi) = \mathe^\xi$:
% \begin{subequations}
\begin{align}
\begin{split}
    \minie\mem \phi^+_{mn}\hnem(z)\mem \wedgetwo{z}{z}{m}{n}
    &\,=\, \mathe^{+i\pounds_y}\, 
    [\hem\minie\mem \phi^+_{mn}\hnem(x)\mem \wedgetwo{x}{x}{m}{n}\hem]
    \,,\\
    \minie\mem \phi^-_{mn}\hnem(\bz)\mem \wedgetwo{\bz}{\bz}{m}{n}
    &\,=\, \mathe^{-i\pounds_y}\, 
    [\hem\minie\mem \phi^-_{mn}\hnem(x)\mem \wedgetwo{x}{x}{m}{n}\hem]
    \,.
\end{split}
\end{align}
% \end{subequations}
Now we redefine the Wilson coefficients with respect to the minimal coupling as
\begin{align}
    \label{eq:Wilson.Cfrelation}
    C^\pm\nem(\xi)
    \mem=:\mem \mathe^\xi \cdot f^\pm\hnem(\xi)
    &\,,\quad
    f^\pm\nem(\xi)
    =\textstyle \sum_{\ell=0}^\infty f^\pm_\ell\mem \xi^\ell
    \,.
\end{align}
Then \eqref{eq:Wilson.Cdef} reads
% \begin{subequations}
\begin{align}
    \label{eq:Wilson.fdef}
\begin{split}
    % \label{eq:Wilson.f+def}
    \omega'^+ &=
    f^+\hnem(+ i\pounds_y)\, 
    [\hem\minie\mem \phi^+_{mn}\hnem(z)\mem \wedgetwo{z}{z}{m}{n}\hem]
    \,,\\
    % \label{eq:Wilson.f-def}
    \omega'^- &=
    f^-\hnem(- i\pounds_y)\, 
    [\hem\minie\mem \phi^-_{mn}\hnem(\bz)\mem \wedgetwo{\bz}{\bz}{m}{n}\hem]
    \,.
\end{split}
\end{align}
% \end{subequations}
An intuitive 
zig-zag 
picture for \eqref{eq:Wilson.fdef} will be 
a set of ``distributed insertions'' on the ribbon
with all types of zig-zag components.

\paragraph{Non-minimal symplectic perturbation}

Direct computation shows that 
% the non-holomorphic analytic continuation of a two-form is given as
\begin{align}
    \label{eq:omega'-arbcompute}
    &
    f(i\pounds_y)\mem
    \big({
        \minie\mem \phi_{mn}\hnem(z)\mem \wedgetwo{z}{z}{m}{n}
    }\big)
    % \,,
    \\
    &=
    \left({
    {\renewcommand{\arraycolsep}{0em}
    \renewcommand{\arraystretch}{1.1}
    \begin{array}{rl}
        &
        \big({
            f\hnem(iy{\mem\cdot\mem}\partial)\mem \phi_{mn}\hnem(z)
        }\big)\mem
        {\textstyle\frac{1}{2}}\mem\wedgetwo{z}{z}{m}{n}    
        \\
        + &\big({
            f'\hnem(iy{\mem\cdot\mem}\partial)\mem \phi_{mn}\hnem(z)
        }\big)\mem
        i\mem\wedgetwo{y}{z}{m}{n}    
        \\
        + &\big({
            f''\hnem(iy{\mem\cdot\mem}\partial)\mem \phi_{mn}\hnem(z)
        }\big)\mem
        {\textstyle\frac{i^2}{2}}\mem\wedgetwo{y}{y}{m}{n} 
    \end{array}}
    }\nem\right)
    =
    \left({
    {\renewcommand{\arraycolsep}{0em}
    \renewcommand{\arraystretch}{1.1}
    \begin{array}{rl}
        &
        \big({
            {^0\mathfrak{S}}[f](iy{\mem\cdot\mem}\partial)\mem \phi_{mn}\hnem(z)
        }\big)\mem
        {\textstyle\frac{1}{2}}\mem\wedgetwo{z}{z}{m}{n}    
        \\
        + &\big({
            {^1\mathfrak{S}}[f](iy{\mem\cdot\mem}\partial)\mem \phi_{mn}\hnem(z)
        }\big)\mem
        \wedgetwo{z}{\bz}{m}{n}    
        \\
        + &\big({
            {^2\mathfrak{S}}[f](iy{\mem\cdot\mem}\partial)\mem \phi_{mn}\hnem(z)
        }\big)\mem
        {\textstyle\frac{1}{2}}\mem\wedgetwo{\bz}{\bz}{m}{n} 
    \end{array}}
    }\nem\right)
    \,,
    \nonumber
\end{align}
where by $y{\mem\cdot\mem}\partial$ we mean $\pounds_y$ acting on scalar functions.
This computation holds regardless of 
the closure of $\minie\mem \phi_{mn}\hnem(z)\mem \wedgetwo{z}{z}{m}{n}$.
We have defined the polynomials
% The $(2,0)$-, $(1,1)$-, and $(0,2)$-components are controlled by 
% \begin{align}
% \begin{split}
%     \label{eq:nonmin-SSigmas}
%     {^0\mathfrak{S}}[f](\xi)
%     &:= 
%     (1 + (d/d\xi) + {\textstyle\frac{1}{4}}\kern0.015em(d/d\xi)^2)
%     \,f\hnem(\xi)
%     \,,\quad
%     {^{0+1}\mathfrak{S}}[f](\xi)
%     := {^0\mathfrak{S}}[f](\xi) 
%     + {^1\mathfrak{S}}[f](\xi)
%     \,,\\
%     {^1\mathfrak{S}}[f](\xi)
%     &:= 
%     ( -{\textstyle\frac{1}{2}}\kern0.015em (d/d\xi) - {\textstyle\frac{1}{4}}\kern0.015em(d/d\xi)^2)
%     \,f\hnem(\xi)
%     \,,\quad
%     \,\,\hhnem
%     {^{1+2}\mathfrak{S}}[f](\xi)
%     := {^1\mathfrak{S}}[f](\xi) 
%     + {^2\mathfrak{S}}[f](\xi)
%     \,,\\
%     {^2\mathfrak{S}}[f](\xi)
%     &:= 
%     ({\textstyle\frac{1}{4}}\kern0.015em(d/d\xi)^2)
%     \,f\hnem(\xi)
%     \,,\quad
%     {^0\mathfrak{S}}[f](\xi) 
%     +
%     2
%     {^1\mathfrak{S}}[f](\xi)
%     +
%     {^2\mathfrak{S}}[f](\xi)
%     = f(\xi)
%     \,.
% \end{split}
% % \\
% %     \label{eq:SSigma-addup}
% %     {^0\mathfrak{S}}[f](\xi) 
% %     &{}+
% %     2
% %     {^1\mathfrak{S}}[f](\xi)
% %     +
% %     {^2\mathfrak{S}}[f](\xi)
% %     = f(\xi)
% %     \,.
% \end{align}
\begin{align}
    \nonumber
    {^0\mathfrak{S}}[f](\xi)
    := 
    (1 + (d/d\xi) + {\textstyle\frac{1}{4}}\kern0.015em(d/d\xi)^2)
    \,f\hnem(\xi)
    &\,,\quad
    {^{0+1}\mathfrak{S}}[f](\xi)
    := {^0\mathfrak{S}}[f](\xi) 
    + {^1\mathfrak{S}}[f](\xi)
    \,,\\
    \label{eq:nonmin-SSigmas}
    {^1\mathfrak{S}}[f](\xi)
    := 
    ( -{\textstyle\frac{1}{2}}\kern0.015em (d/d\xi) - {\textstyle\frac{1}{4}}\kern0.015em(d/d\xi)^2)
    \,f\hnem(\xi)
    &\,,\quad
    {^{1+2}\mathfrak{S}}[f](\xi)
    := {^1\mathfrak{S}}[f](\xi) 
    + {^2\mathfrak{S}}[f](\xi)
    \,,\\
    \nonumber
    {^2\mathfrak{S}}[f](\xi)
    := 
    ({\textstyle\frac{1}{4}}\kern0.015em(d/d\xi)^2)
    \,f\hnem(\xi)
    &\,,\quad
    % {^0\mathfrak{S}}[f](\xi) 
    % +
    % 2\hem
    % {^1\mathfrak{S}}[f](\xi)
    % +
    % {^2\mathfrak{S}}[f](\xi)
    % = f(\xi)
    {^{0+1}\mathfrak{S}}[f](\xi) 
    +
    {^{1+2}\mathfrak{S}}[f](\xi)
    = f(\xi)
    \,.
\end{align}
We emphasize that a Newman-Janis shift of a generic tensor field
must shift not only the components but also the accompanying basis vectors/one-forms.
% Although no much attention has been paid to this point so far,
The $dy$ terms that the Lie derivative $\pounds_y$ induces
are crucial for
a) the closure of \eqref{eq:omega'-arbcompute},
b) incorporating time-varying spin, 
and
c) producing the correct 
``zig-zag vertices'' in 
% the
our
holomorphic/anti-holomorphic
perturbation theory.
Yet, not much attention has been paid to this point
in the literature.

From
% the calculation
\eqref{eq:omega'-arbcompute},
we conclude that \eqref{eq:Wilson.fdef} matches with \eqref{eq:arbomega'+} and \eqref{eq:arbomega'-} as
\begin{align}
\begin{split}
    \label{eq:012omegass}
    {^0\omega'^+_{mn}}
    =
    {^0\mathfrak{S}}[f^+](iy{\mem\cdot\mem}\partial)\mem \phi^+_{mn}\hnem(z)
    \,,\quad
    {^0\omega'^-_{mn}}
    =
    {^0\mathfrak{S}}[f^-](-iy{\mem\cdot\mem}\partial)\mem \phi^-_{mn}\hnem(\bz)
    \,,\\
    {^1\omega'^+_{mn}}
    =
    {^1\mathfrak{S}}[f^+](iy{\mem\cdot\mem}\partial)\mem \phi^+_{mn}\hnem(z)
    \,,\quad
    {^1\omega'^-_{mn}}
    =
    {^1\mathfrak{S}}[f^-](-iy{\mem\cdot\mem}\partial)\mem \phi^-_{mn}\hnem(\bz)
    \,,\\
    {^2\omega'^+_{mn}}
    =
    {^2\mathfrak{S}}[f^+](iy{\mem\cdot\mem}\partial)\mem \phi^+_{mn}\hnem(z)
    \,,\quad
    {^2\omega'^-_{mn}}
    =
    {^2\mathfrak{S}}[f^-](-iy{\mem\cdot\mem}\partial)\mem \phi^-_{mn}\hnem(\bz)
    \,.
\end{split}
\end{align}
Plugging in \eqref{eq:012omegass} to \eqref{eq:eom-nonmin.hybrid}-\eqref{eq:eom-nonmin.vector},
one obtains the generic non-minimal heavenly equation of motion that incorporates all the single-curvature Wilson coefficients.
It is easy to check that 
$f^+\hnem(\xi) = f^-\hnem(\xi) = 1$
derives \eqref{eq:eom-heavenly-components} and \eqref{eq:g=2eom},
while 
$f^+\hnem(\xi) = f^-\hnem(\xi) = \mathe^{-\xi}$
derives \eqref{eq:g=0eom}.
The ``interpolation'' \eqref{eq:omega'-arbitraryg} 
is described by
$f^+\hnem(\xi) = f^-\hnem(\xi) = \mathe^{(g/2-1)\xi}$.

\subsection{Impulse and classical-spin amplitudes}
\label{sec:Flat.amplitude}

Finally, in this last subsection, we obtain
classical-spin
scattering amplitudes from
Born approximation in a monochromatic plane wave background
\begin{align}
    \label{eq:pwH}
    \phi_{\da\db}(z)
    % \omega'^+
    =    
    (i\nu/2)\mem
    \bo_\da\hem \bo_\db
    % \mem (dz \swedge dz)^{\da\db}
    \mem\mathe^{-i\hem\bozo}
    % \,,\quad
    % (dz \swedge dz)^{\da\db}
    % := \e_{\a\b}(\wedgetwo{z}{z}{\da\a}{\db\b})
    \,,
\end{align}
which 
% describes
is interpreted as describing
a positive-frequency field quantum carrying positive helicity \cite{bialynicki1981note,ashtekar1986note}.
Geometrically,
the spinor-helicity variables
$\bo_\da$ and $\o_\a {\mem=\mem} [\bo_\da]^*$
of the field quanta
introduce a principal null direction to spacetime.
Accordingly, we install a Newman-Penrose spin frame \cite{newman1962approachspin}
$\langle\i\o\rangle = \e^{\a\b} \o_\a \i_\b = 1
= \be^{\da\db} \bo_\da \hem \bi_\db = \lsq\bo\bi\rsq$.
Both the ``relativist'' and ``amplitudes'' notations will be used interchangeably:
$\o_\a =: \three_\a$, $\bo_\da =: \bthree_\da$.

\paragraph{Complexified 3pt on-shell kinematics}

\begin{figure}[t]
    \centering
    \includegraphics[width=0.9\linewidth]{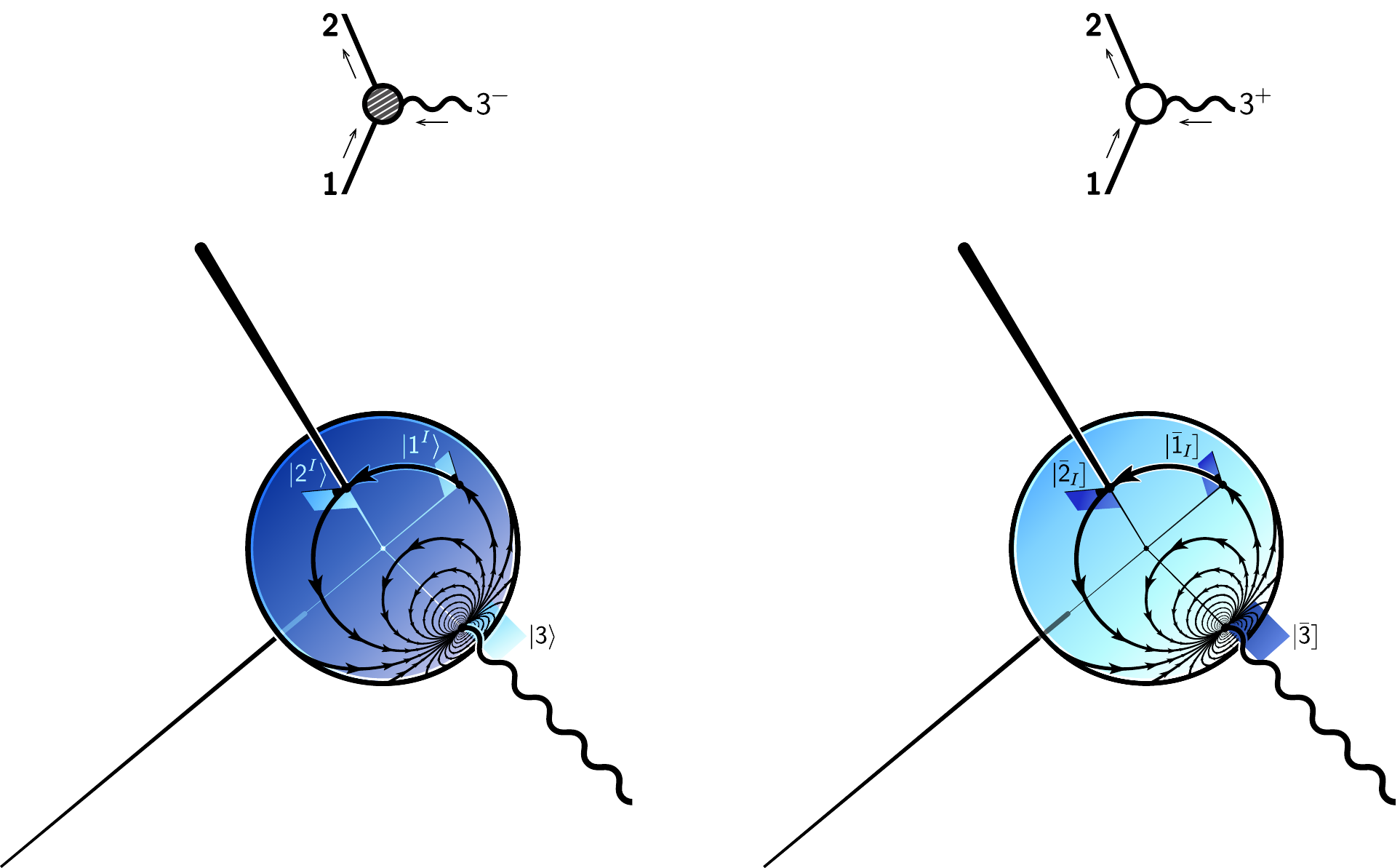}
    \caption{
        Null rotation on the left and right skies.
        As a conformal transformation
        on the celestial sphere, it preserves the angle that 
        the null flag of the spinor being transformed
        makes with the circular flow lines.
        This provides a geometrical way of seeing 
        % that 
        $\langle \three \two^I \rangle = \langle \three \one^I \rangle$
        or
        $[\btwo_I\bthree] = [\bone_I\bthree]$
        \cite{penrose1984spinors1}.
        %% visualization: equal-mass ✓ / colinearity ✗
    }
    \label{fig:celestial}
\end{figure}

As a preliminary, we introduce a one-parameter family of complexified on-shell kinematics
for equal-mass massive-massive-massless amplitudes.
This is relevant to our discussion because
the massive ambitwistor description computes 
not only the momentum impulse $\mathit{\Delta} p_{\a\da}$ but also $\mathit{\Delta}\lambda_\a{}^I$ and $\mathit{\Delta}\rambda_{I\da}$ separately.
Suppose
% if
a particle of fixed mass $m$ gets deflected by absorbing a massless momentum $-\three_\a\bthree_\da$:
\begin{align}
    \label{eq:3kin-conservation}
    |\two^I\rangle \lsq\btwo_I|
    = |\one^I\rangle \lsq\bone_I|
    + |\three\rangle \lsq\bthree|
    % \,\,\text{(conservation)}
    \,,\quad
    |\one^I\rangle [\bone_I \bthree] = m x |\three\rangle
    % \,\,\text{(equal-mass)}
    \,.
\end{align}
The $x$-factor encodes the equal-mass condition.
The real kinematics is given by (e.g., \cite{arkani2020kerr})
\begin{align}
\begin{split}
    \label{eq:3kin-real}
    |\two^I\rangle
    &{}= |\one^I\rangle + (x/2m)\mem |\three\rangle\langle\three\one^I\rangle
    \,,\quad
    \lsq\btwo_I|
    {}= \lsq\bone_I| + (1/2mx)\mem \lsq\bone_I\bthree\rsq\lsq\bthree|
    \,,
\end{split}
\end{align}
which is the unique solution to \eqref{eq:3kin-conservation}
that satisfies the reality condition $[\one_\a{}^I]^* = \bone_{I\da}$, $[\two_\a{}^I]^*$ $= \btwo_{I\da}$.
But, there is a one-parameter family of solutions
if we drop the reality condition:
\begin{align}
\begin{split}
    \label{eq:3kin-g}
    |\two^I\rangle
    &{}= |\one^I\rangle + (1-\vartheta/2)\hem (x/2m)\mem |\three\rangle\langle\three\one^I\rangle
    \,,\\
    \lsq\btwo_I|
    &{}= \lsq\bone_I| + (1+\vartheta/2)\hem (1/2mx)\mem \lsq\bone_I\bthree\rsq\lsq\bthree|
    \,.
\end{split}
\end{align}
The parameter $\vartheta$ controls the ratio between the contributions from $\mathit{\Delta}\lambda_\a{}^I$ and $\mathit{\Delta}\rambda_{I\da}$
that add up to $\mathit{\Delta}p_{\a\da} = -\three_\a\bthree_\da$.
As depicted in the left side of Figure \ref{fig:celestial},
the equation $|\two^I\rangle = |\one^I\rangle + {t\mem |\three\rangle \langle\three\one^I\rangle}$ has a geometric interpretation as a null rotation \cite{penrose1984spinors1} on the celestial sphere
that takes the flagpole direction of the spinor $|\three\rangle$ as the fixed point.
% where $t$ is the null rotation parameter.
% The complexified kinematics \eqref{eq:3kin-g} is possible
% when the celestial spheres of right-handed and left-handed spinors are allowed to transform differently.
Then \eqref{eq:3kin-g} is a complexified Lorentz transformation that
null rotates the celestial spheres of the right-handed and left-handed spinors differently.

In analogy with 
% MHV diagrams
the MHV kinematics
\cite{arkani2012positivegrassmannian}, we consider the following two extreme cases.
\begin{subequations}
    \label{eq:3kin}
\begin{align}
    \label{eq:3kin-plus}
    \includegraphics[scale=1.0,valign=c]{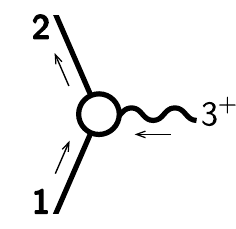}
    &\,\,\,\leftrightarrow\,\,\,
    \left\{
    {\renewcommand{\arraycolsep}{0em}
    \renewcommand{\arraystretch}{1.15}
    \begin{array}{rl}
        |\two^I\rangle
        &{}= |\one^I\rangle
        % \,,
        \phantom{{}+ (x/m)\mem |\three\rangle\langle\three\one^I\rangle}
        \\
        \lsq\btwo_I|
        &{}= \lsq\bone_I| + (1/mx)\mem \lsq\bone_I\bthree\rsq\lsq\bthree|
        % \,.
    \end{array}
    }
    \right.
    % \,\,\,\sim\,\,\,
    % \includegraphics[valign=c]{figs/l-b.pdf}
    % \includegraphics[valign=c]{figs/v+.pdf}
    % \includegraphics[valign=c]{figs/xD-b.pdf}
    \\
    \label{eq:3kin-minus}
    \includegraphics[scale=1.0,valign=c]{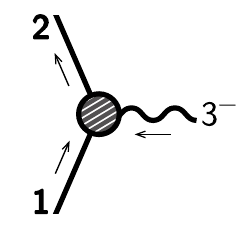}
    &\,\,\,\leftrightarrow\,\,\,
    \left\{
    {\renewcommand{\arraycolsep}{0em}
    \renewcommand{\arraystretch}{1.15}
    \begin{array}{rl}
        |\two^I\rangle
        &{}= |\one^I\rangle + (x/m)\mem |\three\rangle\langle\three\one^I\rangle
        % \,,
        \\
        \lsq\btwo_I|
        &{}= \lsq\bone_I|
        % \,.
        \phantom{{}+ (1/mx)\mem \lsq\bone_I\bthree\rsq\lsq\bthree|}
    \end{array}
    }
    \right.
    % \,\,\,\sim\,\,\,
    % \includegraphics[valign=c]{figs/l.pdf}
    % \includegraphics[valign=c]{figs/v-.pdf}
    % \includegraphics[valign=c]{figs/xD.pdf}
\end{align}
\end{subequations}
These complexified kinematics 
naturally arise from the minimally coupled heavenly equations of motion;
for instance, \eqref{eq:eom-heavenly-components} parallel-transports $\lambda_\a{}^I$ while preserving $\bDelta$.
They are associated with zig-zag vector fields 
(as operators on the $(\lambda,\rambda)$ Lagrange submanifold, say) 
as the following:
\begin{align}
    \label{eq:cqsim}
    \includegraphics[scale=1.0,valign=c]{figs/hkin-.pdf}
    \,\,\,\sim\,\,\,  
    \includegraphics[valign=c]{figs/l.pdf}
    \includegraphics[valign=c]{figs/v-.pdf}
    \includegraphics[valign=c]{figs/xD.pdf}
    \,,\qquad\quad
    \includegraphics[scale=1.0,valign=c]{figs/hkin+.pdf}
    \,\,\,\sim\,\,\,  
    \includegraphics[valign=c]{figs/l-b.pdf}
    \includegraphics[valign=c]{figs/v+.pdf}
    \includegraphics[valign=c]{figs/xD-b.pdf}
    \,.
\end{align}

\paragraph{Minimal impulse and amplitude}

The amplitudes
can be conveniently obtained by
computing the interaction action for a zeroth-order trajectory in an on-shell plane wave background (cf.\,Appendix A of \cite{aoude2021classical}).
However, to keep pursuing the ``symplectic form only'' philosophy, we rather relate the zig-zag expansions to amplitudes by computing the impulses.
% Recall \eqref{eq:pwH}.
The zig-zag vector field \eqref{eq:heavenly-3ptX} is evaluated for \eqref{eq:pwH} as
\begin{align}
\begin{split}
    -
    \smash{
        \includegraphics[valign=c]{figs/l-b.pdf}
        \includegraphics[valign=c]{figs/v+.pdf}
        \includegraphics[valign=c]{figs/xD-b.pdf}
    }
    \,\mem&=\,\, 
        \{\blank,z^{\da\a}\}^\circ\,\mem
        (2/\Delta)\mem p_{\smash{\a\db}\vphantom{\da}}\mem \phi^\db{}_\da(z)
    \,,\\
    \label{eq:3vec+1}
    &=\,\,
    \{\blank,z^{\da\a}\}^\circ\,\mem
    i\big(\mem
        \nu\mem p_{\smash{\a\db}\vphantom{\b}} {\kern0.01em} \bo^\db \bo_{\smash{\da}\vphantom{\a}}
    \hem\big)\mem
    \mathe^{-i\hem\bozo}
    /\Delta
    \,,\\
    &=\,\,
    \{\blank,z^{\da\a}\}^\circ\,\mem
    i\big(\mem
        \hnem\nu
        \langle\i|p|\bo\rsq
        \hem\mathe^{\bo y \o}
        \mem
        \o_\a \bo_\da
        -
        \nu
        \langle\o|p|\bo\rsq
        \hem\mathe^{\bo y \o}
        \mem
        \i_\a \bo_\da
    \hem\big)\mem
    \mathe^{-i\bo x \o}
    /\Delta
    \,.
\end{split}
\end{align}
We have used $\delta_\a{}^\b = \o_\a \i^\b - \i_\a \o^\b$ in the last line.
The heavenly ``ambitwistor impulse'' follows from
integrating \eqref{eq:3vec+1} over time:\footnote{
    One could argue that
    we should rather
    not straightforwardly integrate \eqref{eq:3vec+1} but
    evaluate the impulse on the space of 
    ``boundary'' data.
    % See appendix \ref{app:Shilb} for a more sophisticated treatment.
    % This caveat does not affect the spinorial frame impulse under a constant Regge trajectory.
    Yet, this caveat does not affect
    the constant-mass spinorial frame impulse we are considering here (and $\mathit{\Delta}y^{\da\a}$ as well).
    See \eqref{eq:intpicture}
    % -\eqref{eq:exactimpulse-lrambda}
    for a more accurate treatment.
}
\begin{align}
    \label{eq:twistorimpulse}
    \mathit{\Delta}Z_\rmA{}^I
    = 0
    \,,\quad
    \mathit{\Delta}\bZ_I{}^\rmA
    = \int_{-\infty}^{+\infty} \frac{d\tau}{-2}\,
    \Big\langle
        d\bZ_I{}^\rmA , 
        -
        \smash{
            \includegraphics[valign=c]{figs/l-b.pdf}
            \includegraphics[valign=c]{figs/v+.pdf}
            \includegraphics[valign=c]{figs/xD-b.pdf}
        }
    \Big\rangle
    \,.
\end{align}
% Recall that 
The recipe to obtain the amplitude is 
identifying
% to identify 
the amount of impulse expected from the particle interpretation (absorption of a single quantum) 
and then
reading off the proportionality factor from $\mathit{\Delta} p_{\a\da}$.
In the first-order approximation,
the integration
\eqref{eq:twistorimpulse} 
does nothing other than inducing the delta function $\deltabar(\langle\o|\one|\bo\rsq\hem)$.
Thus, we can just directly read off the ``amplitude times momentum kick'' 
$(\nu m_0x\mem\mathe^{\three\cdot a})(-\three_\a\bthree_\da)$
% \begin{align}
%     (\nu m_0x\mem\mathe^{\three\cdot a})(-\three_\a\bthree_\da)
% \end{align}
from the bracketed term in the last line of \eqref{eq:3vec+1}
by putting $\langle\o|p|\bo\rsq = 0$
and plugging in the scattering data 
$p_{\a\da} {\,=\,} \one_{\a\da}$, $y^{\da\a} {\,=\,} {{-}a^{\da\a}}$.
% (while $-\langle\i|\one|\bo\rsq = mx$).
% 
In this sense, the vector field \eqref{eq:3vec+1} is a ``symplectic avatar'' of the positive-helicity three-point amplitude.
% as we suggested in \eqref{eq:cqsim}.
Note
that it performs a null rotation on $\rambda_{I\da}$, $y^{\da\a}$:
%% "take (z,y) as fundamental" -> \sigma(z,\bz)...
\begin{align}
    \label{eq:zbracket-as-nullrot}
    \{\blank,z^{\da\a}\}^\circ\,\mem
    i\nu\mem p_{\smash{\a\db}\vphantom{\b}} {\kern0.01em} \bo^\db \bo_{\smash{\da}\vphantom{\a}}
    \mem\frac{\mathe^{-i\hem\bozo}}{\Delta}
    \,=\,
    \bigg(\hem{
        \rambda_{I\db} \mem\frac{\partial}{\partial\rambda_{I\da}}
        - y^{\da\b} \hnem\nem\nem\left.\frac{\partial}{\partial y^{\db\b}}\right|_z
    \mem}\bigg)\mem\hem
    i\nu\mem \bo^\db \bo_{\smash{\da}\vphantom{\a}}
    \mem\frac{\mathe^{-i\hem\bozo}}{\Delta}
    \,.
\end{align}

The amplitude $\nu m_0x\mem\mathe^{\three\cdot a}$ is precisely the scalar amplitude $e_h\hem m_0 x^h$ \cite{ahh2017}
times the minimal exponential spin factor \cite{guevara2019scattering,arkani2020kerr}
if 
\begin{align}
    \label{eq:e1e2}
    \nu = e_h\hem x^{h-1}
    \,,
\end{align}
where $e_h$ collectively denotes the coupling constants of Yang-Mills ($h{\mem=\mem}1$) and gravity ($h{\mem=\mem}2$).
We revisit this finding later in section \ref{sec:Wave.caveat}.

\paragraph{Generic impulse and amplitude}
% \paragraph{Non-minimal impulse and amplitude}

For non-minimal couplings, 
the interaction with the self-dual plane wave
alters both $\lambda_\a{}^I$ and $\rambda_{I\da}$.
Consider the maximally non-minimal coupling for an example;
the first two ``avatar'' vector fields \eqref{eq:all3ptX-0}-\eqref{eq:all3ptX-1ramb} contribute to the first-order $\mathit{\Delta}\rambda_{I\da}$,
while \eqref{eq:all3ptX-1lamb}-\eqref{eq:all3ptX-2} contribute to $\mathit{\Delta}\lambda_\a{}^I$.
Generalizing, one finds
\begin{align}
\begin{split}
    \label{eq:nonmin-nullrot+[f]}
    \mathit{\Delta}^{(1)} \lsq\rambda_I|
    &\mem=\, i 
    \Big(\mem\hem
        \nu m_0x\mem\mathe^{\three\cdot a}
        \,{\cdot}\,
        {^{0+1}\mathfrak{S}[f^+](\three{\kern0.02em\cdot\kern0.02em}a)}
    \hem\Big)\,
    \deltabar(2\mem\three{\kern0.02em\cdot\kern0.02em}\one)\,\mem
    {\textstyle\frac{1}{m_0x}}
    \lsq\bone_I\bthree\rsq\lsq\bthree|
    \,,\\
    \mathit{\Delta}^{(1)} |\lambda^I\rangle
    &\mem=\, i 
    \Big(\mem\hem
        \nu m_0x\mem\mathe^{\three\cdot a}
        \,{\cdot}\,
        {^{1+2}\mathfrak{S}[f^+](\three{\kern0.02em\cdot\kern0.02em}a)}
    \hem\Big)\,
    \deltabar(2\mem\three{\kern0.02em\cdot\kern0.02em}\one)\,
    {\textstyle\frac{x}{m_0}} \hem|\three\rangle\langle\three \one^I\rangle
    \,,
\end{split}
\end{align}
while replacing $iy^m\partial/\partial x^m$ with 
$i(-a^m)(i\three_m) = \three{\mem\cdot\mem}a$.
The $x$-factor is defined on the support of $\deltabar(2\mem\three{\mem\cdot\mem}\one)$.
Adding up these two
and using the identity 
${^0\mathfrak{S}[f](\xi)} + 2\mem {^1\mathfrak{S}[f](\xi)} + {^2\mathfrak{S}[f](\xi)}
= f(\xi)$, we find the momentum impulse
\begin{align}
    \mathit{\Delta}^{(1)}p_{\a\da}
    = i 
    \Big(\mem\hem
        \nu m_0x\mem\mathe^{\three\cdot a}
        \,{\cdot}\,
        f^+\hnem(\three{\kern0.02em\cdot\kern0.02em}a)
    \hem\Big)\,
    \deltabar(2\mem\three{\kern0.02em\cdot\kern0.02em}\one)\,
    (-\three_\a\bthree_\da)
    \,.
\end{align}
For concreteness, we also spell out the negative-helicity counterparts.
A positive-frequency field quantum carrying negative helicity is described by
$\bphi_{\a\b}\hnem(\bz) = (i\nu/2)\mem \o_\a\o_\b\mem \mathe^{-i\bo \bz \o}$
(note that it is not the straightforward complex conjugate of \eqref{eq:pwH}).
We find
\begin{align}
\begin{split}
    \label{eq:nonmin-nullrot-[f]}
    \mathit{\Delta}^{(1)} \lsq\rambda_I|
    &\mem=\, i 
    \Big(\mem\hem
        \nu m_0x^{-1}\mem\mathe^{-\three{\mem\cdot\mem}a}
        \,{\cdot}\,
        {^{1+2}\mathfrak{S}[f^-](-\three{\kern0.02em\cdot\kern0.02em}a)}
    \hem\Big)\,
    \deltabar(2\mem\three{\kern0.02em\cdot\kern0.02em}\one)\,\mem
    {\textstyle\frac{1}{m_0x}}
    \lsq\bone_I\bthree\rsq\lsq\bthree|
    \,,\\
    \mathit{\Delta}^{(1)} |\lambda^I\rangle
    &\mem=\, i 
    \Big(\mem\hem
        \nu m_0x^{-1}\mem\mathe^{-\three{\mem\cdot\mem}a}
        \,{\cdot}\,
        {^{0+1}\mathfrak{S}[f^-](-\three{\kern0.02em\cdot\kern0.02em}a)}
    \hem\Big)\,
    \deltabar(2\mem\three{\kern0.02em\cdot\kern0.02em}\one)\,
    {\textstyle\frac{x}{m_0}} \hem|\three\rangle\langle\three \one^I\rangle
    \,,
\end{split}\\
    \mathit{\Delta}^{(1)}p_{\a\da}
    &\mem=\, i 
    \Big(\mem\hem
        \nu m_0x^{-1}\mem\mathe^{-\three{\mem\cdot\mem}a}
        \,{\cdot}\,
        f^-\hnem(-\three{\kern0.02em\cdot\kern0.02em}a)
    \hem\Big)\,
    \deltabar(2\mem\three{\kern0.02em\cdot\kern0.02em}\one)\,
    (-\three_\a\bthree_\da)
    \,.
\end{align}
The helicity flip is effectively implemented by the sign change of the pseudovector $a^m$.

To summarize, we have obtained the minimal and non-minimal amplitudes as follows:
\begin{align}
    \label{eq:born}
    \M^{\pm h}\hnem(\one,a;\three)
    = e_h\hem m_0 x^h \mem\hem \mathe^{\pm\three\cdot a}
    \,,\quad
    \M^{\pm h}\hnem[f^\pm](\one,a;\three)
    = \M^{\pm h}\hnem(\one,a;\three)
    \,{\cdot}\,
    f^\pm\hnem(\pm\three{\kern0.02em\cdot\kern0.02em}a)
    \,.
\end{align}
For example, the ``exponential'' arbitrary-$g$ coupling \eqref{eq:omega'-arbitraryg} leads to
\begin{align}
\begin{split}
    f^\pm\hnem(\xi) = \mathe^{(g/2-1)\xi}
    &\qiq
    \M^{\pm h}\hnem[f^\pm](\one,a;\three)
    =
    \M^{\pm h}\hnem(\one,a;\three)
    \,{\cdot}\,
     \mathe^{\pm(g/2-1)\hem\three{\kern0.02em\cdot\kern0.02em}a}
     \,.
\end{split}
\end{align}
Moreover, we also derived classical complexified kinematics corresponding to the non-minimal amplitudes.
Peeling off the amplitudes $\M^{\pm h}\hnem[f^\pm](\one,a;\three)$
from the impulses \eqref{eq:nonmin-nullrot+[f]} and \eqref{eq:nonmin-nullrot-[f]},
we find that
the ``asymmetry parameter'' $\vartheta$ of \eqref{eq:3kin-g} is given by
\begin{align}
    \pm \vartheta \,=\, 
    \minie\mem 
    {^{0-2}\mathfrak{S}[f^\pm](\pm\three{\kern0.02em\cdot\kern0.02em}a)}
    \mem:=\mem
    \minie\big(\mem{
        {^0\mathfrak{S}[f^\pm](\pm\three{\kern0.02em\cdot\kern0.02em}a)}
        -
        {^2\mathfrak{S}[f^\pm](\pm\three{\kern0.02em\cdot\kern0.02em}a)}
    }\big)
\end{align}
for positive-helicity and negative-helicity amplitudes, respectively.
It is easy to see that \eqref{eq:omega'-arbitraryg} leads to $\vartheta = g$.
Especially, $g=0$ leads to the real kinematics \eqref{eq:3kin-real}.

\paragraph{Matching with definite-spin amplitudes}

Finally, we explicate the connection with the discrete definite-spin amplitudes.
Previous studies \cite{chkl2019,gmoov,guevara2019scattering,guevara2019black,Chung:2019duq,aoude2021classical}
have established a dictionary between the Wilson coefficients $C^\pm_\ell$ and the 
coupling constants $g^\pm_\ell$ appearing in the generic three-point amplitude \cite{ahh2017}:
\begin{subequations}
    \label{eq:3ptAahh}
\begin{align}
    % \label{eq:3ptAahh}
% \begin{split}
    \label{eq:3ptAahh+}
    \mathcal{A}^{+h}_{(2s)}\hnem[g^+](\two{\leftarrow}\one;\three)
    &\mem=\mem
    e_h\hem m_0 x^h
    \mem\sum_{\ell=0}^{2s}\mem
    g^+_\ell\mem
    \langle \two^{-1} \one \rangle^{\kern-0.05em\odot(2s-\ell)}
    {\mem\odot}\nem
    \left(\hnem
        \frac{x}{m_0/\hbar}
        \langle \two^{-1} \three\rangle
        \langle \three \one \rangle
    \nem\right)^{\kern-0.275em\odot\ell}
    \,,\\
    \label{eq:3ptAahh-}
    \mathcal{A}^{-h}_{(2s)}\hnem[g^-](\two{\leftarrow}\one;\three)
    &\mem=\mem
    e_h\hem m_0 x^h
    \mem\sum_{\ell=0}^{2s}\mem
    g^-_\ell\mem
    \lsq \btwo \bone^{-1} \rsq^{\kern-0.05em\odot(2s-\ell)}
    {\mem\odot}\nem
    \left(\hnem
        \frac{1}{(m_0/\hbar)x}
        \lsq \btwo \bthree \rsq
        \lsq \bthree \bone^{-1} \rsq
    \nem\right)^{\kern-0.275em\odot\ell}
    \,.
% \end{split}
\end{align}
\end{subequations}
We have temporarily restored $\hbar$.\footnote{
    Our convention for the $\hbar{\,\neq\,}0$ unit is such that spinor-helicity variables have the dimension of $(\text{length})^{-1/2}$ 
    so that they decompose massive/massless wavenumbers $\one_{\a\da},\two_{\a\da},\three_{\a\da} \sim (\text{length})^{-1}$.
    The mu variables have the dimension of $(\text{length})^{1/2}$,
    and the symplectic form reads
    $\omega^\circ = i\hbar\, d\bZ_I{}^\rmA \swedge dZ_\rmA{}^I$.
    \label{footnote:hbar-restore}
}
We restrict our attention to the diagonal (equal-mass and equal-spin) sector.
% The result of the matching is
As a result of the matching, one finds
% \begin{align}
%     \label{eq:gen-Cfgrelation}
%     C^\pm\hnem(\xi)
%     \,=\, 
%     \mathe^{\xi} {\,\cdot\,} f^\pm\hnem(\xi)
%     \,=\, 
%     \mathe^{\xi} {\,\cdot\,}\hem g^\pm\hnem(-\xi/|\vec{W}|)
%     \,,
% \end{align}
\begin{align}
    \label{eq:gen-Cfgrelation}
    \mathe^{\xi} {\,\cdot\,} f^\pm\hnem(\xi)
    \mem=\mem
    C^\pm\hnem(\xi)
    \mem=\mem
    \mathe^{\xi} {\,\cdot\,}\hem g^\pm\hnem(-\xi/|\vec{W}|)
    \qiq
    f^\pm\hnem(\xi) 
    \mem=\mem
    g^\pm\hnem(-\xi/|\vec{W}|)
    \,,
\end{align}
where the generating function describing the classical limit of $g^\pm_\ell$ is defined as
\begin{align}
    g^\pm\hnem(\xi)
    := \lim_{\hbar\to0} \sum_{\ell=0}^\infty 
    g^\pm_\ell (\hbar\xi)^{\ell}
    % \frac{g^+_\ell (\hbar\xi)^{\ell}}{|\vec{W}|^\ell}
    \,.
\end{align}
Notice that Wilson coefficients on complex worldlines, $f^\pm_\ell$,
are more straightforwardly related to the coupling constants $g^\pm_\ell$.
% than $C^\pm_\ell$.
As a pedagogical demonstration,
we re-derive \eqref{eq:gen-Cfgrelation}
in appendix \ref{app:aoude}
without resorting to the ``real-based'' Wilson coefficients $C^\pm_\ell$.
We briefly discuss how the spin coherent states are implemented in the spherical top framework and then reproduce \cite{aoude2021classical}'s matching calculation with our notations and conventions.
We also observe that using the complexified kinematics \eqref{eq:3kin-plus}-\eqref{eq:3kin-minus},
rather than the real kinematics \eqref{eq:3kin-real},
provides a ``shortcut''
for the calculations.

%
%%[1B]%%%%%%%%%%%%%%%%%%%%%%%%%%%%%%%%%%%%%%%%%%%%
\section{Scattering of \texorpdfstring{{\Kerr}}{√Kerr} and Kerr in Heavenly Plane Waves}
\label{sec:Wave}

In the previous section, we have studied 
``universal'' implications of relating chirality and holomorphy
on the linear-order perturbative physics,
leaving $\phi$ 
% as an ``anonymous field.''
unspecified.
In this section,
we 
finally
specialize $\phi$ to Yang-Mills and gravity.
Still, we
continue pursuing a ``bottom-up'' approach:
we
reverse engineer the {\Kerr} and Kerr actions 
in heavenly backgrounds
from the matching \eqref{eq:e1e2} between the minimal amplitude $e_h\hem m_0 x^h\mem \mathe^{\three\cdot a}$ and the plane wave ansatz \eqref{eq:pwH},
% without 
% invoking a
% having a 
% full-fledged theory.
% A top-down derivation will be developed in 
% the follow-up paper \cite{ambikerr2}.
without having a top-down picture.
Furthermore,
we also exactly solve the classical dynamics of {\Kerr} and Kerr in 
self-dual plane-wave backgrounds of arbitrarily (non-perturbatively) large magnitudes
and re-derive their impulses as exact quantities,
thus examining the validity of the results of Section \ref{sec:Flat.amplitude} 
for {\Kerr} and Kerr.
% (rather than a first-order Born approximation).

% \subsection{Heavenly Actions from Amplitudes}
\subsection{Corrections from symplectivity}
% \subsection{A caveat}
% \subsection{\texorpdfstring{{\Kerr}}{√Kerr} and Kerr in the heaven}
\label{sec:Wave.caveat}

\paragraph{Heavenly plane-wave symplectic perturbations}

The classical-spin amplitudes 
of {\Kerr} and Kerr particles
are given as \cite{arkani2020kerr,aoude2021classical}
\begin{align}
    \label{eq:M+h}
    \M^{+h}\hnem(\one,a;\three)
    = e_h\hem m x^h\mem \mathe^{\three \cdot a}
    \,,\quad
    (e_1,e_2) = (\sqrt{2}g\mem q_ac^a, m{\hhnem/\hnem}\MPl)
    \,,
\end{align}
where 
$g$ and $M_\text{Pl} := (8\pi G)^{-1/2}$ are the Yang-Mills coupling constant and the Planck mass, respectively.
$q_a$ denotes the non-abelian color charge of the {\Kerr} particle,
and $c^a$ denotes the color ``polarization.''
% of the Yang-Mills plane wave.
As already pointed out in \eqref{eq:e1e2},
the amplitude $\nu mx\hem \mathe^{\three\cdot a}$
% re-derived in \eqref{eq:exactimp.py},
boils down to \eqref{eq:M+h} if
$\nu = e_h\hem x^{h-1}$
on the support of $\deltabar(2\mem\three{\mem\cdot\mem}\one)$.
A natural guess is
% \begin{align}
%     \nu = \sqrt{2}g\mem q_ac^a
%     \quad\text{({\Kerr})}
%     \,,\qquad
%     \nu = \MPl^{-1} (-\i^\a p_{\a\da} \bo^\da)
%     \quad\text{(Kerr)}
%     \,.
% \end{align}
% \begin{align}
% \begin{split}
%     \text{\Kerr}:\quad
%     \nu &= \sqrt{2}g\mem q_ac^a
%     \,,\\
%     \text{Kerr}:\quad
%     \nu &= \MPl^{-1} (-\i^\a p_{\a\da} \bo^\da)
%     \,.
% \end{split}
% \end{align}
\begin{align}
\begin{split}
    \label{eq:nuggets}
    \text{\Kerr}:\quad
    \nu = \sqrt{2}g\mem q_ac^a
    \,,\qquad
    \text{Kerr}:\quad
    \nu = \MPl^{-1} (-\i^\a p_{\a\da} \bo^\da)
    \,.
\end{split}
\end{align}
Accordingly,
we can deduce that the {\Kerr} and Kerr particles have the following symplectic perturbations in
the
% self-dual plane-wave
background \eqref{eq:pwH}:\footnote{
    The right-hand sides of \eqref{eq:omega'+1}-\eqref{eq:omega'+2} apparently have mass dimension $[M^{-1}]$
    because we make the mode operator (which carries $[M^1]$) implicit.
    One can assign this ``hidden $[M^1]$'' to $\mathe^{-i\hem\bozo}$, $\deltabar(2\mem\three{\mem\cdot\mem}\one)$,
    etc.
    \label{footnote:mode-operator-dimension}
}
\begin{subequations}
\label{eq:omega'+12}
\begin{align}
    \label{eq:omega'+1}
    \text{\Kerr}:\quad
    \omega'^+
    % {\,=\,\mem} q_aF^a{}_{\da\db}(z)\mem (dz \swedge dz)^{\da\db}
    % {\mem\,=\,\mem} 
    &=
    q_a\mem
    \frac{ig\mem c^a}{\sqrt{2}} \mem \bo_\da\hem \bo_\db\mem (dz \swedge dz)^{\da\db} \mem
    \mathe^{-i\hem\bozo}
    + (\cdots)
    % \qquad\qquad\,\,\mem\text{({\Kerr})}
    \,,\\
    \label{eq:omega'+2}
    \text{Kerr}:\quad
    \omega'^+
    &= 
    % \frac{i}{2M_\text{Pl}}
    % \mem
    % (-\i^\c p_{\c\dc} \bo^\dc)
    % \mem
    % \bo_\da\hem \bo_\db\mem (dz \swedge dz)^{\da\db}
    p_{\c\dc}\mem 
    % \frac{i\hem (-\bo^\dc\i^\c)}{2\MPl}
    \frac{-i\hem \bo^\dc\i^\c}{2\MPl}
    % \frac{i\hem \bo^\dc\i^\c}{(-2)\MPl}
    \mem
    \bo_\da\hem \bo_\db
    \mem(dz \swedge dz)^{\da\db}
    \mem
    \mathe^{-i\hem\bozo} 
    + (\cdots)
    % \quad\text{(Kerr)}
    \,.
    % \,,
\end{align}
\end{subequations}
We have denoted $(dz \swedge dz)^{\da\db} := \e_{\a\b}(\wedgetwo{z}{z}{\da\a}{\db\b})$.

\paragraph{Closure condition}

The astute reader may have noticed that,
while we have worked as if $\nu$ is a constant in previous sections,
$q_ac^a$ and $(-\i^\a p_{\a\da} \bo^\da)$ are actually not.
Instead, they turn out to be constants of motion:
consider how $d\nu \swedge \lsq\bo|dz|\i\rangle \mem\mathe^{-i\hem\bozo}$ contributes to the zig-zag expansion of $\{\nu,\bDelta\}$.
But still, they are not constants ``off-shell,''
% which implies that
so
additional non-$dz\swedge dz$ type terms are needed for
the closure 
$d\omega'^+ = 0$
as indicated already as ellipses in \eqref{eq:omega'+12}.
Noting that the auxiliary spinor $\i_\a$
can be traded off with derivatives as
% \begin{subequations}
\begin{align}
    \label{eq:io-trading}
\begin{split}
    % \label{eq:io-trading+1}
    i\hem\partial_{\b\db}
    \big(
        \bo_\da\hem \i^\b
    \mem \mathe^{-i\hem\bozo} 
    \big)
    &=
    % \big(
        \bo_\da\hem \bo_\db
    \mem \mathe^{-i\hem\bozo} 
    % \big)
    \,,\\
    % \label{eq:io-trading+2}
    i\hem\partial_{\c\dc}
    \big(
    i\hem\partial_{\d\dd}
    \big(
        \bo_\da\hem \bo_\db\hem \i^\c\hem \i^\d
    \mem \mathe^{-i\hem\bozo} 
    \big)
    \big)
    &=
    i\hem\partial_{\d\dd}
    \big(
        \bo_\da\hem \bo_\db\hem \bo_\dc\hem \i^\d
    \mem \mathe^{-i\hem\bozo} 
    \big)
    =
    % \big(
        \bo_\da\hem \bo_\db\hem \bo_\dc\hem \bo_\dd 
    \mem \mathe^{-i\hem\bozo} 
    % \big)
    \,,
\end{split}
\end{align}
% \end{subequations}
we can complete \eqref{eq:omega'+1}-\eqref{eq:omega'+2} as
(notice a double copy structure)
% \footnote{
%     For electromagnetism,
%     one could simply take $\nu = \sqrt{2}q$ as a constant while not enlarging the phase space with $\psi$ and $\bpsi$.
% }
\begin{subequations}
    \label{eq:d-theta'+12}
\begin{align}
    \label{eq:d-theta'+1}
    \text{\Kerr}:\quad
    \omega'^+
    &= d\Big(\mem
        g\mem
        q_a 
            c^a
            (\nem\sqrt{2}\mem\i_\b\hem\bo_{\smash{\db}\vphantom{\b}})\mem
            dz^{\db\b} \mem
            \mathe^{-i\hem\bozo}
    \hem\Big)
    \,,\\
    \label{eq:d-theta'+2}
    \text{Kerr}:\quad
    \omega'^+
    &= d\Big(\hem
        {\textstyle-\frac{1}{2M_\text{Pl}}}\mem
        p_{\a\da}
            (\nem\sqrt{2}\mem\i^\a\hem\bo^\da)
            (\nem\sqrt{2}\mem\i_\b\hem\bo_{\smash{\db}\vphantom{\b}})\mem
            dz^{\db\b} \mem
            \mathe^{-i\hem\bozo}
    \hem\Big)
    \,.
\end{align}
\end{subequations}
We see that the additional terms have the forms $dq \swedge dz$ and $dp \swedge dz$, respectively.
Both are admittable
in the standard of \eqref{eq:decouple-W}.
% although they 
% % surely 
% deviate from the class of symplectic perturbations \eqref{eq:arbomega'+} considered in section \ref{sec:Flat}.

Thankfully,
since $\{q_a,\Delta\}^\circ = \{q_a,\bDelta\}^\circ = 0$
and $\{p_{\a\da},\Delta\}^\circ = \{p_{\a\da},\bDelta\}^\circ = 0$, 
these non-spin-space-time ``correction'' terms
do not spoil the
``zig-zag''
logic of section \ref{sec:Flat} at all:
the $\lambda_\a{}^I$, $\rambda_{I\da}$, $y^{\da\a}$ impulses 
and the resulting amplitudes remain intact.
The only change in the equation of motion is that
a ``gauge drift'' term gets added to the time evolution vector field, contributing to $\dot{q}_a$ or $\dot{x}^{\da\a}$:
\begin{align}
    \label{eq:driftb}
    \{\blank,q_a\}^\circ
    \mem=\mem -q_c\mem f^c{}_{ab} 
    \,\frac{\partial}{\partial q_b}
    \,,\quad
    \{\blank,p_{\a\da}\}^\circ
    \mem=\mem \frac{\partial}{\partial x^{\da\a}}
    \,.
\end{align}

\paragraph{Generic heavenly symplectic perturbations}

The sequences
% \eqref{eq:io-trading+1}-\eqref{eq:io-trading+2}
\eqref{eq:io-trading}
reflect the spin-1 and spin-2 nature of Yang-Mills and gravity.
Identifying the fully gauge-invariant $\bo_\da\hem \bo_\db$ 
as the self-dual Yang-Mills field strength,
its anti-derivative $\i_\a\hem\bo_\da$ is the gauge connection;
identifying $\bo_\da\hem \bo_\db\hem \bo_\dc\hem \bo_\dd$ 
as the self-dual Weyl tensor,
the two anti-derivatives are the self-dual metric (tetrad) perturbation and spin connection.
Indeed, 
as elaborated in 
the appendix of
% the follow-up paper
\cite{ambikerr2},
the plane-wave gauge fields are given in the lightcone gauge as
% \begin{subequations}
%     \label{eq:lcpw}
\begin{align}
    \label{eq:lcpw}
\begin{split}
    % \label{eq:lcpw+1}
    A^a \hnem(z)
        = \sqrt{2}g\mem\hem
        c^a\hem \i_\b\hem\bo_{\smash{\db}\vphantom{\b}}
        \mem dz^{\db\b}
        \mem\mathe^{-i \hem\bozo}
    &\,,\quad
    \textstyle
    F^a \hnem(z)
        = \frac{ig}{\sqrt{2}}\mem\hem
        c^a\hem
        \bo_\da\hem\bo_\db
        \hhhem(dz \swedge dz)^{\da\db}
        \mem\mathe^{-i \hem\bozo}
    \,;\\
    % \label{eq:lcpw+2}
    h^{\dc\c} \hnem(z)
    \textstyle
        = {\frac{-1}{\MPl}}\mem 
        \bo^\dc\i^\c\hhem
        \i_\b\hem\bo_{\smash{\db}\vphantom{\b}}
        \mem dz^{\db\b}
        \mem\mathe^{-i \hem\bozo}
    &\,,\quad
    dh^{\dc\c} \hnem(z)
    \textstyle
        = {\frac{-i}{2\MPl}}\mem 
        \bo^\dc\i^\c\hem
        \bo_\da\hem\bo_\db 
        % \e_{\a\b}(e^{\da\a} \swedge e^{\db\b})
        \hhhem(dz \swedge dz)^{\da\db}
        \mem\mathe^{-i \hem\bozo}
    \,.
\end{split}
\end{align}
% \end{subequations}
Hence, we can propose that
\eqref{eq:d-theta'+1}-\eqref{eq:d-theta'+2} generalize to arbitrary heavenly geometries as
\begin{subequations}
    \label{eq:SDomega'}
\begin{alignat}{2}
    \label{eq:SDomega'+1}
    \text{\Kerr}:\quad
    \omega'^+
    &= d\big(\hem
        q_a A^a\hnem(z)
        % q_a A^a{}_{\a\da}\hnem(z)\mem dz^{\da\a}
    \big)
    % = dq_a \swedge A^a{}_{\a\da}\hnem(z)\mem dz^{\da\a}
    % + q_a F^a{}_{\smash{\da\db}\vphantom{\da}}\hnem(z)\mem 
    % (dz \swedge dz)^{\da\db}
    % \,\,\,\,\,\,
    &&= dq_a \swedge A^a\hnem(z) 
    + q_a dA^a\hnem(z)
    \,,\\
    \label{eq:SDomega'+2}
    \text{Kerr}:\quad
    \omega'^+
    &= d\big(\hem
        p_{\a\da} h^{\da\a}\hnem(z)
    \big)
    &&= dp_{\a\da} \swedge h^{\da\a}\hnem(z)
    + p_{\a\da} \hem dh^{\da\a}\hnem(z)
    \,.
\end{alignat}
\end{subequations}
Here, ``$(z)$'' means that the differential forms are holomorphic in $\mathbb{CE}^{1,3}$.
$A^a\hnem(z)$, $F^a\hnem(z)$,
and $h^{\da\a}\hnem(z)$
denote the Yang-Mills gauge potential, field strength,
and the tetrad perturbation,
analytically continued to the spin-space-time.
% $h^{\da\a}\hnem(z)$ denotes the tetrad perturbation analytically continued to the spin-space-time.
To be clearer,
we mean that
the restriction of
% the holomorphic tetrad
$e^{\da\a}\hnem(z) = dz^{\da\a} + h^{\da\a}\hnem(z)$ 
on the real section
describes a self-dual real spacetime geometry
as a tetrad.
This proposal is indeed consistent with the Newman-Janis shift:
\eqref{eq:SDomega'+1} and \eqref{eq:SDomega'+2} 
respectively follow from
the scalar-particle symplectic structures given in
% \eqref{eq:spt.YM.omega} and \eqref{eq:spt.Grav.omega}
\cite{spt}
upon analytically continuing the fields.

It is now evident that the $dq \swedge dz$ and $dp \swedge dz$ terms
we have deduced from the closure condition
are the plane-wave versions of
$dq_a \swedge A^a\hnem(z)$ and $dp_{\a\da} \swedge h^{\da\a}\hnem(z)$
that covariantize the exterior derivatives 
of the free theory's $i\mem d\bpsi_i \swedge d\psi^i$ and $i\mem d\bZ_I{}^\rmA \swedge dZ_\rmA{}^I$,
respectively.
This again explains the fact
that the equation of motion changes only by the ``gauge drift'' terms.
% Hence, we identify \eqref{eq:omega'+12} without the ellipses 
% as the symplectic perturbation in the covariant perturbation scheme \eqref{eq:pert2-omega}:
% \begin{subequations}
%     \label{eq:D-theta'}
% \begin{align}
%     \label{eq:D-theta'+1}
%     \text{{\Kerr} (covariant scheme)}:\quad
%     \omega'^+
%     &= q_a F^a\hnem(z)
%     \,,\\
%     \label{eq:D-theta'+2}
%     \text{Kerr (covariant scheme)}:\quad
%     \omega'^+
%     &= p_m \tilde{T}^m\hnem(z)
%     \,.
% \end{align}
% \end{subequations}
% This again confirms 
% % the fact
% that the equation of motion changes only by the ``gauge drift'' terms.
% The torsion here is a covariantization of $de^m\hnem(z)$ with an arbitrary flat Lorentz connection
% holomorphic in $\mathbb{CE}^{1,3}$.

To recapitulate, we have inferred
% from the matching between 
% the classical-spin amplitudes \eqref{eq:M+h}
% and
% the monochromatic plane-wave ansatz
% \smash{$\omega'^+ = (i\nu/2)\mem \bo_\da\hem\bo_\db (dz \swedge dz)^{\da\db}\mem \mathe^{-i\hem\bozo}$}
% to the minimal symplectic perturbation
that the Newman-Janis shift of
the {\Kerr} and Kerr particles
% {\Kerr} and Kerr
works by \textit{(anti-)holomorphically extending the gauge fields $A^a$ and $e^m$ to the spin-space-time
in the (anti-)self-dual sector}.
In fact, this already discloses the conclusion
% of the next section,
we arrive at in \cite{ambikerr2}.
Nevertheless,
our understanding
% of \eqref{eq:SDomega'}
is yet far from complete or concrete
because we have not
established the 
general/gauge covariant notion
of curved spin-space-time
nor
% provided a full generalization to earthly configurations
described the earthly/off-shell theory.
We postpone further investigations to \cite{ambikerr2}.

\subsection{Exact solution to equations of motion}
% \subsection{Exact impulses of observables}
% \subsection{Impulse and amplitude revisited}
\label{sec:Wave.revisit}

\paragraph{Exact solution for abelian \texorpdfstring{{\Kerr}}{√Kerr}}
% \paragraph{Exact solution to the equation of motion}
% \paragraph{Exact classical trajectory}

Now, we show that the minimal heavenly equation of motion can be solved \textit{exactly} in a plane-wave background.
We first work with \eqref{eq:eom-heavenly-components}
and \eqref{eq:pwH}
while assuming that $\nu$ is a constant;
% and then consider the $dq \swedge dz$ and $dp \swedge dz$ corrections later.
the $dq \swedge dz$ and $dp \swedge dz$ corrections 
will be discussed later.
% \footnote{
%     This simplified setting can be regarded as describing electromagnetism:
%     for electromagnetism,
%     one could simply take $\nu = \sqrt{2}q$ as a constant while not enlarging the phase space with $\psi$ and $\bpsi$.
% }
This simplified setting can be regarded as describing electromagnetism.\footnote{
    One could simply take $\nu = \sqrt{2}q$ as a constant while not enlarging the phase space with $\psi$ and $\bpsi$.
}

For full generality,
we assume
a non-trivial Regge trajectory
% here.
in this subsection.
Let  $m$ and $\Omega^a$ denote
the mass and 
the 
body-frame components of the angular velocity 
(see \eqref{eq:eom-free}) as constants of motion.

Recall that the monochromatic wave \eqref{eq:pwH} introduces a principal null direction.
The geometrical interpretation of \eqref{eq:zbracket-as-nullrot} as the chiral null rotation implies two simplifications.
First, $\rambda_{I\da}{\kern0.01em}\bo^\da$ is conserved.
% \footnote{
%     In general, $\dot{\rambda}_{I\da}\hem \pi^\da\hnem(z) = 0$ if $\omega'_{\smash{\da\db}}(z)$ is null with principal spinor $\pi^\da\hnem(z)$.
% }
Second, the particle's zitterbewegung ($-{\ast}\phi^m{}_n\mem y^n$)
lies within the self-dual null planes (``$\da$-planes'') of $|\bo\rsq$.
Hence, the ``lightcone'' reparameterization gauge that measures ``time'' along the principal null direction
is ignorant of the interaction
and makes the einbein $(\kappa+\bar{\kappa})/2$ a constant
as in the free theory.
By a rescaling, it coincides with our gauge \eqref{eq:constant-einbein}:
\begin{align}
    \label{eq:LCgauge}
    \bo_\da \frac{dz^{\da\a}}{d\s}
    = \bo_\da \frac{dx^{\da\a}}{d\s}
    = -(\kappa {\mem+\mem} \bar{\kappa})\mem
    \bo_\da u^{\da\a}
    % \mem d\s
    \qiq
    \hem\frac{\lsq\bo|dx|\o\rangle}{\lsq\bo|u|\o\rangle}
    = -(\kappa {\mem+\mem} \bar{\kappa})\mem d\s
    = d\tau
    \,.
\end{align}
We have introduced
\begin{align}
    u^{\da\a}
    :=
    \minime m(p^{-1}\hnem(\t_0))^{\da\a}
    \,,
\end{align}
% $\minime m(p^{-1}\hnem(\t_0))^{\da\a} =: u^{\da\a}$,
where $\t_0$ is an initial time of one's choice.
Integrating \eqref{eq:LCgauge}, we find
% $\bo_\da\hem z^{\da\a}(\t) 
% = \bo_\da\hem z^{\da\a}(\t_0)
% + \bo_\da u^{\da\a} \mem(\t-\t_0)$.
\begin{align}
    \label{eq:zlinear}
    \bo_\da\mem z^{\da\a}(\t) 
    \mem=\mem \bo_\da\mem z^{\da\a}(\t_0)
    + \bo_\da u^{\da\a} \mem(\t-\t_0)
    % \qiq
    % \three{\mem\cdot\mem}z(\t)
    % = \three{\mem\cdot\mem} (z(\t_0)-u\t_0)
    % + (\three{\mem\cdot\mem}u) \t
    \,.
\end{align}
Plugging this back in the plane wave \eqref{eq:pwH},
the equation of motion is solved as
\begin{align}
\begin{split}
    \label{eq:pwH-solution}
    \rambda_{I\da}(\t) = R_I{}^J\nem(\t,\t_0)\,
    \rambda_{\smash{J\db}\vphantom{I}}\hhnem(\t_0)\mem
    L^\db{}_\da(\t_0,\t)
    &\,,\quad
    \lambda_\a{}^I\hnem(\t) = \lambda_\a{}^J\hnem(\t_0)\mem R_J{}^I\hnem(\t_0,\t)
    \,,\\
    z^{\da\a}\hnem(\t) =
    z^{\da\a}\hnem(\t_0) 
    +
    F^\da{}_\db(\t,\t_0)\mem
    % (\minime m(\one^{-1})^{\db\a})
    u^{\db\a}
    &\,,\quad
    y^{\da\a}\hnem(\t) =
    L^\da{}_\db(\t,\t_0)\mem y^{\db\a}\hnem(\t_0)
    \,,
\end{split}
\end{align}
where the internal $\mathrm{SU}(2)$ rotation $R_I{}^J\nem(\t_0,\t)$ and the null rotation $L^\da{}_\db(\t_0,\t)$ are given by
\begin{subequations}
\begin{align}
% \begin{split}
    \label{eq:pwH-rotation}
    % U_I{}^J(\t,\t_0)
    % = \exp\nem\nem\left(
    %     \frac{(\t-\t_0)}{2i}\mem \Omega^a (\s_a)
    %     \nem
    % \right)
    % \nem\nem{\vphantom{\Big|}}_I{\vphantom{\Big|}}^J
    % \,.   
    R_I{}^J\nem(\t_1,\t_2)
    &:= \exp\nem\hnem
    \big(
        (\t_2-\t_1)\hem\Omega^a\hem (\s_a)\hnem/2i
    \hem\big)
    {}_I{}^J
    \,,\\
    \label{eq:pwH-Lorentz}
    L^\da{}_\db(\t_0,\t)
    &:= 
    \exp\nem\nem\left(\hem{
        \frac{i\nu}{2m} |\bthree\rsq\lsq\bthree|\,
        \mathe^{i\three\cdot(z(\t_0)-u\t_0)}
        \nem\nem\int_{\t_0}^{\t}
        \hspace{-0.2em} d\t'\,\mem
        \mathe^{
            i (\three\cdot u)
            \t'
        }
    }\hhem\right)
    \nem\nem{\vphantom{\Big|}}^\da{\vphantom{\Big|}}_\db
    \,.
% \end{split}
\end{align}
\end{subequations}
% In general, $\nu$ can be a constant of motion but not a constant.
Our notation is such that
% \begin{align}
%     R_I{}^K\hnem(\t_1,\t_2) 
%     \, R_K{}^J\hnem(\t_2,\t_1)
%     = \delta_I{}^J
%     \,,\quad 
%     L^\da{}_\dc(\t_0,\t) 
%     \, L^\dc{}_\db(\t,\t_0)
%     = \delta^\da{}_\db
%     \,.
% \end{align}
$
    R_I{}^K\hnem(\t_1,\t_2) 
    \, R_K{}^J\hnem(\t_2,\t_1)
    = \delta_I{}^J
$
and
$
    L^\da{}_\dc(\t_0,\t) 
    \, L^\dc{}_\db(\t,\t_0)
    = \delta^\da{}_\db
$.
Lastly, the matrix $F^\da{}_\db(\t,\t_0)$ in \eqref{eq:pwH-solution} is defined as
\begin{align}
    F^\da{}_\db(\t,\t_0)
    := \int_{\t_0}^{\t}
    d\t'\mem L^\da{}_\db(\t',\t_0)
    \,.
\end{align}
\vspace{-0.8\baselineskip}
% We omit its explicit expression.
% Note that $\nu$ need not be a constant
% but can be a constant of motion, in general.
%% "nu const of motion => sol (4.3) equally applies."

\paragraph{Arbitrary wave profile}

In a strict sense, the scattering problem is not well-defined with \eqref{eq:pwH} because the oscillation of the wave 
does not die off at the past and future timelike infinities.
% Instead, one should consider a ``sandwich'' plane wave.
A generic self-dual plane wave 
with an arbitrary profile
is given by \cite{gibbons1975quantized,adamo2017scattering,adamo2022graviton}
% \begin{align}
%     \label{eq:sdpw}
%     \phi_{\da\db}(z)
%     \mem=\mem
%     (i\nu/2)\mem
%     \bo_\da\hem \bo_\db
%     \mem \Psi'\nem(-i\hem\bozo)
%     \,.
% \end{align}
\begin{align}
    \label{eq:sdpw}
    \phi_{\da\db}(z)
    \mem=\mem
    % -(\nu/2)\mem
    (i\nu/2)\mem
    \bo_\da\hem \bo_\db
    \mem \Psi'\nem(\bozo)
    \,,
\end{align}
and then the solution \eqref{eq:pwH-solution} holds with the Lorentz kernel
\begin{align}
    \label{eq:arbpw-L}
    L^\da{}_\db(\t_0,\t)
    \mem=\mem
    \delta^\da{}_\db 
    -
    \frac{i\nu}{2\mem \three{\mem\cdot\mem}p(\t_0)}\mem
    % +
    % \frac{\nu}{2\mem \three{\mem\cdot\mem}p(\t_0)}\mem
    \bthree^\da\bthree_\db
    \Big(\mem{
        % \Psi\hnem(i\hem \three{\mem\cdot\mem}z(\t))
        \Psi\hnem({-}\three{\mem\cdot\mem}z(\t))
        -
        % \Psi\hnem(i\hem \three{\mem\cdot\mem}z(\t_0))
        \Psi\hnem({-}\three{\mem\cdot\mem}z(\t_0))
    }\mem\Big)
    % \,,
    \,.
\end{align}
The (classical and quantum) S-matrix is well-defined
% when
if
the function $\Psi'\nem(\xi)$ is compactly supported on the real line $\xi{\,\in\,}\mathbb{R}$.
Then
the monochromatic plane wave 
% $\Psi'\nem(\xi) = {-i\mem} \mathe^{-i\xi}$ 
$\Psi'\nem(\xi) = \mathe^{-i\xi}$ 
shall be thought of as the $\e {\mem\to\mem} 0^+$ limit of its Gaussian regularization
% $\Psi'\nem(\xi) = {-i\mem} \mathe^{-i\xi - \e\mem \xi^2\hnem/2}$.
$\Psi'\nem(\xi) = \mathe^{-i\xi - \e\mem \xi^2\hnem/2}$.
With this understanding, we continue to work with \eqref{eq:pwH} for simplicity.

\paragraph{Classical interaction picture}
% \paragraph{Exact impulse and amplitude}

% See appendix \ref{app:shilb} for a construction in terms of Dirac bracket.
% The subtlety is in that,
% unlike in non-relativistic quantum mechanics where time is absolute,
% our parameter $\t$ is an internal, worldline-wise measure of time
% so that the equal-time Cauchy slices of free and interacting theories
% are not guaranteed to coincide.

To define the impulse properly,
one should implement the classical analog of the \textit{interaction picture}.
The covariant phase space 
of the free theory
is the space of gauge orbits in $\ma_\diamond$ 
generated by the time evolution
vector field
$\{\blank, (\bDelta {\mem+\mem} \Delta) {\mem-\mem} 2m(\vec{W}^2)\}^\circ$.
This quotient space, which we denote as $\ma_*$,
realizes the space of asymptotic in- and out-states.
Then the interaction picture
describes an interacting-theory trajectory
by the evolution of its ``shadow'' on $\ma_*$.
Identifying $\ma_*$ with a particular gauge slice $\chi + \bchi = p_m x^m = 0$ in $\ma_\diamond$,
the ``shadow'' of \eqref{eq:pwH-solution} is given by\footnote{
    It would be interesting if
    the description of the ``shadow''
    % trajectory
    can be further formalized
    % in terms of
    by
    Dirac bracket.
}
\begin{subequations}
    \label{eq:intpicture}
\begin{align}
    \label{eq:intpicture-a}
    \smash{\underset{\adjustbox{raise=0.2em,scale=0.8}{$*$}}{\lambda}{}_\a{}^I\hnem(\t)}
    :={}& \lambda_\a{}^J\hnem(\t)\mem R_J{}^I\hnem(\t,\t_*)
    = \lambda_\a{}^J\hnem(\t_0)\mem
    R_J{}^I\hnem(\t_0,\t_*)
    \,,\\
    \label{eq:intpicture-b}
    \smash{\underset{\adjustbox{raise=0.2em,scale=0.8}{$*$}}{\rambda}{}_{I\da}(\t)}
    :={}& R_I{}^J\nem(\t_*,\t)\mem \rambda_{J\da}(\t)
    = R_I{}^J\nem(\t_*,\t_0)\mem 
    \rambda_{\smash{J\db}\vphantom{\da}}(\t_0)\mem
    L^\db{}_\da\hnem(\t_0,\t)
    \,,\\
    \label{eq:intpicture-c}
    \smash{\underset{\adjustbox{raise=0.2em,scale=0.8}{$*$}}{y}{}^{\da\a}(\t)}
    :={}& y^{\da\a}\hnem(\t) 
    = L^\da{}_\db(\t,\t_0)\mem y^{\db\a}\hnem(\t_0)
    \,,\\
    \label{eq:intpicture-d}
\begin{split}
    \smash{\underset{\adjustbox{raise=0.2em,scale=0.8}{$*$}}{z}{}^{\da\a}(\t)}
    :={}& z^{\da\a}\hnem(\t) 
    % + \minime\mem (p^{-1}\hnem(\t))^{\da\a}\mem 
    % p_{\b\db}(\t)\mem z^{\db\b}(\t)
    - \minime\mem m(p^{-1}\hnem(\t))^{\da\a}\mem (\t-\t_*)
    \,,\\
    ={}& \smash{
        z^{\da\a}\hnem(\t_0)
        + ((\t_* {-\mem} \t_0)\hem L^\da{}_\db(\t,\t_0)
        - G^\da{}_\db(\t,\t_0))\mem u^{\db\a}
        \,,
    }
\end{split}
\end{align}
\end{subequations}
where $\t_*$ is determined as a function of $\t_0$ and $\t$ by the equation
\begin{align}
    \label{eq:t*equation}
    -m(\t-\t_*) 
    \mem&=\mem
    p_{\a\da}(\t)\mem z^{\da\a}\hnem(\t)
    \,,\\
    \iq
    \label{eq:timedifference}   
    \t_*(\t_0,\t)
    \mem&=\mem
    \smash{
        \t_0 \mem+\mem
        z^{\da\a}\hnem(\t_0)\mem p_{\smash{\a\db}\vphantom{\da}}(\t_0)\mem L^\db{}_\da(\t_0,\t) /m
    }
    \,.
\end{align}
That is, we evolve backward in time by the amount $(\t-\t_*)$
along the free theory trajectory
such that $\smash{\underset{\adjustbox{raise=0.2em,scale=0.8}{$*$}}{\smash{p}\vphantom{x}}}{}_m(\t)\mem \smash{\underset{\adjustbox{raise=0.2em,scale=0.8}{$*$}}{x}}{}^m(\t) = 0$.
We have introduced
\begin{align}
\begin{split}
    G^\da{}_\db(\t,\t_0)
    \mem:=\,{}&
        (\t-\t_0)\mem L^\da{}_\db(\t,\t_0) - F^\da{}_\db(\t,\t_0)
    % \,,\\
    % ={}& 
    \,=
        % \nem\nem
        \int_{\t_0}^{\t} \hspace{-0.2em}d\t'\,
        (\t'{-}\t_0)\mem \frac{\partial}{\partial\t'} L^\da{}_\db(\t',\t_0)
    \,.
    % ={}&
    %     \frac{\nu}{2m}\mem \bthree^\da \bthree_\db\mem
    %     \mathe^{i\hem \three\cdot z(\t_0)}
    %     \nem\nem\left[\hem{
    %         \frac{\partial}{\partial\alpha}\nem\hnem
    %         \int_{\t_0}^{\t} \hspace{-0.2em}d\t'\,
    %         \mathe^{i\alpha(\t'-\t_0)}
    %     }\hem\right]_{\alpha = 3\cdot u}\nem
    % \,.
\end{split}
\end{align}
% \vspace{-0.8\baselineskip}
% From the monochromatic wave, it is given by
% \begin{align}
%     G^\da{}_\db(\t,\t_0)
%     =
%     \frac{\nu}{2m}\mem \bthree^\da \bthree_\db\mem
%         \mathe^{i\hem \three\cdot z(\t_0)}
%         \nem\nem\left[\hem{
%             \frac{\partial}{\partial\alpha}\nem\hnem
%             \int_{\t_0}^{\t} \hspace{-0.2em}d\t'\,
%             \mathe^{i\alpha(\t'-\t_0)}
%         }\hem\right]_{\alpha = 3\cdot u}\nem
% \end{align}

\paragraph{Precise definition of impulse}

Finally, the impulse of an observable $f$ is defined as
\begin{align}
    \label{eq:int-impulse}
    \smash{\mathit{\Delta}f := \underset{\adjustbox{raise=0.2em,scale=0.8}{$*$}}{f}({+}\infty) - \underset{\adjustbox{raise=0.2em,scale=0.8}{$*$}}{f}({-}\infty)}
    \vphantom{\underset{\cdot}{0}}
    \,.
\end{align}
The interaction-picture initial state at infinite past
defines the scattering data:
\begin{align}
\begin{split}
    \smash{\underset{\adjustbox{raise=0.2em,scale=0.8}{$*$}}{\lambda}{}_\a{}^I\hnem({-}\infty)}
    =: \one_\a{}^I
    &\,,\quad
    \smash{\underset{\adjustbox{raise=0.2em,scale=0.8}{$*$}}{\rambda}{}_{I\da}\hnem({-}\infty)}
    =: \bone_{I\da}
    \,,\quad
    % \one_{\a\da}
    % := -\one_\a{}^I \bone_{I\da}
    % \,,\\
    \smash{\underset{\adjustbox{raise=0.2em,scale=0.8}{$*$}}{x}{}^{\da\a}\hnem({-}\infty)} =: b^{\da\a}
    % &\,,\quad
    \,,\quad
    \smash{\underset{\adjustbox{raise=0.2em,scale=0.8}{$*$}}{y}{}^{\da\a}\hnem({-}\infty)} =: -a^{\da\a}
    \,.
\end{split}
\end{align}
These of course satisfy the constraints
$\one^2 {\hem+\mem} m^2 = 0$ and
$\one{\mem\cdot\mem}a {\,=\,} 0 {\,=\,} \one{\mem\cdot\mem}b$.
Indeed, one finds from \eqref{eq:intpicture-b}-\eqref{eq:timedifference} that $a^m$ and $b^m$ are the ``orbital and spin'' impact parameters, respectively:
\begin{align}
    \label{eq:impactparams}
    z^m\hnem({-}\infty) + u^m\mem u{\mem\cdot\mem} z({-}\infty)
    = -i(a+ib)^m
    \,.
\end{align}
We have taken $\t_0 = {-}\infty$
so that
$p_{\a\da}({-}\infty) = \smash{\underset{\adjustbox{raise=0.2em,scale=0.8}{$*$}}{p}{}_{\a\da}}({-}\infty) = \one_{\a\da}$
and 
$u^{\da\a} = \minime m(\one^{-1})^{\da\a}$.

\begin{figure}[t]
    \centering
    \includegraphics[scale=0.46,valign=c]{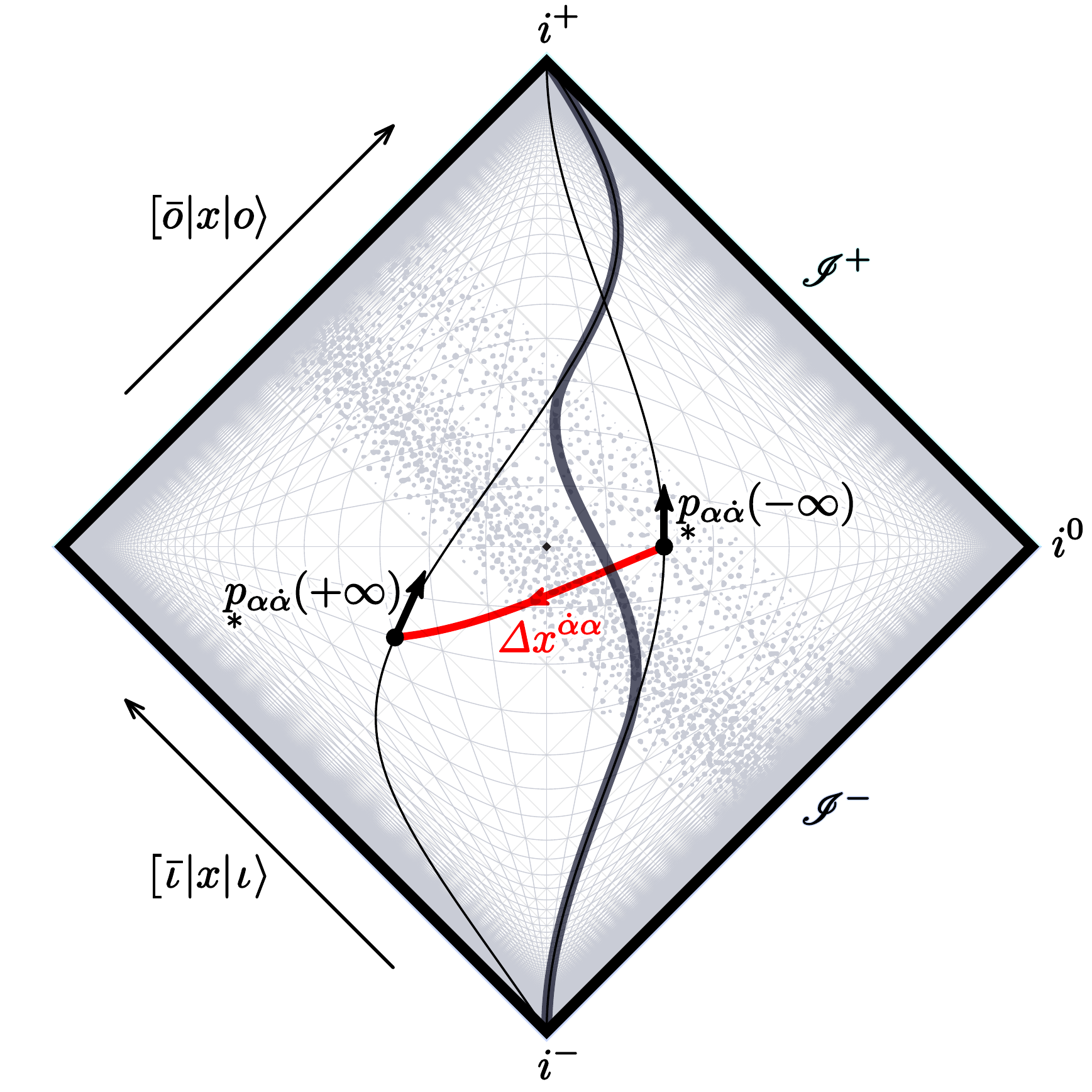}
    \caption{
        Scattering of a massive particle in a ``sandwich'' plane wave geometry.
        The interacting trajectory is matched with free-theory worldlines in the in- and out-regions.
        The ``classical S-matrix'' is then defined as a map 
        % on the covariant phase space of the free theory.
        acting on the representative free-particle states.
    }
    \label{fig:diamond}
\end{figure}

\paragraph{``Exact amplitude''}

Now we are ready to compute \eqref{eq:int-impulse} for various classical observables.
Using \eqref{eq:impactparams}, 
the infinite time limit of \eqref{eq:pwH-Lorentz} is found as
% one finds that \eqref{eq:pwH-Lorentz} reproduces the heavenly three-point kinematics:
\begin{align}
\begin{split}
    \label{eq:inftylimit-L}
    L^\da{}_\db({-}\infty, {+}\infty)
    % \,=\, 
    % \delta^\da{}_\db +
    % i\nu\mem \bthree^\da\bthree_\db
    % \mem \mathe^{\three\cdot(a+ib)}
    % \mem \deltabar(2\mem\three{\mem\cdot\mem}\one)
    \,=\,
    \delta^\da{}_\db
    + i\hem \M^+\mem
    \deltabar(2\mem\three{\mem\cdot\mem}\one)\mem
    (\bthree^\da\bthree_\db/mx)
    \,,\quad
    \M^+ := \nu mx\mem\hem \mathe^{\three\cdot(a+ib)}
    \,,
\end{split}
\end{align}
from which it follows that
the minimal momentum impulse and amplitude found in Section \ref{sec:Flat.amplitude} are reproduced as exact quantities.
Besides,
as expected from $\dot{W}^a = 0$,
we find that $y^{\da\a}$ experiences the same chiral null rotation as $p_{\a\da}$:
\begin{subequations}
    \label{eq:exactimp.py}
\begin{align}
    \label{eq:exactimp.p}
    \mathit{\Delta}p_{\a\da}
    &
    = i\hem \M^+\mem
    \deltabar(2\mem\three{\mem\cdot\mem}\one)\,
    \one_{\smash{\a\db}\vphantom{\da}}\hem (\bthree^\db \bthree_\da/mx)
    = i\hem \M^+\mem
    \deltabar(2\mem\three{\mem\cdot\mem}\one)\,
    \three_{\a\da}
    \,,\\
    \label{eq:exactimp.y}
    \mathit{\Delta}y^{\da\a}
    &
    = i\hem \M^+\mem
    \deltabar(2\mem\three{\mem\cdot\mem}\one)\,
    (\bthree^\da \bthree_\db/mx)\mem a^{\db\a}
    % \,,\quad
    % \M^+ := \nu mx\mem\hem \mathe^{\three\cdot(a+ib)}
    \,.
\end{align}
\end{subequations}
For the spinorial frame impulse, however,
there are additional $\mathrm{SU}(2)$ factors
due to $\delta\t_* := \t_*({-}\infty,{+}\infty) - \t_*({-}\infty,{-}\infty)$:
\begin{align}
\begin{split}
    \label{eq:exactimp.lrambda}
    \smash{\underset{\adjustbox{raise=0.2em,scale=0.8}{$*$}}{\lambda}{}_\a{}^I\hnem({+}\infty)}
    &= 
    \one_\a{}^J
    \exp\nem\hnem
    \big(\hem{
        \delta\t_*\mem
        \Omega^a (\s_a)\hnem/2i
    }\hem\big)
    {}_J{}^I
    \,,\\
    \smash{\underset{\adjustbox{raise=0.2em,scale=0.8}{$*$}}{\rambda}{}_{I\da}\hnem({+}\infty)}
    &= 
    \exp\nem\hnem
    \big(\hem{
        -\delta\t_*\mem
        \Omega^a (\s_a)\hnem/2i
    }\hem\big)
    {}_I{}^J\mem
    \bone_{\smash{J\db}\vphantom{I\da}}\mem
    L^\db{}_\da({-}\infty,{+}\infty)
    \,.
\end{split}
\end{align}
% A constant Regge trajectory ($\Omega^a = 0$)
Putting $\Omega^a = 0$
recovers the result of Section \ref{sec:Flat.amplitude}.
One finds from \eqref{eq:timedifference} that
% Specifically, one finds from \eqref{eq:timedifference} that
\begin{align}
    \label{eq:timed*}
    \delta\t_*
    &\mem=\mem \M^+\hem
        % (\three{\mem\cdot\mem}(a{\mem+\mem}ib)/m)\mem\hem
        \deltabar(2\mem\three{\mem\cdot\mem}\one)\,
        (\three{\mem\cdot\mem}(a{\mem+\mem}ib)/m)
    \,.
\end{align}

\paragraph{Angular momentum impulses}

The orbital and spin angular momenta of the particle 
are given as
$L_{mn} := 2x_{[m}p_{n]}$ and
$S_{mn} := \ve_{mnrs}{\hem}y^r{\hnem}p^s$.
In the spinor notation, the total angular momentum 
$J_{mn} := L_{mn} + S_{mn}$ is given by
\begin{align}
    \label{eq:def-JAM}
    % J_{\da\db} = z^{\dc\c} p_{\smash{\c(\db}\vphantom{\da)}} \be_{\da)\dc}
    % \,,\quad
    % J^{\a\b} = p_{\c\dc} \bz^{\dc(\b} \e^{\a)\c}
    % \,.
    J^\da{}_\db
    = - z^{\da\a} p_{\a\db} + \minie\mem \delta^\da{}_\db\hem (z^{\dc\c}p_{\c\dc})
    \,,\quad
    \bar{J}_\a{}^\b
    = - p_{\a\da} \bz^{\da\b} + \minie\mem \delta_\a{}^\b\hem \hnem(p_{\c\dc}\bz^{\dc\c})
    \,.
\end{align}
Since $L_{mn}$, $S_{mn}$, and $J_{mn}$ are all conserved quantities in the free theory, we can use the interaction-picture expressions for computing their impulse.

First, we find from \eqref{eq:exactimp.py} that
\begin{align}
    \label{eq:exactimp.S}
% \begin{split}
{\renewcommand{\arraycolsep}{0em}
\renewcommand{\arraystretch}{1.1}
\begin{array}{rl}
    \mathit{\Delta}\bar{S}_\a{}^\b = 0
    \,,\quad
    \mathit{\Delta}S^\da{}_\db
    &{}= \M^+\,
    \deltabar(2\mem\three{\mem\cdot\mem}\one)\,
    % \big({
    %     (a^{\da\a}\one_{\a\dc}) (\bthree^\dc \bthree_\db/mx)
    %     - (\bthree^\da \bthree_\dc/mx) (a^{\dc\c}\one_{\c\db})
    % }\big)    
    \big(
        |\bthree\rsq\lsq\bthree| a|\one| 
        - |a|\one |\bthree\rsq\lsq\bthree| 
    \big){}^\da{}_\db/mx
    \,,\\
    &{}= \M^+\,
    \deltabar(2\mem\three{\mem\cdot\mem}\one)\,
    (
        \three^{\da\a}a_{\smash{\a\db}\vphantom{\b}}
        - a^{\da\a}\three_{\smash{\a\db}\vphantom{\b}} 
    )
    \,,\\
    &{}= i\hem \M^+\,
    \deltabar(2\mem\three{\mem\cdot\mem}\one)\,
    (-\ve_{mnrs} \hem a^r\hnem \three^s)\hem 
    (\bs^{mn})^\da{}_\db
    \,,
\end{array}}
% \end{split}
\end{align}
which is consistent with the chiral Lorentz transformation interpretation.

Next, we compute $\mathit{\Delta}J^\da{}_\db$.
The 
$(\t_* {-\mem} \t_0)\mem L^\da{}_\db(\t,\t_0)\hem u^{\db\a}
\propto (p^{-1}\hnem(\t))^{\da\a}$
term in \eqref{eq:intpicture-d} 
does not boil down to a particularly neat expression in the infinite time 
limit, but it 
% limit but
drops out when 
one computes
$p_{\a\da}\hnem(\t)\mem \smash{\underset{\adjustbox{raise=0.2em,scale=0.8}{$*$}}{z}{}^{\da\b}(\t)}$
or
$\smash{\underset{\adjustbox{raise=0.2em,scale=0.8}{$*$}}{z}{}^{\da\a}(\t)}\mem p_{\smash{\a\db}\vphantom{\a\da}}\hnem(\t)$
and symmetrizes the 
two spinor 
indices.
As a result,
we find
\begin{align}
\begin{split}
    \label{eq:exactimp.am}
    \mathit{\Delta}J^{\smash{\da\db}\vphantom{(\da}}
    \mem&=\mem 
    i\hem \M^+\mem \deltabar(2\mem\three{\mem\cdot\mem}\one)\mem
        % z_{\smash{*}}^{(\da\a} 
        % \three_\a \bthree\vphantom{z}^{\smash{\db)}\vphantom{(\da}}
        \big({
            - z_{\smash{*}}^{(\da\a} 
            \hem \three_\a\vphantom{z}^{\smash{\db)}\vphantom{(\da}}
        }\big)
    \,,\\
    \mathit{\Delta}L^{\smash{\da\db}\vphantom{(\da}}
    \mem&=\mem 
    i\hem \M^+\mem \deltabar(2\mem\three{\mem\cdot\mem}\one)\mem
        % (z_{\smash{*}} \hem{+}\mem 2ia)^{(\da\a}
        % \three_\a \bthree\vphantom{z}^{\smash{\db)}\vphantom{(\da}}
        \big({
            - (z_{\smash{*}} \hem{+}\mem 2ia)^{(\da\a}
            \hem \three_\a\vphantom{z}^{\smash{\db)}\vphantom{(\da}}
        }\big)
    \,,\\
    \mathit{\Delta}\bar{J}^{\smash{\a\b}\vphantom{(\da}}
    \mem=\mem
    \mathit{\Delta}\bar{L}^{\smash{\a\b}\vphantom{(\da}}
    \mem&=\mem 
    i\hem \M^+\mem \deltabar(2\mem\three{\mem\cdot\mem}\one)\,
        \big({
            - \three^{(\a}{}_\da 
            \hem z_{\smash{*}}^{\da\b)}
        }\big)
    \,,
    % \,.
\end{split}
\end{align}
noting that
\begin{align}
    % G^\da{}_\db(\t,\t_0)
    % =
    % \frac{\nu}{2m}\mem \bthree^\da \bthree_\db\mem
    %     \mathe^{i\hem \three\cdot z(\t_0)}
    %     \nem\nem\left[\hem{
    %         \frac{\partial}{\partial\alpha}\nem\hnem
    %         \int_{\t_0}^{\t} \hspace{-0.2em}d\t'\,
    %         \mathe^{i\alpha(\t'-\t_0)}
    %     }\hem\right]_{\alpha = 3\cdot u}\nem
    G^\da{}_\db({+}\infty,{-}\infty)
    = (2/x)\mem \M^+\hem
    \deltabar'\hnem(2\mem\three{\mem\cdot\mem}\one)\,
    \bthree^\da \bthree_\db\mem
        % \deltabar'\hnem(2\mem\three{\mem\cdot\mem}\one)
    = (2/m)\mem \M^+\hem
    \deltabar'\hnem(2\mem\three{\mem\cdot\mem}\one)\,
    \one^{\da\a} \three_{\smash{\a\db}\vphantom{\da}}
        % \mem\deltabar'\hnem(2\mem\three{\mem\cdot\mem}\one)
    \,.
\end{align}
We have defined
\begin{align}
    \label{eq:deflection-intersection}
    z_{\smash{*}}^{\da\a}
    \mem:=\mem (b-ia)^{\da\a} 
    + u^{\da\a}\mem T
    % + u^{\da\a} 
    % \nem\hnem
    % \left(
    %     \frac{-im}{\three{\mem\cdot\mem}\one}
    %     - u \cdot z({-}\infty)
    % \right)
    % \,.
    \,,\quad
    T:= 
        % - u \cdot z({-}\infty)
        % \frac{-im}{\three{\mem\cdot\mem}\one}
        -im/(\three{\mem\cdot\mem}\one)
        - u \cdot z({-}\infty)
    % \,.
    \,,
\end{align}
% \begin{align}
%     \label{eq:exactimp.J}
%     \mathit{\Delta} J^{\da\db}
%     &= i\hem \M^+
%     \Big(\mem{
%         \deltabar'\hnem(2\mem\three{\mem\cdot\mem}\one)\mem
%         (i\mem\one^{(\da\a} \three_\a{}^{\smash{\db)}\vphantom{(\da}})
%         +
%         \deltabar(2\mem\three{\mem\cdot\mem}\one)\mem
%         (-(b-ia)^{(\da\a} \three_\a{}^{\smash{\db)}\vphantom{(\da}})
%     }\Big)
%     \,,\\
%     \label{eq:exactimp.L}
%     \mathit{\Delta} L^{\da\db}
%     &= i\hem \M^+
%     \Big(\mem{
%         \deltabar'\hnem(2\mem\three{\mem\cdot\mem}\one)\mem
%         (i\mem\one^{(\da\a} \three_\a{}^{\smash{\db)}\vphantom{(\da}})
%         +
%         \deltabar(2\mem\three{\mem\cdot\mem}\one)\mem
%         (-(b+ia)^{(\da\a} \three_\a{}^{\smash{\db)}\vphantom{(\da}})
%     }\Big)
%     \,.
% \end{align}
% \begin{align}
%     \label{eq:exactimp.Lleft}
%     \mathit{\Delta} \bar{L}^{\a\b}
%     = \mathit{\Delta} \bar{J}^{\a\b}
%     = i\hem \M^+
%     \Big(\mem{
%         \deltabar'\hnem(2\mem\three{\mem\cdot\mem}\one)\mem
%         (i\mem
%         \three^{(\a}{}_\da \one^{\da\b)})
%         +
%         \deltabar(2\mem\three{\mem\cdot\mem}\one)\mem
%         (-\three^{(\a}{}_\da (b-ia)^{\da\b)})
%     }\Big)
%     \,.
% \end{align}
which has a natural interpretation as
the intersection point of the initial and final asymptotic \textit{holomorphic worldlines}.
The existence of such a point is not a trivial fact
and is a consequence of the special geometry of this scattering problem.

To sum up, 
\eqref{eq:exactimp.am} boils down to
% these impulses translate to
% the vectorial language as
% \begin{subequations}
    % \label{eq:amimpulses}
\begin{align}
\begin{split}
    \label{eq:amimpulses}
    % \label{eq:amimpulse.S}
    \mathit{\Delta}S^{mn} &= 
    i\hem \M^+\mem \deltabar(2\mem\three{\mem\cdot\mem}\one)\mem
    \big(\mem{
        \hnem- 2i\mem a^{[m} \three^{n]}
        - \ve^{mnrs} a_r \three_s 
    }\big)
    \,,\\
    % \label{eq:amimpulse.L}
    \mathit{\Delta}L^{mn} &= 
    i\hem \M^+\mem \deltabar(2\mem\three{\mem\cdot\mem}\one)\mem
    \big(\mem{
        2\mem \one^{[m} \three^{n]} (T/m)
        + 2b^{[m} \three^{n]} 
        + \ve^{mnrs} a_r \three_s
    }\big)
    \,,\\
    % \label{eq:amimpulse.J}
    \mathit{\Delta}J^{mn} &=
    i\hem \M^+\mem \deltabar(2\mem\three{\mem\cdot\mem}\one)\mem
    \big(\mem{
        2\mem \one^{[m} \three^{n]} (T/m)
        + 2b^{[m} \three^{n]} 
        - 2i\mem a^{[m} \three^{n]}
    }\big)
    \,.
\end{split}
\end{align}
% \end{subequations}
% % The bracketed term in \eqref{eq:amimpulse.S} 
% % can be interpreted
% % as the self-dual spin angular momentum carried by the single quantum that the wave \eqref{eq:pwH} describes. % a?
% % Also note that, if
% If the $\deltabar'\hnem(2\mem\three{\mem\cdot\mem}\one)$ terms are ignored,
% only $\mathit{\Delta}L_{mn}$ is real-valued
% (after peeling off $i\hem \M^+\mem \deltabar(2\mem\three{\mem\cdot\mem}\one)$).
% % only $\mathit{\Delta}L_{mn}$ is ``real-valued''
% % (after peeling off $i\hem \M^+\mem \deltabar(2\mem\three{\mem\cdot\mem}\one)$):
% % the bracketed term of 
% % \eqref{eq:exactimp.L} and \eqref{eq:exactimp.Lleft}
% % are complex conjugate to each other,
% % while $\mathit{\Delta}L_{\smash{\a\da\b\db}\vphantom{\b}} = \e_{\a\b} L_{\smash{\da\db}\vphantom{\b}} + \bar{L}_{\a\b}\hem \be_{\smash{\da\db}\vphantom{\b}}$.
% Interestingly,
% there is no such $\smash{\underset{\adjustbox{raise=0.2em,scale=0.8}{$*$}}{x}{}^m}$ such that
% $\mathit{\Delta}L_{mn}
% = 2\smash{\underset{\adjustbox{raise=0.2em,scale=0.8}{$*$}}{x}}{}_{[m} \mathit{\Delta}p_{n]}$ for nonzero $a^m$,
% so a local particle interpretation is not possible in real spacetime.
% To have a picture of a particle receiving a single quantum carrying the momentum $\three_m$,
% one should move to the holomorphic or anti-holomorphic complex spacetimes.
Note that only $\mathit{\Delta}J^{mn} = i\hem\M^+\mem \deltabar(2\mem\three{\mem\cdot\mem}\one)\, z_{\smash{*}}^{[m} \three^{n]}$ is ``polar.''
The other two have ``axial'' components,
i.e., terms involving the epsilon tensor.
% As a result, a local particle interpretation is not possible in real spacetime:
% to have a picture of a particle getting instantaneously deflected at the intersection point $z_{\smash{*}}^m$
% by receiving a single field quantum carrying the momentum $\three_m$,
% one should describe in the complex spacetime
% unless $a^m = 0$.
Especially, the orbital angular momentum impulse $\mathit{\Delta}L^{mn}$ is not polar.
One may interpret this fact
as a failure of a local particle interpretation
in real spacetime
(and the necessity of complex spacetime).

\paragraph{Exact \texorpdfstring{{\Kerr}}{√Kerr} impulses}

Finally, let us re-solve the equation of motion with \eqref{eq:d-theta'+12}.
We start off with {\Kerr}.
For clarity's sake, we spell out the entire symplectic form:
\begin{align}
    \label{eq:pws.sqrt-Kerr}
{\renewcommand{\arraystretch}{1.1}
\begin{array}{rl}
    \omega_{\nem{\Kerr}}^{\text{SDPW}} 
    \,=
    {}&{} 
        i\mem d\bZ_I{}^\rmA \swedge dZ_\rmA{}^I
        + i\hem d\bpsi_i \swedge d\psi^i
    \\
    {}&{}
        {}+ \sqrt{2}g\, dq_a c^a \swedge \lsq\bo|dz|\i\rangle
        \mem \mathe^{-i\hem\bozo}
        + {\textstyle\frac{ig}{\sqrt{2}}}\mem q_a c^a \mem
        \bo_\da\hem\bo_\db (dz \swedge dz)^{\da\db}
        \mem \mathe^{-i\hem\bozo}
    \,.
\end{array}}
\end{align}
% One can use either the non-covariant scheme or the covariant scheme to obtain the equation of motion.
% % As a result, one finds 
% % an imaginary-deviated Wong's equation:
The resulting equation of motion boils down to
\begin{subequations}
    \label{eq:xeom+1}
\begin{align}
\begin{split}
    \label{eq:xeom+1a}
    \dot{\lambda}_\a{}^I 
    = {\textstyle\frac{1}{2i}} (\lambda \hem \s_a)_\a{}^I
    &\,,\quad
    \dot{\rambda}_{I\da} 
    = (i\nu{\hnem/}2m)\mem  
    \rambda_{\smash{I\db}\vphantom{I}} \mem \bo^\db \bo_\da
    \mem\mathe^{-i\hem\bozo}
    - {\textstyle\frac{1}{2i}} (\s_a \rambda)_{I\da}
    \,,\\
    \dot{z}^{\da\a} = \minime m(p^{-1})^{\da\a}
    &\,,\quad
    \dot{y}^{\da\a}
    = -(i\nu{\hnem/}2m)\mem 
    \bo^\da \bo_\db \mem y^{\db\a}
    \mem\mathe^{-i\hem\bozo}
    \,,
\end{split}\\
    \label{eq:xeom+1b}
    \dot{\psi}^i
    = -({\textstyle\frac{g}{\sqrt{2}}}\tilde{x}\hem c^a)\mem
     (t_a)^i{}_j \hhem \psi^j \hem
    % \mem(\langle\i|p|\bo\rsq/m)
    \mem\mathe^{-i\hem\bozo}
    &\,,\quad
    \dot{\bar{\psi}}_i
    = ({\textstyle\frac{g}{\sqrt{2}}}\tilde{x}\hem c^a)\mem
     \bpsi_j \hhnem (t_a)^j{}_i \hem
    % \mem(\langle\i|p|\bo\rsq/m)
    \mem\mathe^{-i\hem\bozo}
    % \,,
    \,.
\end{align}
\end{subequations}
The last line implies
$\dot{q}_a
= (q_b f^b{}_{ca}\hhem c^c)\mem 
{\textstyle\frac{g}{\sqrt{2}}}\tilde{x}
\,\mathe^{-i\hem\bozo}$.
We have defined the ``off-shell $x$-factor'' 
% $\tilde{x} := -\langle\i|p|\bo\rsq/|\Delta|$,
\begin{align}
    \tilde{x} := -\langle\i|p|\bo\rsq/|\Delta|
    \,,
\end{align}
which is a constant of motion.
Note that \textit{both} ``holomorphic'' and ``anti-holomorphic'' color degrees of freedom,
$\psi^i$ and $\bpsi_i$,
are localized at the \textit{holomorphic} position $z^{\da\a}$,
being parallel-transported by the holomorphic gauge connection.
% given in \eqref{eq:lcpw+1}
% Indeed, otherwise $q_a F^a\hnem(z) = (i\hem \bpsi \hem t_a \psi) F^a\hnem(z)$ cannot be gauge invariant.
Indeed, 
% every color degrees of freedom carried by the particle
both $\psi^i$ and $\bpsi_i$
should couple to the holomorphic connection
for $q_a F^a\hnem(z) = (i\hem \bpsi \hem t_a \psi) F^a\hnem(z)$ to be gauge invariant.

We can understand \eqref{eq:xeom+1} as a combination of
an infinitesimal self-dual null rotation, color rotation, and internal $\mathrm{SU}(2)$ rotation
with fixed points $\bo^\da$, $c^a$, and $\Omega^a$, 
respectively.\footnote{
    We believe that it is clear from the context whether
    the indices $a,b,c,\cdots$ denote the color indices or the little group indices.
}
Hence, $\nu = \sqrt{2}g\mem q_ac^a$ is conserved.

Now, since \eqref{eq:xeom+1a} is totally unaffected by the $dq \swedge dz$ correction,
it follows that
the solution \eqref{eq:pwH-solution}
and the impulses \eqref{eq:exactimp.py}-\eqref{eq:amimpulses}
are all valid.
The only additional piece of information 
is the color time evolution and impulse
that follow from \eqref{eq:xeom+1b}:
\begin{subequations}
\begin{align}
    \psi^i\hnem(\t)
    &= U^i{}_j(\t,\t_0)\mem \psi^j\hnem(\t_0)
    \,,\quad
    \bpsi_i(\t)
    = \bpsi_j(\t_0)\mem U^j{}_i(\t_0,\t)
    \,,\\
    % \stepcounter{parentequation} \gdef\theparentequation{\arabic{section}.\arabic{parentequation}} \setcounter{equation}{0}
    % \mathit{\Delta}\psi^i
    % = 
    % &\,,\quad
    \label{eq:pwH-colorphase}
    U^i{}_j(\t_2,\t_1)
    &= 
    \exp\nem\nem\left(\hem{
        \hnem\frac{g\tilde{x}}{\sqrt{2}}\mem c^a(t_a)\,
        \mathe^{i\three\cdot(z(\t_0)-u\t_0)}
        \nem\nem\int_{\t_1}^{\t_2}
        \hspace{-0.2em} d\t'\,\mem
        \mathe^{
            i (\three\cdot u)
            \t'
        }
    }\hhem\right)
    \nem\nem{\vphantom{\Big|}}^i{\vphantom{\Big|}}_j
    \,,\\
    \label{eq:inftylimit-U}
    U^i{}_j({+}\infty,{-}\infty)
    &= 
    \exp\nem\nem\left(\mem{
        i\hem \M^{+1}\nem(\one,a;\three)\mem
        \mathe^{i\three\cdot b}\mem
        \deltabar(2\mem\three{\mem\cdot\mem}\one)
        \mem c^a\mem i(t_a)
    }\hhem\right)
    \nem\hnem{\vphantom{\big|}}^i{\vphantom{\big|}}_j
    \,.
\end{align}
\end{subequations}
Note that $\smash{\underset{\adjustbox{raise=0.2em,scale=0.8}{$*$}}{
    \psi
}{\vphantom{\psi}}^i\hnem(\t)} = \psi^i\hnem(\t)$
and 
$\smash{\underset{\adjustbox{raise=0.2em,scale=0.8}{$*$}}{
    \bpsi
}{\vphantom{\bpsi}}_i(\t)} = \bpsi_i(\t)$,
since $\dot{\psi}^i = 0$ and $\dot{\bar{\psi}}_i = 0$ in the free theory.
% \begin{align}
%     \smash{\underset{\adjustbox{raise=0.2em,scale=0.8}{$*$}}{
%         \psi
%     }{\vphantom{\psi}}^i\hnem(\t)}
%     = U^i{}_j(\t_*,\t_0)\mem 
%     \psi^j\hnem(\t_0)
%     \,,\quad
%     \smash{\underset{\adjustbox{raise=0.2em,scale=0.8}{$*$}}{
%         \bpsi
%     }{\vphantom{\bpsi}}_i(\t)}
%     = \bpsi_j(\t_0)\mem
%     U^j{}_i(\t_0,\t_*)
%     \,.
% \end{align}
The color rotation \eqref{eq:pwH-colorphase} may be thought of as the ``square root'' of 
the chiral Lorentz transformation \eqref{eq:pwH-Lorentz}
or the 
% ``gauge'' translation
additional null translation
\eqref{eq:pwH-X} 
which we derive shortly.
%% "gauge" ~ zitterbewegung in the $|\bo\rsq$-plane is "invisible" ~ residual diff in the LC gauge spacetime v_\da(q) \partial/\partial p_\da (?)
%% splitting $x^{\da\a}$ = "grav gauge-charge" + "kinematics" --- ambiguou
%%
%% how many "independent" pieces can be put together? maximal?
%% null zitterbewegung of z / internal rotation / null rotation / color rotation
The time evolution and the impulse of the color charge $q_a$
easily follow from translating \eqref{eq:pwH-colorphase}-\eqref{eq:inftylimit-U}
to the adjoint representation.
One may expand the exponential and find
\begin{align}
\begin{split}
    \label{eq:colorimpulses}
    \mathit{\Delta}\psi^i
    \mem&=\mem i\hem \M^{+1}\nem(\one,a;\three)\mem
        \mathe^{i\three\cdot b}\mem
        \deltabar(2\mem\three{\mem\cdot\mem}\one)\,
        (\hhem (t_c)^i{}_j\hem \psi^j \hhem)\mem ic^c 
    + \mathcal{O}(g^2)
    \,,\\
    \mathit{\Delta}\bpsi_i
    \mem&=\mem i\hem \M^{+1}\nem(\one,a;\three)\mem
        \mathe^{i\three\cdot b}\mem
        \deltabar(2\mem\three{\mem\cdot\mem}\one)\,
        (\hhem -\bpsi_j (t_c)^j{}_i \hhem)\mem ic^c 
    + \mathcal{O}(g^2)
    \,,\\
    \mathit{\Delta}q_a
    \mem&=\mem i\hem \M^{+1}\nem(\one,a;\three)\mem
        \mathe^{i\three\cdot b}\mem
        \deltabar(2\mem\three{\mem\cdot\mem}\one)\,
        (\hhem -q_b\hem f^b{}_{ca} \hhem)\mem ic^c 
    + \mathcal{O}(g^2)
    \,.
\end{split}
\end{align}
\vspace{-0.8\baselineskip}

% \paragraph{Kerr exact scattering}
\paragraph{Exact Kerr impulses}

For Kerr in the self-dual gravitational plane wave, 
% we have
the symplectic form is
given by
\begin{align}
\begin{split}
    \label{eq:pws.Kerr}
    {\renewcommand{\arraystretch}{1.1}
    \renewcommand{\arraycolsep}{0em}
    \begin{array}{rl}
        \omega_\text{Kerr}^\text{SDPW}
        \,=\,\mem\hhem
        {}&{}
        i\mem d\bZ_I{}^\rmA \swedge dZ_\rmA{}^I
        \\
        {}&{}\displaystyle
        - \frac{1}{\MPl}\mem
            \langle\i|dp|\bo\rsq {\hem\wedge\hem} \lsq\bo|dz|\i\rangle
            \mem\mathe^{-i\hem\bozo}
        + \frac{im\tilde{x}}{2\MPl}\mem
            \bo_\da\hem\bo_\db\hhem(dz \swedge dz)^{\da\db}
            \mem\mathe^{-i\hem\bozo}
        \,.
    \end{array}}
\end{split}
\end{align}
% for the symplectic form.
The equation of motion reads
\begin{align}
\begin{split}
    \label{eq:xeom+2}
    \dot{\rambda}_{I\da} 
    = (i\nu{\hnem/}2m)\mem  
    \rambda_{\smash{I\db}\vphantom{I}} \mem \bo^\db \bo_\da
    \mem\mathe^{-i\hem\bozo}
    - {\textstyle\frac{1}{2i}} (\s_a \rambda)_{I\da}
    &\,,\quad
    \dot{\lambda}_\a{}^I 
    = {\textstyle\frac{1}{2i}} (\lambda \hem \s_a)_\a{}^I
    \,,\\
    % \dot{z}^{\da\a} 
    % = \minime m(p^{-1})^{\da\a} 
    % + (\tilde{x}\mem{\bo^\da\i^\a}\nem\nem/2\MPl)
    % \mem\mathe^{-i\hem\bozo}
    \dot{z}^{\da\a} 
    = \minime\big(\hem{
        m(p^{-1})^{\da\a} 
        {\hem-\mem} (\nu{\hnem/\hhnem}m)\mem{\bo^\da\i^\a}
        \mem\mathe^{-i\hem\bozo}
    }\mem\big)
    &\,,\quad
    \dot{y}^{\da\a}
    = -(i\nu{\hnem/}2m)\mem 
    \bo^\da \bo_\db \mem y^{\db\a}
    \mem\mathe^{-i\hem\bozo}
    \,.
\end{split}
\end{align}
The additional ``zitterbewegung'' of $z^{\da\a}$ lies within the $\da$-plane of $|\bo\rsq$.
As a result,
$\nu {\mem=\mem}$ $(m{\hhnem/\hnem}\MPl)\hem\tilde{x}$ is conserved, and
\eqref{eq:zlinear} is valid.
This means that the solution \eqref{eq:pwH-solution} holds except for $z^{\da\a}\hnem(\t)$.
The revised solution for $z^{\da\a}\hnem(\t)$ is given as
\begin{align}
    \label{eq:zsol-fluc}
    z^{\da\a}\hnem(\t)
    \mem=\mem z^{\da\a}\hnem(\t_0) 
    + F^\da{}_\db(\t,\t_0)\mem u^{\db\a}
    + X^{\da\a}\hnem(\t,\t_0)
    \,,
\end{align}
where the new contribution $X^{\da\a}\hnem(\t,\t_0)$ is given by
% \begin{subequations}
\begin{align}
    \label{eq:pwH-X}
    X^{\da\a}\hnem(\t,\t_0)
    \mem:={}&\mem
    \frac{\tilde{x} \mem\bo^\da\i^\a}{2\MPl}\mem
        \mathe^{i\three\cdot(z(\t_0)-u\t_0)}
        \nem\nem\int_{\t_0}^{\t}
        \hspace{-0.2em} d\t'\,\mem
        \mathe^{
            i (\three\cdot u)
            \t'
        }
    \,,\\
    \label{eq:Delta-X}
    \iq
    X^{\da\a}\hnem({+}\infty,{-}\infty)
    \mem={}&\mem
        \M^{+2}\hnem(\one,a;\three)\mem
        \mathe^{i\three\cdot b}\mem
        \deltabar(2\mem\three{\mem\cdot\mem}\one)\,
        (\bo^\da\i^\a\nem\hnem/mx)
    \,.
\end{align}
% \end{subequations}
% 
Accordingly, we have 
\begin{align}
    \label{eq:z*-fluc}
    % \iq
    \smash{\underset{\adjustbox{raise=0.2em,scale=0.8}{$*$}}{z}{}^{\da\a}\hnem(\t)}
    \mem&=\mem
    z^{\da\a}\hnem(\t_0)
    + ((\t_* {-\mem} \t_0)\hem L^\da{}_\db(\t,\t_0)
    - G^\da{}_\db(\t,\t_0))\mem u^{\db\a}
    + X^{\da\a}\hnem(\t,\t_0)
    \,.
\end{align}
Finally, it follows
from \eqref{eq:zsol-fluc} and \eqref{eq:t*equation} that
% \begin{subequations}
% \begin{align}
% \begin{split}
%     \label{eq:timedifference-fluc}
%     \t_*(\t_0,\t)
%     \,&=\, 
%     \t_0 
%     \mem+\mem
%     z^{\da\a}\hnem(\t_0)\mem p_{\a\db}(\t_0)\mem L^\db{}_\da(\t_0,\t) /m
%     \,,\\
%     &
%     \phantom{{}=\,\t_0}
%     \mem-\mem
%     \frac{\tilde{x}^2}{2\MPl}\mem
%         \mathe^{i\three\cdot(z(\t_0)-u\t_0)}
%         \nem\nem\int_{\t_0}^{\t}
%         \hspace{-0.2em} d\t'\,\mem
%         \mathe^{
%             i (\three\cdot u)
%             \t'
%         }
%     \,,
% \end{split}\\
%     \iq
%     \t_*({-}\infty,{+}\infty) - \t_*({-}\infty,{-}\infty)
%     \,&=\, 
% \end{align}
% \end{subequations}
\begin{align}
    \label{eq:timedifference-fluc}
    \t_*(\t_0,\t)
    = \t_0 
    + \frac{1}{m}\hem
    z^{\da\a}\hnem(\t_0)\mem p_{\a\db}(\t_0)\mem L^\db{}_\da(\t_0,\t)
    - \frac{\tilde{x}^2}{2\MPl}\mem
    \mathe^{i\three\cdot(z(\t_0)-u\t_0)}
    \nem\nem\int_{\t_0}^{\t}
    \hspace{-0.2em} d\t'\,\mem
    \mathe^{
        i (\three\cdot u)
        \t'
    }
    \,,
\end{align}
from which we obtain
\begin{align}
    \label{eq:timed*-fluc}
    \delta\t_*
    \mem&=\mem
    \M^{+2}\hnem(\one,a;\three)\mem \mathe^{i\three\cdot b}\mem
    \deltabar(2\mem\three{\mem\cdot\mem}\one)\,
        ((\three{\mem\cdot\mem}(a{\mem+\mem}ib){\mem-\mem}1)/m)
        % \mem\hhem
    % \,,\\  
    % \label{eq:z*-fluc}
    % % \iq
    % \smash{\underset{\adjustbox{raise=0.2em,scale=0.8}{$*$}}{z}{}^{\da\a}\hnem(\t)}
    % \mem&=\mem
    % z^{\da\a}\hnem(\t_0)
    % + ((\t_* {-\mem} \t_0)\hem L^\da{}_\db(\t,\t_0)
    % - G^\da{}_\db(\t,\t_0))\mem u^{\db\a}
    % + X^{\da\a}\hnem(\t,\t_0)
    \,.
\end{align}

The additional term $X^{\da\a}\hnem(\t,\t_0)$ does not affect
$\mathit{\Delta}p_{\a\da}$ and $\mathit{\Delta}y^{\da\a}$,
but changes occur in $\mathit{\Delta}\lambda_\a{}^I$, $\mathit{\Delta}\rambda_{I\da}$, and $\mathit{\Delta}z^{\da\a}$.
First, the expression \eqref{eq:exactimp.lrambda} 
for the spin frames
holds with \eqref{eq:timed*-fluc} instead of \eqref{eq:timed*}.
Second, the angular momentum impulses \eqref{eq:exactimp.am} change to
\begin{align}
    \label{eq:exactimp.am-X}
    \mathit{\Delta}J^{\smash{\da\db}\vphantom{(\da}}
    \mem=\mem {}&{}
    i\hem \M^{+2}\hnem(\one,a;\three)\mem \mathe^{i\three\cdot b}\mem \hhem\deltabar(2\mem\three{\mem\cdot\mem}\one)\mem
        \big({
            - z_{\smash{*}}^{(\da\a} 
            \hem \three_\a\vphantom{z}^{\smash{\db)}\vphantom{(\da}}
    + i\mem \bo^{\smash{(\da}\vphantom{(\da}}\bi^{\smash{\db)}\vphantom{(\da}}
    + K^{\smash{\da\db}\vphantom{(\da}}\hhem
        }\big)
    % - X^{\da\a}\hnem({+}\infty,{-}\infty)\mem p_{\smash{\a\db}\vphantom{\da}}({+}\infty)
    \,,\\
    \nonumber
    \mathit{\Delta}L^{\smash{\da\db}\vphantom{(\da}}
    \mem=\mem {}&{}
    i\hem \M^{+2}\hnem(\one,a;\three)\mem \mathe^{i\three\cdot b}\mem \hhem\deltabar(2\mem\three{\mem\cdot\mem}\one)\mem
        \big({
            - (z_{\smash{*}} \hem{+}\mem 2ia)^{(\da\a}
            \hem \three_\a\vphantom{z}^{\smash{\db)}\vphantom{(\da}}
    + i\mem \bo^{\smash{(\da}\vphantom{(\da}}\bi^{\smash{\db)}\vphantom{(\da}}
    + K^{\smash{\da\db}\vphantom{(\da}}\hhem
        }\big)
    % - X^{\da\a}\hnem({+}\infty,{-}\infty)\mem p_{\smash{\a\db}\vphantom{\da}}({+}\infty)
    \,,\\
    \nonumber
    \mathit{\Delta}\bar{J}^{\smash{\a\b}\vphantom{(\da}}
    \mem=\mem
    \mathit{\Delta}\bar{L}^{\smash{\a\b}\vphantom{(\da}}
    \mem=\mem {}&{}
    i\hem \M^{+2}\hnem(\one,a;\three)\mem \mathe^{i\three\cdot b}\mem \hhem\deltabar(2\mem\three{\mem\cdot\mem}\one)\,
        \big({
            - \three^{(\a}{}_\da 
            \hem z_{\smash{*}}^{\da\b)}
    % + \bar{K}^{\smash{\a\b}\vphantom{(\da}}\hhem
    -i\mem \o^{\smash{(\a}\vphantom{(\da}}\i^{\smash{\b)}\vphantom{(\da}}
        }\big)
    % - p_{\smash{\a\da}\vphantom{\da}}({+}\infty)\mem X^{\da\b}\hnem({+}\infty,{-}\infty)
    \,,\\
    \label{eq:Kam}
    K^{\smash{\da\db}\vphantom{(\da}}
    \mem:=\mem {}&{}
    % i\mem \bo^{\smash{(\da}\vphantom{(\da}}\bi^{\smash{\db)}\vphantom{(\da}}
    % + 
    (\bo^{\smash{\da}\vphantom{(\da}}\bo^{\smash{\db}\vphantom{(\da}}\nem\nem/mx)\mem
    \big(\hem{
        \M^{+2}\hnem(\one,a;\three)\mem \mathe^{i\three\cdot b}\mem \hhem\deltabar(2\mem\three{\mem\cdot\mem}\one)
        +i\mem \langle\i|\one|\bi\rsq
    }\mem\big)
    % \,,\\
    % \bar{K}^{\smash{\a\b}\vphantom{(\da}}
    % &= 
    % -i\mem \o^{\smash{(\a}\vphantom{(\da}}\i^{\smash{\b)}\vphantom{(\da}}
    \,.
\end{align}
% where
% \begin{align}
% \begin{split}
%     \label{eq:Kam}
%     K^{\smash{\da\db}\vphantom{(\da}}
%     &\mem:=\mem 
%     % i\mem \bo^{\smash{(\da}\vphantom{(\da}}\bi^{\smash{\db)}\vphantom{(\da}}
%     % + 
%     (\bo^{\smash{\da}\vphantom{(\da}}\bo^{\smash{\db}\vphantom{(\da}}\nem\nem/mx)\mem
%     \big(\hem{
%         \M^{+2}\hnem(\one,a;\three)\mem \mathe^{i\three\cdot b}\mem \hhem\deltabar(2\mem\three{\mem\cdot\mem}\one)
%         +i\mem \langle\i|\one|\bi\rsq
%     }\mem\big)
%     % \,,\\
%     % \bar{K}^{\smash{\a\b}\vphantom{(\da}}
%     % &= 
%     % -i\mem \o^{\smash{(\a}\vphantom{(\da}}\i^{\smash{\b)}\vphantom{(\da}}
%     \,.
% \end{split}
% \end{align}
Since the $\t_*$ term drops out in the angular momenta, 
the definition \eqref{eq:deflection-intersection} of $z_{\smash{*}}^{\da\a}$ still applies to \eqref{eq:exactimp.am-X}.
% Lastly, \eqref{eq:amimpulses} changes to
% \begin{align}
% \begin{split}
%     \label{eq:amimpulses-X}
%     % % \label{eq:amimpulse.S}
%     % \mathit{\Delta}S_{mn} &= 
%     % i\hem \M^+\mem \deltabar(2\mem\three{\mem\cdot\mem}\one)\hem
%     % \big(\mem{
%     %     \hnem- 2i\mem a_{[m} \three_{n]}
%     %     - \ve_{mnrs} a^r \three^s 
%     % }\big)
%     % \,,\\
%     % \label{eq:amimpulse.L}
%     \mathit{\Delta}L_{mn} &=
%     i\hem \M^{+2}\hnem(\one,a;\three)\mem \mathe^{i\three\cdot b}\mem \hhem\deltabar(2\mem\three{\mem\cdot\mem}\one)\mem
%     \big(\mem{
%         2\mem \one_{[m} \three_{n]} (T/m)
%         + 2b_{[m} \three_{n]} 
%         + \ve_{mnrs} a^r \three^s
%     }\big)
%     \,,\\
%     % \label{eq:amimpulse.J}
%     \mathit{\Delta}J_{mn} &=
%     i\hem \M^{+2}\hnem(\one,a;\three)\mem \mathe^{i\three\cdot b}\mem \hhem\deltabar(2\mem\three{\mem\cdot\mem}\one)\mem
%     \big(\mem{
%         2\mem \one_{[m} \three_{n]} (T/m)
%         + 2b_{[m} \three_{n]} 
%         - 2i\mem a_{[m} \three_{n]}
%     }\big)
%     \,.
% \end{split}
% \end{align}
However, the intersection point
no longer exists
because $K^{\smash{\da\db}\vphantom{(\da}}$ makes $\mathit{\Delta}J^{mn}$ no longer polar.

Overall, the heavenly plane-wave scattering problem of {\Kerr}/Kerr
is a maximally extended system 
of interlocking commuting transformations:
self-dual null rotation,
holomorphic self-dual null translation,
and
color and little group rotations
with a fixed point.

\section{Summary and Outlook}
\label{sec:Discussion}

%% Elevator in 10 minutes
%% "Perturbation theory can be reformulated"
%% "apply to this particular problem of a massive twistor"
%% what we have computed & new results
Let us recapitulate the main ideas and results we have presented.
The inputs of this paper were
a)
symplectic perturbation theory \cite{spt},
b)
the massive twistor description of a free massive spinning particle in four dimensions \cite{ambikerr0}.
\begin{itemize}
    \item
        The ``symplectic perturbation theory'' of a particle \cite{spt}
        % implements
        understands
        coupling 
        to background fields
        as perturbations on the
        % particle's
        Poisson bracket
        while retaining the same Hamiltonian.
        From the geometric series expansion \eqref{eq:pert1-series},
        one perturbatively 
        % obtains
        computes
        the
        Poisson bracket and
        the Hamiltonian equation of motion
        in the orders of the 
        % ``field strength.''
        symplectic perturbation.
        The amplitude 
        is then
        obtained
        from 
        the Born approximation.
    \item
        % Lorentz symmetry and little group symmetry
        % uniquely determine
        % the geometry of the free theory's physical phase space \cite{ambikerr0},
        % i.e., 
        % the kinematics of a massive spinning particle.
        Lorentz symmetry and little group symmetry
        uniquely determine
        the kinematics of a massive spinning particle:
        the geometry of the free theory's physical phase space \cite{ambikerr0}.
        % The massive twistor variables 
        % offer an interesting complex-geometrical way of describing the phase space,
        % % grouping 
        % unifying spin and spacetime into ``spin-space-time''
        % and redescribing
        % momentum and the body frame with massive spinor-helicity variables.
        % In turn, it reveals a hidden 
        % ``zig-zag'' (K\"ahler) structure of massive spinning particles.
        % % The massive twistor description reveals a hidden ``zig-zag'' structure of massive spinning particles.
        A hidden K\"ahler (``zig-zag'')
        structure of the phase space
        manifests in the massive twistor description
        % which 
        that
        unifies spin and spacetime into ``spin-space-time''
        and redescribes
        momentum and the body frame with massive spinor-helicity variables.
\end{itemize}
Applying symplectic perturbation theory
to the massive ambitwistor space,
we aimed to construct a formulation
of interacting massive spinning particles
that respects the remarkable K\"ahler property of the free theory.
It turned out that
the zig-zag structure of the free theory (``kinematics'')
tells us a lot about the interacting theory (``dynamics'').
\begin{itemize}
    \item
        Via the zig-zag structure,
        spin precession behavior
        constrains
        the spin-space-time part of the symplectic perturbation.
    \item 
        ({Newman-Janis shift derived})
        Especially,
        the fact that the minimal spin precession equation of motion under a self-dual background
        is given by $\dot{\lambda}_\a{}^I\nem = 0$
        implies that self-dual field strengths continue holomorphically into the 
        complexified Minkowski space
        if the particle is minimally coupled.
    \item
        ({Geometrical interpretation of the Wilson coefficients})
        For 
        % generic
        non-minimal
        couplings,
        Levi and Steinhoff \cite{Levi:2015msa}'s 
        spin multipole
        Wilson coefficients
        % ---redefined on complex worldlines---
        control
        non-holomorphic continuations 
        of the self-dual ``field strength''
        into the spin-space-time.
    \item 
        The zig-zag diagrammatics
        % vividly
        manifests the minimal nature of the Kerr-Newman coupling
        as well as the non-minimal nature of generic couplings.
        The number of terms in the zig-zag representation of the perturbed 
        % Poisson
        bracket
        at $n$\textsuperscript{th} order
        is
        given as the following.
        % \begin{itemize}[leftmargin=1.86em,label={\adjustbox{scale=0.6,valign=c}{$\bullet$}}]
        %     \item
        %         (Earthly background) 
        %         $2^{2n+1}$ for generic couplings,
        %         but only $2$ for the minimal and the ``antipodal minimal'' couplings
        %     \item
        %         (Heavenly background)
        %         $2^{n+1}$ for generic couplings,
        % \end{itemize}
        \begin{center}
            \begin{tabular}{r|cc}
                \hlinewd{1pt}
                    {} 
                    & generic couplings 
                    & 
                        minimal or 
                        antipodal minimal 
                \\\hline
                    Earthly background
                    & $2^{2n+1}$
                    & $2$       
                \\\hline
                    \,\,\,Heavenly background 
                    & $2^{n+1}$
                    & expansion truncates
                \\
                \hlinewd{1pt}
            \end{tabular}
        \end{center}
\end{itemize}
Not only gaining these insights,
we also have obtained new results:
a) a complete generalization of the TBMT and MPTD equations up to all orders in spin
and its matching with the three-point on-shell amplitude,
b) exact symplectic structures of the {\Kerr} and Kerr particles in self-dual backgrounds
and exact solutions to their equations of motion.
\begin{itemize}
    \item
        A dictionary between
        $\mathcal{O}((\text{curvature})^1{\hem\cdot\mem}(\text{spin})^\infty)$ 
        spin precession
        equations of motion,
        spin frame impulses,
        % impulses of classical observables,
        and 
        % on-shell
        classical-spin
        three-point amplitudes
        is
        constructed
        by taking the
        spin-space-time
        symplectic perturbation as a common root.
    % \item
    %     When redefined on complex worldlines, 
    %     the Wilson coefficients get 
    %     straightforwardly related to the coupling constants 
    %     that appear in \cite{ahh2017}'s on-shell amplitude.
    \item
        The non-spin-space-time part of the symplectic perturbation can be deduced from demanding
        % symplectivity.
        the closure of the symplectic form.
    \item
        Exact expressions of
        {\Kerr} and Kerr symplectic structures
        in self-dual plane-wave backgrounds
        are bootstrapped
        from the matching \eqref{eq:e1e2}
        and the symplectivity requirement.
        The resulting equations of motion
        can be
        analytically solved.
        In turn,  
        % impulses of classical observables
        % and amplitudes
        amplitudes and impulses
        are obtained as exact quantities.
    \item
        Further,
        symplectic structures of
        {\Kerr} and Kerr 
        in generic self-dual backgrounds
        are deduced from
        the derivative counting \eqref{eq:io-trading}:
        \begin{align}
        \begin{split}
            \label{eq:disc.Kerrs}
            \text{{\Kerr}}:\quad
            \omega
            &\mem=\mem
                i\mem d\bZ_I{}^\rmA \swedge dZ_\rmA{}^I
                + i\mem d\bpsi_i \swedge d\psi^i
                % + d(\hhem 
                %     q_a\hem A^a{}_m\hnem(z)\mem dz^m
                % \hhem)
                + d\Big(\hem\mem{
                    i\hem\bpsi_i(t_a)^i{}_j\psi^j\hem
                    A^a{}_m\hnem(z)\mem dz^m
                }\mem\Big)
            \,,\\
            \text{Kerr}:\quad
            \omega
            &\mem=\mem
                i\mem d\bZ_I{}^\rmA \swedge dZ_\rmA{}^I
                + d\Big(\mem{
                    -\rambda_{I\da} (\bs_m)^{\da\a} \lambda_\a{}^I\hem
                    h^m{}_n\hnem(z)\mem dz^n
                }\mem\Big)
            \,.
            % \,,
        \end{split}
        \end{align}
\end{itemize}

% For the ``dictionary,''
For the reader's convenience,
let us provide a summary of the ``dictionary'' here.
When the spin-space-time part of the self-dual symplectic perturbation is given as
\begin{align}
    \label{eq:disc.omega-generic}
    \omega'^+
    {}&{}\hhem\,\supset\,\,
    {^0\omega'^+_{mn}}\mem
        (\minie\mem \wedgetwo{z}{z}{m}{n})
    + 
    {^1\omega'^+_{mn}}\mem
        \wedgetwo{z}{\bz}{m}{n}
    +
    {^2\omega'^+_{mn}}\mem
        (\minie\mem \wedgetwo{\bz}{\bz}{m}{n})
    % \,.
    \,,
\end{align}
where each component is controlled by the Wilson coefficients as \eqref{eq:012omegass},
the
first-order
Poisson brackets, 
equation of motion,
and
spin frame impulses
are
respectively given by
\begin{align}
    &
    % \left\{
    {\renewcommand{\arraycolsep}{0.05em}
    \renewcommand{\arraystretch}{1.2}
    \begin{array}{rl}
        \{
            \rambda_{I\da}
            ,
            \rambda_{\smash{J\db}\vphantom{\b}}
        \}
        &{}\simeq\,{}
            2\hhem\Delta^{-1}\e_{IJ}
            \cdot
            {^0\omega'_{\smash{\da\db\vphantom{\b}}}}
        \,,\\
        \{
            \rambda_{I\da}
            ,
            \lambda_\b{}^J
        \}
        &{}\simeq\,{}
            2(\lambda^{-1}\e)_{I\b}
            \cdot
            {^1\omega'_{\smash{\da\dd\vphantom{\b}}}}\mem
            (\rambda^{-1})^{\dd J}
        \,,\\
        \{
            \lambda_\a{}^I
            ,
            \lambda_\b{}^J
        \}
        &{}\simeq\,{}
            2\e_{\a\b}
            \cdot
            {^2\omega'_{\smash{\dc\dd\vphantom{\b}}}}\mem
            (\rambda^{-1})^{\dc I}\hhnem
            (\rambda^{-1})^{\dd J}
        \,,
        % \adjustbox{raise=-2ex}{\vphantom{0}}
    \end{array}}
    % \right.
    \\[0.3ex]
    &
    % \left\{
    {\renewcommand{\arraycolsep}{0.05em}
    \renewcommand{\arraystretch}{1.2}
    \begin{array}{rl}
        m_0\hem \dot{\rambda}_{I\da}
        &{}\simeq\,{}
        \rambda_{\smash{I\db}\vphantom{\b}}\hem
        (\hem^{0+1}\hnem\omega')^\db{}_\da
        \,,\\
        m_0\hem \dot{\lambda}_\a{}^I
        &{}\simeq\,{}
        {-(^{1+2}\mathllap{\adjustbox{raise=-0.2ex}{$_\mathcal{I}\hem$}}\omega')_\a{}^\b
        \hem\lambda_\b{}^I}
        =
        % p_{\a\da}\mem (^{1+2}\hnem\omega')^\da{}_\db (p^{-1})^{\db\b}
        % \hem\lambda_\b{}^I
        \lambda_\a{}^J\, \rambda_{J\da}\mem (^{1+2}\hnem\omega')^\da{}_\db\mem (\rambda^{-1})^{\db I}
        % \,.
        \,,
        % \adjustbox{raise=-2.05ex}{\vphantom{0}}
    \end{array}}
    % \right.
    \\[0.38ex]
    &
    % \left\{
    {\renewcommand{\arraycolsep}{0.05em}
    \renewcommand{\arraystretch}{1.2}
    \begin{array}{rl}
        \lsq\two_I|
        &{}\simeq\,{}
        \lsq\one_I| 
        \,+\,
        i\hem (e_h\hem m_0 x^h \hem\mathe^{\three\cdot a})\,
        {^{0+1}\mathfrak{S}[f^+](\three{\kern0.02em\cdot\kern0.02em}a)}\,
        \deltabar(2\mem\three{\kern0.02em\cdot\kern0.02em}\one)\,\mem
        {\textstyle\frac{1}{m_0x}}
        \lsq\bone_I\bthree\rsq\lsq\bthree|
        \,,\\
        |\two^I\rangle
        &{}\simeq\,{}
        |\one^I\rangle 
        \,+\,
        i\hem (e_h\hem m_0 x^h \hem\mathe^{\three\cdot a})\,
        {^{1+2}\mathfrak{S}[f^+](\three{\kern0.02em\cdot\kern0.02em}a)}\,
        \deltabar(2\mem\three{\kern0.02em\cdot\kern0.02em}\one)\,
        {\textstyle\frac{x}{m_0}} \hem|\three\rangle\langle\three \one^I\rangle
        % \,.
        % \,,
    \end{array}}
    % \right.
\end{align}
% at linear order in $\omega'$.
in a self-dual plane wave background 
with spinor-helicity variables $\three_\a$ and $\bthree_\da$.

Penrose 
has emphasized
the power and elegance of complex geometry \cite{penrose2005road,penrose1975aims,penrose1973twistor,penrose1984spinors1,penrose1987origins,Penrose:1999sz}.
The central message of 
our ``zig-zag''
approach
% to the physics of
% interacting
% massive spinning particles
is that the physics of interacting massive spinning particles
should be understood in a way that
fully appreciates and respects
the complex-geometrical structure
that is already inherent in the free theory.
Analyzing
everything in the holomorphic/anti-holomorphic
basis
has
made
the formulation of classical physics
more consistent with the 
amplitudes-level understanding,
% \cite{chkl2019,gmoov,guevara2019scattering,guevara2019black,Chung:2019duq,aoude2021classical,ahh2017},
provided new geometrical insights on the physics of spin precession, 
revealed properties
% facts and connections between ideas
that were obscure in the conventional approach,
and
% motivated 
% led to
% new ways of doing calculations.
motivated a new way of doing calculations.

%% Comments on the main text

A few comments are in order.
It was
% also
crucial in our analysis
that we have complexified the massive twistor space
to the massive ambitwistor space.
In particular,
for the heavenly equation of motion 
to admit a solution,
it is necessary to ``unlink''
$\lambda_\a{}^I$ with $\rambda_{I\da}$
and $z^{\da\a}$ with $\bz^{\da\a}$:
the scattering kinematics is complexified, 
and both 
% the
spacetime coordinates $x^{\da\a}$ and 
the
spin length $y^{\da\a}$ become complex variables.
Otherwise,
the plane-wave scattering test in definite-helicity plane-wave backgrounds
considered in Section \ref{sec:Wave}
will make no sense.

We also clarify
the role of pure self-duality in our discussion.
% There is a caveat in saying that earthly equations of motion simply follow from taking ``{\hhem$2\hem\Re$}'' to
% heavenly equations
% when there exists a non-spin-space-time part in the symplectic perturbation.
% Such a term can be hidden in the covariant symplectic perturbation theory \cite{ambikerr2},
% % as done in our follow-up paper \cite{ambikerr2},
% % but one still does not have a full-fledged theory.
% but it still requires extra work for making the Newman-Janis shift manifest.
% As we briefly remarked in Section \ref{sec:Pre.spt},
% an electric-magnetic dual formulation seems to be crucial 
% in resolving this issue.
%% -> first boil down A(z) in the heavenly eom to A(x) + ...
There seems
to be a fundamental obstruction 
in having an exact complex-worldline description
in earthly backgrounds,
which can be traced back to the 
``riddle'' of Kerr black hole:
the existence of the opposite-helicity contact vertex \cite{aoude2022searching}.
As we elaborate further in \cite{ambikerr2},
the
all-order exact
Newman-Janis shift 
of the particle's symplectic structure
% can only be manifested
manifests only
in heavenly backgrounds.
To pursue 
a complex-worldline description
in the earthly setting,
it seems that
the background has to be linearized/abelianized.
% In the earthly setting,
% a complex-worldline description is possible 
% if one linearizes/abelianizes the background.
For instance,
\cite{gmoov} was able to explore the Newman-Janis property 
in the earthly setting
by taking the two-potential approach
to self-dual and anti-self-dual field strengths
at the linearized level.
In this paper, we have put our focus on the heavenly case
% to obtain exact results.
to discuss exact results.

There are still a lot more topics to explore.
For instance,
the exact solutions in Section \ref{sec:Wave} 
can also be obtained for arbitrary wave profiles.
It would be 
more inspiring
if we could 
discuss the memory effect
% \cite{Taub:1948zza,sengupta1949scattering,bieri2013electromagnetic,Ilderton:2013dba,Ilderton:2013tb,Susskind:2015hpa,Pasterski:2015zua,strominger2016gravitational,pate2017color,Zhang:2017rno,Strominger:2017zoo,Shore:2018kmt,Campoleoni:2019ptc,adamo2022classical} 
% \cite{Taub:1948zza,sengupta1949scattering,bieri2013electromagnetic,Ilderton:2013dba,strominger2016gravitational,pate2017color,Zhang:2017rno,Strominger:2017zoo,adamo2022classical} 
\cite{Taub:1948zza,sengupta1949scattering,strominger2016gravitational,pate2017color,Strominger:2017zoo,adamo2022classical} 
and asymptotic symmetries of heavenly geometries \cite{Campiglia:2021srh}
in such a setting
and, in turn,
give a clearer physical interpretation 
of
% to
the impulses.
Also,
we could consider shockwave geometries
\cite{Penrose:1968me,Penrose:1968ar,penrose1972geometry}
and provide a symplectic perturbation theory reincarnation of the Hamiltonian-based approach of
% the classic papers
\cite{penrose1973twistor,tod1976two}.
Going further, one can relax various assumptions taken in our discussion.
For example, 
one can go beyond three-point amplitudes
or 
% go beyond 
% the
test-particle limit.
Also,
one can go beyond the equal-mass sector and
% try to 
incorporate more general interactions
such as torque effects \cite{costa2016spacetime}.
% in terms of symplectic perturbations.
Working in terms of symplectic perturbations
might help 
classifying such interactions
systematically.
% The symplectic perturbation language
% might enable a systematic classification
% of such interactions.

Finally, let us end with a few
further
suggestions for future directions.

\paragraph{Quadratic order in fields}

Regarding the zig-zag expansions,
we have consistently ignored
terms from the second order.
Yet,
we could have computed ``four-point'' zig-zag diagrams
as well.
For instance, suppose
a complexified background
$\omega' = \omega'^{\hem\four} + \omega'^{\hem\three}$
with both 
self-dual and anti-self-dual modes
such that
\begin{align}
    \label{eq:Compton-scheme}
    % \omega' = \omega'_{\four} + \omega'_\three
    % \,,\quad
    \omega'^{\hem\four} =
    \frac{i\nu_\four}{2}\mem 
    \bfour_\da \bfour_\db\mem 
    (dz \swedge dz)^{\da\db}
    \,
    \mathe^{-i\hem \bfour z \four}
    \,,\quad
    \omega'^{\hem\three} =
    \frac{i\bar{\nu}_\three}{2}\mem 
    \three_\a \three_\b\mem 
    (d\bz \swedge d\bz)^{\a\b}
    \,
    \mathe^{-i\hem \bthree \bz \three}
    \,.
\end{align}
Then 
we can consider zig-zag vector fields such as
\begin{align}
% \begin{split}
%     \label{eq:4vec+-}
%     \smash{
%         \rlap{\makebox[20.00pt][l]{}\adjustbox{raise=15.760pt,scale=0.85}{$\four$}}
%         \rlap{\makebox[39.24pt][l]{}\adjustbox{raise=-16.515pt,scale=0.85}{$\three$}}
%         \includegraphics[valign=c]{figs/l-b.pdf}
%         \includegraphics[valign=c]{figs/v+.pdf}
%         \includegraphics[valign=c]{figs/v-.pdf}
%         \includegraphics[valign=c]{figs/xD.pdf}
%     }
%     &= \{\blank,z^{\da\a}\}^\circ\,
%     \frac{8i}{\bDelta}\,
%     % (-8i/\bDelta)\,
%     \omega'^{\hem\four}_{\smash{\da\db}\vphantom{\b}}(z)
%     \, y^{\db\b} \,
%     \bomega'^{\hem\three}_{\b\a}(\bz)
%     \,,\\
%     &= \{\blank,z^{\da\a}\}^\circ\,\mem
%     i
%     \Big(
%         {-2\nu_\four\bar{\nu}_\three\mem [\bfour|y|\three\rangle}
%         \mem
%         (-\three_\a\bfour_\da)
%         \hnem
%     \Big)
%     \,e^{-i \bthree \bz \three -i\bfour z\four}
%     /\bDelta
%     \,,
%     % \,.
% \end{split}
%     \\
\begin{split}
    \label{eq:4vec-+}
    \smash{
        \rlap{\makebox[39.24pt][l]{}\adjustbox{raise=15.760pt,scale=0.85}{$\four$}}
        \rlap{\makebox[20.00pt][l]{}\adjustbox{raise=-16.515pt,scale=0.85}{$\three$}}
        \includegraphics[valign=c]{figs/l.pdf}
        \includegraphics[valign=c]{figs/v-.pdf}
        \includegraphics[valign=c]{figs/v+.pdf}
        \includegraphics[valign=c]{figs/xD-b.pdf}
    }
    &= \{\blank,\bz^{\da\a}\}^\circ\,
    \frac{-8i}{\Delta}\, 
    % (+8i/\Delta)\,
    \bomega'^{\hem\three}_{\a\b}(\bz)
    \, y^{\db\b} \,
    \omega'^{\hem\four}_{\smash{\db\da}\vphantom{\b}}(z)
    \,,\\
    &= \{\blank,\bz^{\da\a}\}^\circ\,\mem
    i
    \Big(\hem\hhhem
        {2\nu_\four\bar{\nu}_\three\mem \lsq\bfour|y|\three\rangle}
        \mem
        (-\three_\a\bfour_\da)
        \hnem
    \Big)
    \,e^{-i\hem \bthree \bz \three -i\hem \bfour z\four}
    /\Delta
    % \,.
    \,,
\end{split}
\end{align}
% Such ``higher-point'' zig-zag vector fields
% appear only if $\omega'$ has both
% holomorphic and anti-holomorphic components.
which contributes to the impulse at the quadratic order.

As 
the combinations $\lsq\bfour|y|\three\rangle$
and $(-\three_\a\bfour_\da)$
naturally 
arise
when studying the notorious Compton amplitude of Kerr
\cite{aoude2022searching},
na{\umi}vely
we can 
speculate on
a relationship between
zig-zag diagrams such as \eqref{eq:4vec-+}
and the
four-point amplitudes.
Higher-order zig-zag diagrams indeed seem to be 
a unique feature of the \textit{spinning} particle,
as for the spin-less case
$\omega'^{\hem\four}$ and $\omega'^{\hem\three}$
cannot cascade
because $\{x^{\da\a},x^{\db\b}\}^\circ = 0$.
However, 
it remains to be 
questionable
whether symplectic perturbations can be directly related to amplitudes
at higher orders;
as mentioned in \cite{spt} as well,
we only na{\umi}vely expect that
a ``dequantization'' of the S-matrix will be
a time-ordered action of the deformed Hamiltonian vector fields
as infinitesimal diffeomorphisms on the physical phase space,
\`a la geometric prequantization \cite{woodhouse1997geometric}.
Besides,
further $\mathcal{O}(\phi^2)$
terms can
in principle
enter into the symplectic perturbation
when there are both self-dual and anti-self-dual modes.
% \eqref{eq:4vec-+} may 
% % capture only the piece from the on-shell gluing.
% % be insufficient to 
% not
% capture all contributions from the contact vertices.
% 
% Overall,
% % whether zig-zag diagrams can 
% % illuminate the four-point physics of Kerr
% whether zig-zag diagrams will be successful also in illuminating the four-point physics of Kerr
% needs further investigation.
Overall, whether zig-zag diagrams will also be successful in illuminating Kerr's four-point physics needs further investigation.

\paragraph{Physical interpretation of zig-zag diagrams}

Now we would like to ask 
even
bigger questions.
Let us
remark on the inspirations behind 
the zig-zag diagram notation
and envision further developments of the idea.
% Zig-zag diagrams started as a 
% simplification of tensor graphs.
Zig-zag diagrams started from denoting
$\{z^{\da\a},\bz^{\db\b}\}^\circ =$ $-2iy^{\da\b} (\rambda^{-1}\nem\lambda^{-1})^{\db\a}$ and $\phi^\pm_{mn}$ 
in the Penrose graphical notation 
\cite{Penrose:1956tensormethods,penrose1971negdim,penrose1984spinors1,Kim:2019eab}:
\begin{align}
\begin{split}
    \label{eq:Penrosegraphical}
    \{z^{\da\a},\bz^{\db\b}\}^\circ = 
    \,
    \includegraphics[valign=c,scale=1.2]{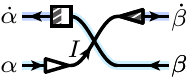}
    &\,,\quad
    \{\bz^{\da\a},z^{\db\b}\}^\circ = 
    \,
    \includegraphics[valign=c,scale=1.2]{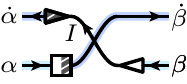}
    \,,\\
    \phi^+_{mn}\hnem(z)\mem
    \e_{\a\b}\hspace{0.12em}(\bs^{mn})_{\smash{\da\db}\vphantom{\b}} =
    \includegraphics[valign=c,scale=1.2]{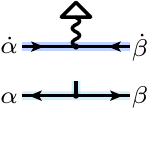}
    &\,,\quad
    \phi^- _{mn}\hnem(\bz)\hem
    (\s^{mn})_{\a\b}\mem\be_{\smash{\da\db}\vphantom{\b}} =
    \includegraphics[valign=c,scale=1.2]{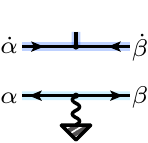}
    \,.
    % \,,
\end{split}
\end{align}
Simplifying,
these 
% motivated us to introduce
evolved to
the notations
$\,\smash{\includegraphics[valign=c,scale=1.0]{figs/z-bz.pdf}}\,$,
$\,\smash{\includegraphics[valign=c,scale=1.0]{figs/bz-z.pdf}}\,$
and 
$\,\smash{\includegraphics[valign=c,scale=1.0]{figs/insert+.pdf}}\,$,
$\,\smash{\includegraphics[valign=c,scale=1.0]{figs/insert-.pdf}}\,$.
% The notations
% $\,\smash{\includegraphics[valign=c,scale=1.0]{figs/l.pdf}\includegraphics[valign=c,scale=1.0]{figs/xD-b.pdf}}\,$
% and
% $\,\smash{\includegraphics[valign=c,scale=1.0]{figs/l-b.pdf}\includegraphics[valign=c,scale=1.0]{figs/xD.pdf}}\,$
% also 
% came from $\{z^{\da\a},\bDelta\}^\circ = \bDelta (p^{-1})^{\da\a}$
Thus,
one can regard
the double-line format
of zig-zag diagrams
as representing
the $\mathrm{SL}(2,\mathbb{C})$ index flows.
% Hence, the double-line format of the ``ribbons'' represent the $\mathrm{SL}(2,\mathbb{C})$ index flows.

% Curiously,
Interestingly,
such a double-line
% ---or ``ribbon''---
(or ``ribbon'')
design is reminiscent of
the ``worldsheet'' of 
Guevara, Maybee, Ochirov, O'Connell, and Vines
\cite{gmoov}.
% \cite{gmoov}'s ``worldsheet.''
In fact, the reader might have noticed that 
we pretended as if the white/black filling
somehow represents an actual surface connecting 
$\s {\,\mapsto\,} z^m\hnem(\s)$ and $\s {\,\mapsto\,} \bz^m\hnem(\s)$
(which may be called 
the
``GMOOV worldsheet'' \cite{gmoov}):
\begin{align}
    \omega'^+
    \,\,\,\,=\,\,\,\,
    \left\{\,\,\mem
    {\renewcommand{\arraycolsep}{0.3em}
    \renewcommand{\arraystretch}{1.5}
    \begin{array}{cccccccl}
            \includegraphics[valign=c,scale=1.0]{figs/insert+.pdf}
            & &
            & &
            & &
            &\,\,\,\text{(minimal)}
            \,,
        \\
            \includegraphics[valign=c,scale=1.0]{figs/insertg+ww.pdf}
            &+&
            \includegraphics[valign=c,scale=1.0]{figs/insertg+wb.pdf}
            &+&
            \includegraphics[valign=c,scale=1.0]{figs/insertg+bw.pdf}
            &+&
            \includegraphics[valign=c,scale=1.0]{figs/insertg+bb.pdf}
            &\,\,\,\text{(``exponential arbitrary-$g$'')}
            \,,
        \\
            \includegraphics[valign=c,scale=1.0]{figs/insert0+ww.pdf}
            &+&
            \includegraphics[valign=c,scale=1.0]{figs/insert0+wb.pdf}
            &+&
            \includegraphics[valign=c,scale=1.0]{figs/insert0+bw.pdf}
            &+&
            \includegraphics[valign=c,scale=1.0]{figs/insert0+bb.pdf}
            &\,\,\,\text{(maximally non-minimal)}
            \,,
        \\
            {}
            & &
            {}
            & &
            {}
            & &
            \includegraphics[valign=c,scale=1.0]{figs/insertA+.pdf}
            &\,\,\,\text{(antipodal minimal)}
            \,.
    \end{array}
    }
    \right.
\end{align}
The dot can be thought of as indicating the incident point of a self-dual ``non-linear quantum.''
The coupling is minimal
when the incident point lies
on the holomorphic worldline.

Thus, thinking of the ``graphical realism'' of Feynman diagrams,\footnote{
    a natural graphical representation of a perturbative series is believed to represent
    % actual
    space-time
    % ``physical''
    processes
}
one might ask, 
``do zig-zag diagrams represent an actual physical
% worldsheet?''
entity, such as the worldsheet?''
% The ``graphical realism'' of Feynman diagrams is that
% a natural diagrammatic representation of a perturbative expansion
% is believed to represent (physical) space-time processes.
% 
% While we avoid to answer such philosophical questions,
% We try to avoid answering
% such philosophical questions here,
It seems too early to answer such philosophical questions.
Yet,
na{\umi}vely
the association of 
zig-zag vector fields with amplitudes
such as 
in \eqref{eq:cqsim} or \eqref{eq:4vec-+}
suggests that
% \eqref{eq:disc.dequant-ampl} suggests that
the graphical elements of zig-zag diagrams 
might be
modules for building up various 
% ``vertex operators'' of
``vertices'' for
the
% a massive 
spinning particle.

\begin{figure}[t]
    \centering
    \includegraphics[scale=1.25, valign=c]{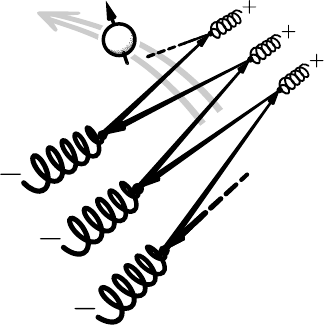}
    \caption{
        Penrose's zig-zag electron, reimagined.
        (Cf.\,Fig.\,25.1b of \cite{penrose2005road}.)
    }
    \label{fig:electron}
\end{figure}

At least,
we should stress that
zig-zag diagrams do not represent
% the worldline
spacetime trajectories
of a massive spinning particle,
although we 
once
made an analogy between zig-zag and Feynman diagrams
by imagining $\omega^\circ{}^{-1}$ as ``propagators''
and $\omega'$ as ``vertices.''
Our zig-zag theory
is not the same as Penrose's original concept of the zig-zag electron;
instead, it is a certain remake of it.
The zigs and zags
referred to
holomorphic and anti-holomorphic objects in the massive twistor space,
which in turn get associated with
% self-dual and anti-self-dual
left and right
chiralities of the background field that the particle interacts with.
% for the minimal coupling
If one wants to associate
% To associate
the image of ``zig-zag'' with spacetime trajectories,
schematically 
% we
one
can imagine a minimally coupled massive twistor particle
moving in an earthly background
as depicted in
Figure \ref{fig:electron}
(while ignoring the contact vertices).

\paragraph{Massive twistor diagrams?}

Our explorations
on the 
``diagrammar''
of 
the massive twistor particle
% finally 
eventually
extend to
Hodges diagram
\cite{arkani2010s-matrix,penrose1973twistor,penrose1975aims,hodges1980twistor,hodges1982twistor,hodges1983moller,hodges1983compton,hodges1985regularization,hodges1985mass,hodges1990feynman,huggett1993cohomology,penrose1998geometric,hodges2005a,hodges2005b},
which was indeed another inspiration
behind the development of zig-zag diagrams.

Let us reimagine zig-zag diagrams
in a Hodges-like
notation.
% We present 
% a Hodges-like version of zig-zag diagrams.
The implementation 
% of diagrams
is again by Penrose graphical notation,
but with 
the ``differentiation balloon'' of Penrose 
% Penrose's ``differentiation balloon''
as well as
birdtracks/trace diagram notations
for the $\mathrm{SU}(2)$ epsilon tensor
(see \cite{Kim:2019eab} and references therein).
The basic elements are
\begin{align}
    Z_\rmA{}^I
    =
    \includegraphics[valign=c,scale=1.0]{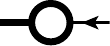}
    \,,\quad
    \bZ_I{}^\rmA 
    =
    \includegraphics[valign=c,scale=1.0]{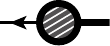}
    \,,\quad
    \e_{IJ} = \includegraphics[valign=c,scale=1.0]{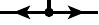}
    \,,\quad
    \e^{IJ} = \includegraphics[valign=c,scale=1.0]{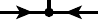}
    \,.
\end{align}
The difference in thickness distinguishes 
% the dimensions of
the vector spaces.
Also, let us denote
differential forms by ``highlighted'' dots:
% The one-forms are denoted as ``highlighted'' dots:
\begin{align}
    dZ_\rmA{}^I
    =
    \includegraphics[valign=c,scale=1.0]{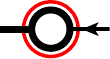}
    \,,\quad
    d\bZ_I{}^\rmA 
    =
    \includegraphics[valign=c,scale=1.0]{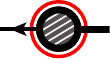}
    \,.
\end{align}
The free theory Poisson bivector $\omega^\circ{}^{-1}$
is suggestive of the Fourier transform 
% from the twistor space to the dual twistor space,
between the twistor and dual twistor spaces,
which is drawn like squiggly ``propagators'' in the Hodges notation.
In this light, we denote
\begin{align}
    i\includegraphics[valign=c]{figs/l.pdf}\includegraphics[valign=c]{figs/r-b.pdf}
    \,\,\,\,\text{or}\,\,\,\,
    {-i}\includegraphics[valign=c]{figs/l-b.pdf}\includegraphics[valign=c]{figs/r.pdf}
    \quad\rightsquigarrow\quad
    \includegraphics[valign=c]{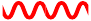}
\end{align}
so that
\begin{align}
    \frac{\partial}{\partial Z_\rmA{}^I}
    =
    \includegraphics[valign=c]{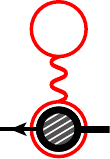}
    \,,\qquad
    \frac{\partial}{\partial\bZ_I{}^\rmA}
    =
    \includegraphics[valign=c]{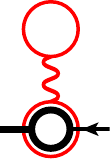}
    \,,\qquad\,
    \includegraphics[valign=c]{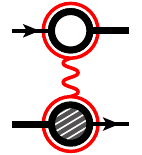}
    =
    \includegraphics[valign=c]{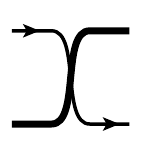}
    % =
    % \delta_I{}^J \delta_\rmA{}^\rmB
    \,.
\end{align}
Now consider the self-dual plane wave symplectic perturbation for electromagnetism:
\begin{align}
    \omega'^+
    \mem=\mem
    % =
    % \frac{iq}{\sqrt{2}}\mem \bo_\da \bo_\db
    % (dz \swedge dz)^{\da\db}
    % \,\mathe^{-i\bozo}
    % =
     -\frac{iq}{\sqrt{2}\Delta}\mem
    (\bthree_\dc\rho^{\dc\rmA})(\bthree_\dd\rho^{\dd\rmB})
    \hem(dZ_\rmA{}^I\nem\nem\wedge dZ_\rmB{}^J)\mem
    \e_{IJ}
    \,\mathe^{-i\hem\bthree{z}\three}
    \,.
\end{align}
We can half-Fourier transform this into the photon's dual twistor space,
$\bW^A = \big( -i\bar{\varpi}^\a \,\,\mem \bthree_\da \big)$:
\begin{align}
    \label{eq:disc.MTem-2}
    \omega'^+
    \mem=\mem
    -\frac{iq}{\sqrt{2}\Delta}\mem
    (\bW^\rmA dZ_\rmA{}^I) {\mem\wedge\mem} (\bW^\rmB dZ_\rmB{}^J)\mem
    \e_{IJ}
    \,\delta^{(2)}\hnem(\bar{\varpi}-\bthree z)
    \,.
\end{align}
Curiously,
% Intriguingly,
the photon's massless \textit{dual} twistor is incident at the massive particle's \textit{holomorphic} position $z^{\da\a}$.
This hints that the photon's propagator is also ``zig-zag''
in the field theory of Newman-Janis shifted fields in spin-space-time:
Zwanziger electromagnetism 
\cite{zwanziger1968quantum,zwanziger1968exactly,zwanziger1971local,brandt1978remarks,blagojevic1988quantum}.
Anyway,
% dropping the incidence delta function
% for simplicity, 
% we can denote \eqref{eq:disc.MTem-2} as
if the 
% incidence
delta function is dropped
for simplicity, 
\eqref{eq:disc.MTem-2} appears as
\begin{align}
    \omega'^+
    \mem=\mem
    -\frac{iq}{\sqrt{2}\Delta}\,\mem
    \includegraphics[valign=c]{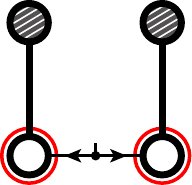}
    \,.
\end{align}
In turn,
the vector field 
$-\includegraphics[valign=c]{figs/l-b.pdf}
 \includegraphics[valign=c]{figs/v+.pdf}
 \includegraphics[valign=c]{figs/xD-b.pdf}$
is graphically computed as
\begin{align}
\begin{split}
    \frac{i\sqrt{2}q}{\Delta}\,
    \includegraphics[valign=c]{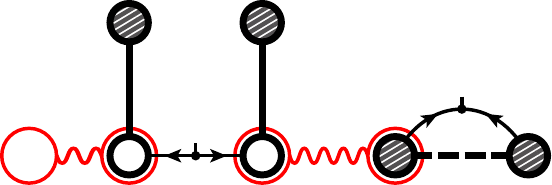}
    \,=\,\mem
    \frac{i\sqrt{2}q}{\Delta}\,
    \includegraphics[valign=c]{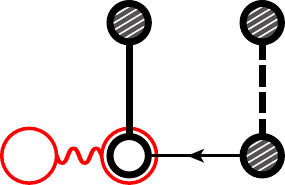}
    \,.
\end{split}
\end{align}
Following \cite{arkani2010s-matrix},
the infinity twistors are denoted as dashed lines.

While Guevara \cite{guevara2021reconstructing} has half-Fourier transformed the classical-spin amplitudes 
to the twistor space of the massless particle,
it is tempting to have 
% a ''fully twistorial'' description of amplitudes
% a for both massless and massive particles.
a description of the amplitudes
that is
fully twistorial
for both massless and massive particles.
In this context,
it would be interesting to see if
half-Fourier transformed symplectic perturbations and zig-zag vector fields
% can define
can lead to
a Hodges diagram for massive particles.

For 
% more than
over
fifty years,
twistor theory has provided 
a unique perspective
on the physics of 
massless particles and fields in four dimensions.
We hope
that
our zig-zag symplectic perturbation theory 
in the massive twistor space
could signal the dawn of
``massive twistor theory.''

{
\acknowledgments
    We are grateful to 
        Clifford Cheung,
        Julio Parra-Martinez,
        % Clifford Cheung,
        Andreas Helset,
        % Tim Adamo,
            and
        Jung-Wook Kim
    for discussions and comments on the manuscript.
    JHK would like to thank
        Clifford Cheung,
        Julio Parra-Martinez,
        Andreas Helset,
        Jordan Wilson-Gerow,
            and
        Zander Moss
    for
    enlightening conversations and
    helpful discussions.
    The work of JHK is supported
    in part by Ilju Academy and Culture Foundation.
    The work of SL is supported in part by 
    the National Research Foundation of Korea grant NRF-2019R1A2C2084608.
}

\newpage
\appendix

\section{TBMT and MPTD equations}
\label{app:BMT-MPD}

\paragraph{TBMT}
The Thomas-Bargmann-Michel-Telegdi equation
% \cite{thomas1926motion,thomas1927kinematics,bargmann1959precession,frenkel1926spinning,frenkel1926elektrodynamik,jackson2008examples,hushwater2014discovery,jackson2014jackson,rafelski2018relativistic}
\cite{thomas1927kinematics,bargmann1959precession,rafelski2018relativistic,frenkel1926spinning,frenkel1926elektrodynamik,thomas1926motion}
describes a spinning particle moving in a background electromagnetic field
in special relativity.
In this paper,
by a massive spinning particle is governed by the TBMT equation with a certain $g$ value 
we mean that its momentum and spin length pseudovector
obey the following equations
in the covariant spin gauge with the gauge fixing $-p_m\hem y^m = 0$ of the residual redundancy:
\begin{align}
\begin{split}
    \label{eq:BMT}
    0
    &{}\,=\,{} 
        \dot{p}_m 
        + p_n\mem \phi^n{}_m/m
        % + \mathcal{O}(y^1\phi^1,\phi^2)
        + \mathcal{O}(y^1)
    \,,\\
    0
    &{}\,=\,{} 
        \dot{y}^m
        - \frac{1}{m} \bigg(\mem{
            \frac{g}{2}\, \phi^m{}_n
            - \Big(1{\mem-\mem}\frac{g}{2}\mem\Big)\,
            \frac{\hhem{p^m p_l}\hhem}{m^2}\, \phi^l{}_n
        }\bigg)
        \mem y^n
        % + \mathcal{O}(y^1\phi^1,\phi^2)
        + \mathcal{O}(y^2)
    \,.
\end{split}
\end{align}
% $\phi_{mn}$ is an antisymmetric tensor that can depend on other variables of the spinning particle.
$\phi_{mn}$ is an antisymmetric tensor.
% this displays our definition of the ``$g$ factor.''
We use these terminologies
``TBMT equation'' and ``$g$ factor''
in a broad, general sense
that does not restrict 
$\phi_{mn}$
to be the field strength of electromagnetism.
For instance,
it can be something depending on other dynamical variables of the spinning particle
such as $q_a F^a{}_{mn}\hnem(x)$ of Yang-Mills
or $2\hem p_r\gamma^r{}_{[mn]}\hnem(x)$ of gravity.

\paragraph{MPTD}
The Mathisson-Papapetrou-Tulczyjew-Dixon equation \cite{mathisson1937neue,papapetrou1951spinning,tulczyjew1959motion,dixon1964covariant,dixon1965classical,Trautman:2002zz,costa2016spacetime,costa2015ssc}
describes
the dynamics of
a spinning particle in curved spacetime.
It is extended by a parameter $\kappa$,
which is
called the  ``gravimagnetic ratio''
in the literature
\cite{Khriplovich:1989ed,yee1993equations,khriplovich1997equations,khriplovich2000equations,pomeranskii2000spinning,deriglazov2016ultrarelativistic,deriglazov2017mathisson}.
Our version
reads
% is as the following:
% By a massive spinning particle is governed by the $\kappa$-MPTD equation,
% we mean that the time evolution of its momentum and spin length vector is given as
\begin{align}
\begin{split}
    \label{eq:MPD}
    0
    &{}\,=\,{} 
        m\mem e^m{}_\m\hnem(x)\mem \dot{x}^\m
        - p^m
        + \mathcal{O}(y^1)
    \,,\\
    0 
    &{}\,=\,{} 
        \dot{p}_m
        - p_n\mem \gamma^n{}_{ms}\hnem(x)\mem e^s{}_\s\hnem(x)\mem \dot{x}^\s
        % - p_n\mem \gamma^n{}_{m\s}\hnem(x)\mem \dot{x}^\s
        - \frac{g}{2}\, p_n\mem {\star R}^n{}_{mrs}\hnem(x)\mem y^r\hem e^s{}_\s\hnem(x)\mem \dot{x}^\s
        + \mathcal{O}(y^2)
    \,,\\
    0 
    &{}\,=\,{} 
        \dot{y}^m
        + \bigg(\mem{
            \frac{g}{2}\, \gamma^m{}_{ns}\hnem(x)
            - \Big(1{\mem-\mem}\frac{g}{2}\mem\Big)\,
            \frac{\hhem{p^m p_l}\hhem}{m^2}\, \gamma^l{}_{ns}\hnem(x)
        }\bigg)
        \mem y^n\mem e^s{}_\s\hnem(x)\mem \dot{x}^\s
    \\
    &{}\phantom{{}\,=\,{} 
        \dot{y}^m
    }{}
        + \bigg(\mem{
            \kappa\, {\star R}^m{}_{nrs}\hnem(x)
            - \Big(\hem\frac{g}{2}{\mem-\mem}\kappa\Big)\,
            \frac{\hhem{p^m p_l}\hhem}{m^2}\, {\star R}^l{}_{nrs}\hnem(x)
        }\bigg)
        \mem y^n\mem y^r\mem e^s{}_\s\hnem(x)\mem \dot{x}^\s
        + \mathcal{O}(y^3)
    \,,
    % \,.
\end{split}
\end{align}
% assuming the covariant gauge and
% the constraint
% $-p_m\hem y^m = 0$.
with the covariant gauge and
the constraint
$-p_m\hem y^m = 0$.
$e^m{}_\s\hnem(x)$,
$\gamma^m{}_{ns}\hnem(x)$,
$R^m{}_{nrs}\hnem(x)$
denote the 
tetrad,
spin connection coefficients,
and Riemann tensor.
As in \eqref{eq:BMT},
the momentum equation 
% serves as
% tells us
sets
the ``zero point''
of non-minimal parameters:
the precessions of $p_m$ and $y^m$ are synchronized when
% $g{\,=\,}2$ and $\kappa{\,=\,}1$.
$1 = g/2 = \kappa$.
By including $g$ in \eqref{eq:MPD}, we are being maximally pedantic,
intending to maximally mimic 
% \cite{ahh2017}'s amplitudes.
the amplitude.
% % % intending  to make contact with amplitudes 
Of course, it is \textit{fixed} to $g{\,=\,}2$ by 
the general covariance of the spin vector:
the spin vector should behave as the same as
two infinitesimally deviated points.
For more details,
see Appendix B of \cite{chkl2019}.

Either of the two bracketed terms in the $\dot{y}^m$ equation
can be viewed as a ``double copy'' of the bracketed term in \eqref{eq:BMT}.
This is because there exist two ways of drawing analogies between electromagnetism/Yang-Mills and gravity.
The first term,
which defines the ``gravitational gyromagnetic ratio'' $g$,
follows from \eqref{eq:BMT} by 
taking
$\phi_{mn} = -m\mem \gamma_{mn\r}\hnem(x)\mem \dot{x}^\r
\,(\sim p_r\hem 2\gamma^r{}_{[mn]}\hnem(x))$.
Then one is comparing $A$ to $e^m$ and $F$ to $\gamma^m{}_n$
as in Kaluza-Klein reduction \cite{kol2008classical,kol2008non,kol2011comparing,kaluza1921dawning,Klein:1926tv} or the gravitoelectromagnetism formulation of linearized gravity \cite{zee1985gravitomagnetic,maartens1998gravito,wald2010general,forward1961general,moller1972theory,braginsky1977laboratory,thorne1986black,thorne1988gravitomagnetism,harris1991analogy,damour1991general,jantzen1992many,clark2000gauge,mashhoon2001gravitomagnetism,mashhoon2001gravitoelectromagnetism,mashhoon2003gravitoelectromagnetism,ehlers2006newtonian,adler2010gravitomagnetism,adler2012gravitomagnetism},
which realizes the ``connection as field strength'' argument in \cite{spt}.
This parallel is suggested from 
the usual formulation of flat-space graviton scattering amplitudes in terms of metric perturbations
and also from
the ``exact sequences of electromagnetism and gravity'' diagram in \cite{spt}.
% This parallel is suggested from the diagram \eqref{eq:parallel-2} as it is drawn and also from the usual formulation of flat-space graviton scattering amplitudes in terms of metric perturbations.
More poetically,
we could say 
``Aharanov-Bohm double copies to Sagnac'' \cite{sakurai1980comments}:
the Coriolis force is the gravitomagnetic force.
On the other hand, the second term follows from \eqref{eq:BMT} by
 taking 
$\phi_{mn} = -m \,{\star R}_{mnrs}\hnem(x)\mem y^r\hem e^s{}_\s\hnem(x)\mem \dot{x}^\s$
while
putting $g {\mem=\mem} 2$
and replacing $g/2$ with $\kappa$.
Then
one is comparing $A$ to $\gamma^m{}_n$ and $F$ to $R^m{}_n$.
% as tetradic general relativity is a Lorentz gauge theory.
This way of thinking is more natural in a background independent/non-perturbative set-up
and is encouraged by
Ashtekar variables \cite{ashtekar1986new,ashtekar1987new,ashtekar1991lectures,henneaux1989derivation,capovilla1991self,celada2016bf,ashtekar2021gravitational},
non-linear graviton \cite{penrose1976curvedtwistor,penrose1976nonlinear}, Weyl double copy \cite{luna2019type,godazgar2021weyl,white2021twistorial,chacon2021weyl},
covariant color-kinematics duality \cite{cheung2021cck},
% and so on.
etc.
% % % In this covariant sense, $\kappa$ is the gravitomagnetic coupling, controlling the interaction with the magnetic component of the Riemann tensor in the $\dot{p}_m$ equation.

Both of these two parallels 
emerge in our discussion on 
the
symplectic perturbation theory
of spinning particles:
the former from the non-covariant scheme
employed in this paper
and the latter from the covariant scheme
employed in the follow-up paper \cite{ambikerr2}.

\paragraph{\texorpdfstring{Generic $\mathcal{O}(\phi^1)$ spin precession equation of motion}{Generic O(ϕ¹) spin precession equation of motion}}

In this paper, we have proposed a generic spin precession equation of motion, all orders in spin and first order in the formal ``field strength'' $\phi$.
When $\phi$ is self-dual, it reads
\begin{align}
    \label{eq:prec-eq.vec}
    \begin{split}
    0
    &{}\,=\,{} 
        \dot{p}_m
        + \frac{1}{m}\,
        p_n
        \Bigg[\mem{
            \sum_{\ell=0}^{\infty}\mem
            \frac{
                (C_\ell {\mem+\mem} \tilde{C}_\ell \mem\star^{-1})
            }{\ell!}\mem
            \big({
                (y{\mem\cdot\mem}\partial)^\ell\mem {\star^\ell\hnem \phi}(x)
            }\big)\hhnem{}^n{}_m
        }\Bigg]
    \mem+\mem \mathcal{O}(\phi^2)
    \,,\\
    0
    &{}\,=\,{} 
        \dot{y}^m
        - \frac{1}{m}\nem\hnem
        \left[{
        \nem
        \begin{array}{l}
            \displaystyle
            \sum_{\ell=0}^{\infty}
            \frac{
                (C_{\ell+1} {\mem+\mem} \tilde{C}_{\ell+1} \mem\star^{-1})
            }{\ell!}\mem
            \big({
                (y{\mem\cdot\mem}\partial)^\ell\mem {\star^\ell\hnem \phi}(x)
            }\big)\hhnem{}^m{}_n
            \\
            \displaystyle
            {}-
            \frac{p^mp_r}{m^2}\mem
            \sum_{\ell=0}^{\infty}
            \frac{
                (C_\ell {\mem+\mem} \tilde{C}_\ell \mem\star^{-1})
                -
                (C_{\ell+1} {\mem+\mem} \tilde{C}_{\ell+1} \mem\star^{-1})
            }{\ell!}\mem
            \big({
                (y{\mem\cdot\mem}\partial)^\ell\mem {\star^\ell\hnem \phi}(x)
            }\big)\hhnem{}^r{}_n
        \end{array}
        \hhnem\hnem\nem
        }\right]\hnem
        y^n
    \\
    &\quad \mem+\mem \mathcal{O}(\phi^2)
    % \,.
    \,,
    \end{split}
\end{align}
where
we have also included magnetic Wilson coefficients
$\tilde{C}_\ell$
for full generality.
% For full generality,
% we have included magnetic Wilson coefficients
% $\tilde{C}_\ell$ as well.
% where $C^\pm_\ell = C_\ell {\mem\mp\mem} i \tilde{C}_\ell$.
One obtains
\eqref{eq:BMT} and \eqref{eq:MPD}
by truncating \eqref{eq:prec-eq.vec}
with having
$C_1 {\,=\,} g/2$, $C_2 {\,=\,} \kappa$,
and $\tilde{C}_\ell {\,=\,} 0$.

As pointed out recently in \cite{gmoov},
$\dot{p}_m$ and $\dot{y}^m$
can be described together in a more unified manner
in terms of the spin frames
as $\dot{\lambda}{}_\a{}^I$, $\dot{\rambda}{}_{I\da}$.\footnote{
    This idea can be traced back to
    the works \cite{Bette:1989zt,Buitrago:2007zz,Bette:2004cs,Buitrago:1999yb}.
}
This is best illustrated
if
the body-frame components
% $W_a$
of the spin 
% angular momentum
pseudovector
are constant,
which happens in our setting
when
the particle has a constant Regge trajectory.
The spin frame version of \eqref{eq:prec-eq.vec} reads
\begin{align}
\begin{split}
    \label{eq:prec-eq.frames}
    m_0\mem \dot{\rambda}{}_{I\da}
    &{}\,=\,{} 
    \rambda_{\smash{I\db}\vphantom{I\da}}
    \mem
    \Bigg[\mem{
        \sum_{\ell=0}^{\infty}\mem
        \frac{1}{\hhem\ell!\hhem}
        \frac{
            C^+_\ell\nem
            {\mem+\mem}
            C^+_{\ell+1}
        }{2}\mem
        (iy{\mem\cdot\mem}\partial)^\ell
        \phi^\db{}_\da\hnem(x)
    }\Bigg]
    \mem+\mem \mathcal{O}(\phi^2)
    \,,\\
    m_0\mem \dot{\lambda}{}_\a{}^I
    &{}\,=\,{} 
    \lambda_\a{}^J\mem
    \mem
    \rambda_{\smash{J\db}\vphantom{I\da}}
    \Bigg[\mem{
        \sum_{\ell=0}^{\infty}\mem
        \frac{1}{\hhem\ell!\hhem}
        \frac{
            C^+_\ell\nem
            {\mem-\mem}
            C^+_{\ell+1}
        }{2}\mem
        (iy{\mem\cdot\mem}\partial)^\ell
        \phi^\db{}_\da\hnem(x)
    }\Bigg]
    (\rambda^{-1})^{\da I}
    \mem+\mem \mathcal{O}(\phi^2)
    % \,.
    \,,
\end{split}
\end{align}
where $C^\pm_\ell = C_\ell {\,\mp\,} i\hem \tilde{C}_\ell$.
% propose to take this as a field basis invariant definition of
% the Wilson coefficients of the classical particle
One can clearly see that
$\dot{\lambda}{}_\a{}^I = 0 + \mathcal{O}(\phi^2)$
in self-dual backgrounds
implies $C^+_\ell {\,=\,} C^+_0 {\,=\,} 1$
and vice versa:
\textit{all} three-point spin multipole couplings
(not only dipole and quadrupole, to emphasize)
to the self-dual field
are Kerr-Newman.\footnote{
    % By ``Kerr-Newman coupling,''
    % we mean $C^\pm_\ell = 1$
    % for both helicities $1$ and $2$.
    % This is not a misnomer, as
    % the electromagnetic coupling of the Kerr-Newman black hole coincides with
    % the {\Kerr} coupling \cite{chung2020kerr}. 
    We clarify that
    by ``Kerr-Newman coupling''
    we mean $C^\pm_\ell {\,=\,} 1$
    for both helicities $1$ and $2$.
    The electromagnetic coupling of the Kerr-Newman black hole coincides with
    the {\Kerr} coupling \cite{chung2020kerr}. 
}

These equations
were
already 
implied throughout
\eqref{eq:eom-nonmin.hybrid},\,\eqref{eq:eom-nonmin.vector},\,\eqref{eq:Wilson.Cfrelation},\,\eqref{eq:nonmin-SSigmas},\,\eqref{eq:012omegass},
% in essence,
but
we have 
restated 
them
in terms of
the real-worldline Wilson coefficients
for reference.
% Also,
% taking a ``square root'' to the spin precession equation of motion
% traces back to Bette \cite{Bette:1989zt}
% and further to
% the insight that the dynamics of a massive particle
% can be understood as
% a succession of spacetime-dependent infinitesimal Lorentz transformations
% \cite{Buitrago:1999yb,Buitrago:2007zz,Bette:2004cs}.

Lastly, we spell out
the spin frame versions of 
\eqref{eq:BMT} and \eqref{eq:MPD}:
\begin{align}
\begin{split}
    \label{eq:frames-BMT}
    \dot{\rambda}{}_{I\da}
    &{}\mem=\mem \textstyle
        \frac{1+g/2}{2}\,\hem
        \rambda_{\smash{I\db}\vphantom{I\da}}\mem\mem \phi^\db{}_\da\hnem(x) 
        /m_0
    + \mathcal{O}(y^1)
    \,,\\
    \dot{\lambda}{}_\a{}^I
    &{}\mem=\mem \textstyle
        \frac{1-g/2}{2}\,\hem
        \lambda_\a{}^J\,
        \rambda_{\smash{J\db}\vphantom{I\da}}\mem
        \phi^\db{}_\da\hnem(x)\hhem
        (\rambda^{-1})^{\da I}
        \nem\hnem
        /m_0
    + \mathcal{O}(y^1)
    \,,
    % \vphantom{\Big]}
\end{split}\\[0.2ex]
\begin{split}
    \label{eq:frames-MPD}
    \dot{\rambda}{}_{I\da}
    &{}\mem=\mem \textstyle
        -
        \rambda_{\smash{I\db}\vphantom{I\da}}\mem
        % \mem\minie\hnem
        \Big[\hem{
            \frac{1+g/2}{2}\mem\hhem
            \gamma^\db{}_{\da s}\hnem(x)
            +
            \frac{g/2+\kappa}{2}\hem
            R^\db{}_{\da rs}\hnem(x)\mem 
            y^r
            + \mathcal{O}(y^2)
        }\hhem\Big]
        \, e^s{}_\s\hnem(x)\mem \dot{x}^\s
    \,,\\[-0.2ex]
    \dot{\lambda}{}_\a{}^I
    &{}\mem=\mem \textstyle
        -
        \lambda_\a{}^J\,
        \rambda_{\smash{J\db}\vphantom{I\da}}\mem
        % \mem\minie\hnem
        \Big[\hem{
            \frac{1-g/2}{2}\mem\hhem
            \gamma^\db{}_{\da s}\hnem(x)
            +
            \frac{g/2-\kappa}{2}\hem
            R^\db{}_{\da rs}\hnem(x)\mem 
            y^r
            + \mathcal{O}(y^2)
        }\hhem\Big]
        (\rambda^{-1})^{\da I}
        \, e^s{}_\s\hnem(x)\mem \dot{x}^\s
    \,,
    % \nonumber
\end{split}
\end{align}
% Taking a ``square root'' to the TBMT equation
% % Also, taking a ``square root'' to the spin precession equation of motion 
% traces back to the works \cite{Bette:1989zt,Buitrago:2007zz,Bette:2004cs,Buitrago:1999yb}.

% \section{Classical interaction picture}
% \label{app:shilb}

\section{Matching Calculation}
\label{app:aoude}

\paragraph{Classical spin from coherent states}
% Recently,
Aoude and Ochirov
\cite{aoude2021classical} 
have
explicated the procedure of obtaining massive spinning amplitudes in the classical spin limit
in terms of spin coherent states.
We briefly review their methodology in our notation.
See also \cite{Cangemi:2022bew} for a succinct review.

Modeling the spinning degrees of freedom
by the coadjoint orbit \cite{kirillov1999merits,kirillov2012elements,lyakhovich1998massive,rempel2016interaction,gracia2018kirillov},
the classical value of spin angular momentum is identified with its coherent-state expectation value in the first-quantized theory as
\begin{align}
    \label{eq:psiW}
    W_a = (\hbar/2)\mem \bpsi^I (\s_a)_I{}^J \psi_J
    \,,\quad
    % W_0 = -\minie\mem \bpsi^I \psi_I = -|\vec{W}|
    |\vec{W}| = (\hbar/2)\mem \bpsi^I \psi_I
    \,.
\end{align}
The classical-spin amplitude then follows from the 
coherent-state
expectation value of all definite-spin amplitudes in the $\hbar\to0$ limit with $|\vec{W}|$ fixed:
% \begin{subequations}
\begin{align}
    \label{eq:Aoudeformula}
\begin{split}
    % \label{eq:Aoudeformula-1}
    \mathcal{M}\hnem(\one,a;\three)
    {\phantom{{}:={}}\mathllap{{}={}}}{}& \lim_{\substack{\hbar\to0 \\ \bpsi\psi = 2|\vec{W}|/\hbar}}
    \sum_{2s=0}^\infty
    \mathcal{M}_{(2s)}\hnem(\one,\bpsi,\psi;\three)
    \,,\\
    % \label{eq:Aoudeformula-2}
    \mathcal{M}_{(2s)}\hnem(\one,\bpsi,\psi;\three)
    :={}& 
    \frac{\mathe^{-\bpsi\psi}}{(2s)!}\,\mem
    \bpsi^{J_1}\cdots\bpsi^{J_{2s}}
    \mem
    (\mathcal{A}_{(2s)}\hnem(\two{\leftarrow}\one;\three))_{J_1\cdots J_{2s}}{}^{I_1\cdots I_{2s}}
    \,
    \psi_{I_1}\cdots\psi_{I_{2s}}
    \,,
\end{split}
\end{align}
% \end{subequations}
% the generating function of all definite-spin amplitudes.
where $\mathe^{-\bpsi\psi}$ is the measure normalizing coherent states.
For example, the spin-tensor $\identity^{\nem\odot 2s}$ of the plus-helicity minimal amplitude 
$\smash{
    \mathcal{A}^{+h}_{(2s)}\hnem(\two{\leftarrow}\one;\three)
    = e_h\hem m_0 x^h\mem \identity^{\nem\odot 2s}
}$
boils down to 
\begin{align}
\begin{split}
    \label{eq:expfactor}
    % \text{\eqref{eq:3kin-real}}
    % \qiq
    % 
    % \lim_{\bpsi\psi \to \infty}
    \mathe^{-\bpsi\psi} 
    \exp\nem\Big(
        \bpsi^J\nem\hnem\langle\two^{-1}_J\one^I\rangle\psi_I
    \nem\Big)
    % = \lim_{\bpsi\psi \to \infty} e^{\three\cdot a}
    \mem=\mem \mathe^{\three\cdot a}
    \,,
\end{split}
\end{align}
upon plugging in the real on-shell kinematics given in \eqref{eq:3kin-real}
\cite{arkani2020kerr,aoude2021classical}.
Note that
\begin{align}
\begin{split}
    \label{eq:3a-identity}
    \langle\two^{-1}_J\one^I\rangle
    &= \delta_J{}^I
    - \minie\hem \langle\one^{-1}_I\three\rangle \lsq\bthree\bone^{-1}{}^J\rsq
    \,,\\
    &= \delta_J{}^I
    + \three_m
    \Big(\hnem{
        \textstyle
        \frac{\hbar}{2m_0}\hem (\L(\one))^m{}_a\hem (\s^a)_I{}^J
        + \frac{1}{2}\hem (\one^{-1})^m\hem \delta_I{}^J
    }\kern-0.02em\Big)
    \,,\\
    &= \delta_J{}^I
    + 
    {\textstyle\frac{\hbar}{2m_0}}\mem
    \three_m (\L(\one))^m{}_a\hem (\s^a)_I{}^J
    % \,,\\
    % \iq
    % \bpsi \langle\two^{-1}\one\rangle \psi - \bpsi\psi
    % &= \three \cdot a(\one) / \hbar
    % \,.
\end{split}
\end{align}
on the support of $\deltabar(2\mem\one{\mem\cdot\mem}\three)$.
In turn, it follows that
\begin{align}
    \bpsi^J \langle\two^{-1}_J\one^I\rangle \psi_I - \bpsi^I\psi_I
    \mem=\mem
    \three{\mem\cdot\mem} a(\one)
    \,.
\end{align}
% % \begin{align}
% % \begin{split}
% %     % \lim_{\bpsi\psi \to \infty}
% %     \mathe^{-\bpsi\psi} 
% %     \exp\nem\Big(
% %         \bpsi^J\nem\hnem\langle\two^{-1}_J\one^I\rangle\psi_I
% %     \nem\Big)
% %     % = \lim_{\bpsi\psi \to \infty} e^{\three\cdot a}
% %     = \mathe^{\three\cdot a}
% %     \,,
% % \end{split}
% % \end{align}
% upon plugging in the real on-shell kinematics given in
% % \eqref{eq:nullrot-arbitraryg} 
% \eqref{eq:3kin-real}
% \cite{arkani2020kerr,aoude2021classical}.
% We record that
% % Note that
% the definition of the body frame in \eqref{eq:solve-constraints} implies 
% % the following, 
% % while $\three_m (\frac{1}{2}\hem (\one^{-1})^m\hem \delta_I{}^J) = 0$:
% % $\three_m (\one^{-1})^m = 0$:
% \begin{align}
%     \label{eq:3a-identity}
%     \textstyle
%     \langle\two^{-1}_J\one^I\rangle
%     = \delta_J{}^I
%     {-}\mem \frac{1}{2}\hem \langle\one^{-1}_I\three\rangle \lsq\bthree\bone^{-1}{}^J\rsq
%     = \delta_J{}^I
%     + \three_m
%     \nem\left(\hnem
%     % \big(
%         \frac{1}{2m_0}\hem (\L(\one))^m{}_a\hem (\s^a)_I{}^J
%         + \frac{1}{2}\hem (\one^{-1})^m\hem \delta_I{}^J
%     % \big)
%     \kern-0.02em\right)
%     \,,
% \end{align}
% while the last term vanishes: $\three_m (\frac{1}{2}\hem (\one^{-1})^m\hem \delta_I{}^J) = 0$.

In our spherical top setting,
such classical-spin states
are realized in the $(\lambda,\rambda)$ polarization as
polynomials of $\lambda_\a{}^I$ and $(\rambda^{-1})^{\da I}$
saturated by a common spinor $\psi_I$;
i.e., they are the ``right'' coherent states with respect to the $\mathrm{SU}(2)$ group action on $\mathbb{MA}$.
The unoccupied ``left'' $\mathrm{SL}(2,\mathbb{C})$ indices contract to amplitudes as spacetime spinor-tensors.

\paragraph{Matching}
Applying \eqref{eq:Aoudeformula} with the real kinematics \eqref{eq:3kin-real}, 
% the corresponding plus-helicity classical-spin amplitude reads
the coherent-state expectation value of the plus-helicity spinning amplitude \eqref{eq:3ptAahh+} 
is given as
\begin{align}
\begin{split}
    % \label{eq:M+2scontrib}
    % \mathcal{M}^+_{(2s)}\hnem[g^+](\psi;\bpsi;\three)
    % &= e_h\hem m_0 x^h
    % \mem\mathe^{-\bpsi\psi}
    % \mem\sum_{\ell=0}^{2s}\mem
    % \frac{g^+_\ell}{(2s)!}\mem
    % \big(\bpsi\langle \two^{-1} \one \rangle \psi\big)^{2s-\ell}
    % \left(\hnem
    %     \frac{x}{m_0}
    %     \bpsi
    %     \langle \two^{-1} \three\rangle
    %     \langle \three \one \rangle
    %     \psi
    % \nem\right)^{\kern-0.23em\ell}
    % \,,\\
    % &= e_h\hem m_0 x^h
    % \mem\mathe^{-\bpsi\psi}
    % \mem\sum_{\ell=0}^{2s}\mem
    % \frac{g^+_\ell}{(2s)!}\mem
    % \big(\bpsi\psi + \three{\kern0.02em\cdot\kern0.02em}a \big)^{2s-\ell}
    % (-2\mem\three{\kern0.02em\cdot\kern0.02em}a)^\ell
    % \,,
    \label{eq:M+2scontrib}
    % & \mathcal{M}^{+h}\hnem[g^+](x,a;\three)
    % \\
    % &=
    &\phantom{=}\,\,\mem\mem\hem
    % \lim_{\bpsi\psi\to\infty}
    e_h\hem m_0 x^h
    \mem\mathe^{-\bpsi\psi}
    \mem\sum_{2s=0}^{\infty} \sum_{\ell=0}^{2s}
    \frac{\hem g^+_\ell \hnem \hbar^\ell \hem}{(2s)!}\mem
    \big(\bpsi\psi + \three{\mem\cdot\mem}a\big)^{2s-\ell}
    (-2\mem\three{\mem\cdot\mem}a)^\ell
    \,,\\
    &= 
    % \lim_{\bpsi\psi\to\infty}
    e_h\hem m_0 x^h
    \mem\mathe^{-\bpsi\psi}
    \mem\sum_{\ell=0}^{\infty} 
    \mem g^+_\ell \hnem \hbar^\ell
    \nem\hnem \left(
        \frac{-2\mem\three{\mem\cdot\mem}a}{\vphantom{\adjustbox{raise=0.2em}{$\bpsi$}}\bpsi\psi + \three{\mem\cdot\mem}a}
    \right)^{\kern-0.17em\ell}
    \sum_{2s=\ell}^{\infty}
    \frac{1}{(2s)!}\hem
    \big(\bpsi\psi + \three{\mem\cdot\mem}a\big)^{2s}
    \,,\\
    &= 
    % \lim_{\bpsi\psi\to\infty}
    e_h\hem m_0 x^h
    \mem\mathe^{\three\cdot a}
    \mem\sum_{\ell=0}^{\infty} 
    \mem g^+_\ell \hnem \hbar^\ell
    \nem\hnem \left(
        \frac{-2\mem\three{\mem\cdot\mem}a}{\vphantom{\adjustbox{raise=0.2em}{$\bpsi$}}\bpsi\psi + \three{\mem\cdot\mem}a}
    \right)^{\kern-0.17em\ell}
    \left(\hnem
        1 - \frac{1}{(\ell-1)!}\mem \Gamma\big(\ell\hem\big|\hem\bpsi\psi + \three{\mem\cdot\mem}a\big)
    \nem\kern-0.015em\right)
    % \,,
    \,.
\end{split}
\end{align}
% using the holomorphic representation of the three-point amplitude in \eqref{eq:3ptAahh}.
As argued in \cite{aoude2021classical} around equations (3.22)-(3.27),
for large $\bpsi\psi$ this has the limit
\begin{align}
\begin{split}
    \label{eq:M+coherent}
    & \mathcal{M}^{+h}\hnem[g^+](x,a;\three)
    = 
    e_h\hem m_0 x^h
    \mem\mathe^{\three\cdot a}
    \hnem
    \left(
        \lim_{\hbar\to0}
        \mem\sum_{\ell=0}^{\infty} 
        \mem g^+_\ell \hnem \hbar^\ell\mem
        \nem \bigg({
            -\frac{\three{\mem\cdot\mem}a}{\hem|\vec{W}|/\hbar\hem}
        }\bigg)^{\kern-0.17em\ell}
    \,\right)
    \,.
\end{split}
\end{align}
Therefore, provided that $g^+_\ell$ scales as $\mathcal{O}(\hbar^{-\ell})$,
we conclude that
\begin{align}
    \label{eq:gen-fgrelation}
    % f^+_\ell = 
    % \frac{1}{(-|\vec{W}|)^\ell}
    % \lim_{\hbar\to0} g^+_\ell\hbar^\ell
    f^+_\ell 
    = (-|\vec{W}|)^{-\ell} \cdot \lim_{\hbar\to0} g^+_\ell \hnem \hbar^\ell
    \qiq
    f^+\hnem(\xi)
    = g^+\hnem(-\xi/|\vec{W}|)
\end{align}
by comparing \eqref{eq:M+coherent} with 
$\mathcal{M}^{+h}\hnem[f^+](\one,a;\three)
= 
\mathcal{M}^{+h}\hnem(\one,a;\three)
\,{\cdot}\, f^+\hnem(\three{\mem\cdot\mem}a)$
of \eqref{eq:born}.

% The minus sign in \eqref{eq:gen-fgrelation} is reasonable because 
% % $g=0$ corresponds to $g^+_1>0$.
% $g^+_1>0$ maps to $g=0$.
% For example, the $g=0$ Proca amplitude has $g^+_1 = 1$:
% \begin{align}
%     \langle \two^{-1}_{(J_1} \one^{(I_1} \rangle
%     \lsq \btwo_{J_2)} \bone^{-1}{}^{I_2)} \rsq
%     = 
%     \langle \two^{-1}_{(J_1} \one^{(I_1} \rangle
%     \langle \two^{-1}_{J_2)} \one^{I_2)} \rangle
%     + (x/m_0)\mem
%     \langle \two^{-1}_{(J_1} \one^{(I_1} \rangle
%     \langle \two^{-1}_{J_2)} \three \rangle \langle \three \one^{I_2)} \rangle
%     \,.
% \end{align}

\paragraph{A trick}

Lastly, we demonstrate a slightly quicker derivation.
As shown in \eqref{eq:expfactor},
the real kinematics \eqref{eq:3kin-real} has induced the minimal exponential spin factor $\mathe^{\three\cdot a}$.
However, we find that we can either apply the complexified kinematics \eqref{eq:3kin-plus} and put $\mathe^{\three\cdot a}$ by hand.
For minimal coupling, it works as
\begin{align}
\begin{split}
    % \text{\eqref{eq:3kin-plus}}
    \one_\a{}^I \hnem=\mem \two_\a{}^I
    \qiq
    \M(\one,a;\three)
    \mem&=\mem
    e_h\hem m_0 x^h\mem
    e^{\three\cdot a}
    \left(
    \mathe^{-\bpsi\psi} 
    \exp\nem\Big(
        \bpsi^J\nem\hnem\langle\two^{-1}_J\one^I\rangle\psi_I
    \nem\Big)
    \right)
    \,,\\
    \mem&=\mem
    e_h\hem m_0 x^h\mem
    e^{\three\cdot a}
    \,.
\end{split}
\end{align}
This trick may be thought of as 
an analog of
taking the holomorphic complex worldline as the ``zero point'':
the 
Newman-Janis shifted
amplitude
is taken as default.
One can check that it also works in the non-minimal case:
\begin{align}
\begin{split}
    &\phantom{=}\,\,\mem\mem\hem
    \mathe^{\three\cdot a}\left(
        e_h\hem m_0 x^h
        \mem\mathe^{-\bpsi\psi}
        \mem\sum_{2s=0}^{\infty} \sum_{\ell=0}^{2s}
        \frac{\hem g^+_\ell \hnem \hbar^\ell\hem}{(2s)!}\mem
        (\bpsi\psi)^{2s-\ell}
        (-2\mem\three{\mem\cdot\mem}a)^\ell
    \right)
    \,,\\
    &= 
    \mathe^{\three\cdot a}\left(
        e_h\hem m_0 x^h
        \mem\sum_{\ell=0}^{\infty} 
        \mem g^+_\ell \hnem \hbar^\ell
        \nem\hnem \left(
            \frac{-2\mem\three{\mem\cdot\mem}a}{\vphantom{\adjustbox{raise=0.2em}{$\bpsi$}}\bpsi\psi}
        \right)^{\kern-0.17em\ell}
        \left(\hnem
            1 - \frac{1}{(\ell-1)!}\mem \Gamma\big(\ell\hem\big|\hem\bpsi\psi + \three{\mem\cdot\mem}a\big)
        \nem\kern-0.015em\right)
    \nem\nem\right)
    \,.
\end{split}
\end{align}
We have plugged in \eqref{eq:3kin-plus} to \eqref{eq:3ptAahh+}
and then placed $\mathe^{\three\cdot a}$ in front of it by hand
while omitting $e_h\hem m_0 x^h\mem$ to avoid clutter.
In the large $\bpsi\psi$ limit, this agrees with the last line of \eqref{eq:M+2scontrib}
so that \eqref{eq:M+coherent}-\eqref{eq:gen-fgrelation} are reproduced.

It is not difficult to 
understand
% check
why this trick works
at the level of computation.
For a deeper reason,
it 
might be plausible
% is tempting
to relate it to the fact that
the quantum amplitude of the minimally coupled particle, derived in \eqref{eq:can-amp.3},
is supported on $\one_\a{}^I = \two_\a{}^I$.

\section{Hamiltonian Perturbation Theory}
% \section{Hamiltonian perturbation theory and KMO'C}
\label{app:canonical-pert}

Finally, we would like to comment that 
it is also possible to develop the textbook perturbation theory 
(which perturbs the Hamiltonian, not the symplectic structure)
for the massive twistor particle.
The purpose of this digression
is to complement our symplectic perturbation 
% (Feynman bracket)
approach
and point out possibilities of future applications,
e.g., applying the KMO'C formalism \cite{kosower2019amplitudes}.
We only 
demonstrate
the simplest example: 
scattering of a minimally coupled 
fixed-mass
particle in the plus-helicity electromagnetic plane wave.

\paragraph{Canonical quantization}

We first-quantize the massive twistor particle
in the reparameterization gauge \eqref{eq:constant-einbein}
where the Hamiltonian is $H = \minime (\bDelta {\mem+\mem} \Delta) + m_0$
and treat it as a standard quantum-mechanical system
with ``absolute time.''
% We assume constant mass.
In order to perform canonical quantization, 
one should figure out what the canonical coordinates are.
A holomorphic electromagnetic field $A_{\a\da}\hnem(z)\mem dz^{\da\a}$
translates to $-i\hem
(\lambda^{-1})_I{}^\a
A_{\smash{\a\dc}\vphantom{\da}}\hnem(z)\mem \rho^{\dc\rmA}\hem dZ_\rmA{}^I$
in the notation \eqref{eq:hingedef}.
Taking $i\mem \bZ_I{}^\rmA dZ_\rmA{}^I$ for the free-theory symplectic potential,
we find that the ``kinetic'' and ``canonical'' twistor variables are related as
\begin{align}
    \label{eq:kinZcanZ}
    (Z^{\text{kin}})_\rmA{}^I
    \mem=\mem (Z^{\text{can}})_\rmA{}^I 
    \,,\quad
    (\bZ^{\text{kin}})_I{}^\rmA
    \mem=\mem (\bZ^{\text{can}})_I{}^\rmA
    \mem+\mem
    q(\lambda^{\text{kin}}{}^{-1})_I{}^\a
    \mem A_{\smash{\a\dc}\vphantom{\da}}\hnem(z^{\text{kin}})\,
    (\rho^{\text{kin}})^{\dc\rmA}
    \,.
\end{align}

From now, we exclusively work with the canonical variables
(unlike in the main text)
and drop the ``can'' subscript to avoid clutter.
Then \eqref{eq:kinZcanZ} boils down to
\begin{align}
\begin{split}
    \lambda_\a{}^I 
    &= (\lambda^\text{kin})_\a{}^I
    \,,\quad
    \mu^{\da I}
    = (\mu^\text{kin})^{\da I}
    \,,\\
    \rambda_{I\da}
    &= (\rambda^\text{kin})_{I\da}
    - q(\lambda^{-1})_I{}^\b A_{\smash{\b\da}\vphantom{\da}}\hnem(z)
    \,,\\
    \bmu_I{}^\a
    &= (\bmu^\text{kin})_I{}^\a
    - q(\lambda^{-1})_I{}^\b A_{\smash{\b\da}\vphantom{\da}}\hnem(z)\mem
    z^{\da\a}
    % \,,
    \,.
\end{split}
\end{align}
While $z^{\da\a} {\mem=\mem} (\mu\lambda^{-1})^{\da\a} {\mem=\mem} (z^{\text{kin}})^{\da\a}$,
$\bz^{\da\a} {\mem=\mem} (\rambda^{-1}\hhnem\bmu)^{\da\a}$ differs from $(\bz^{\text{kin}})^{\da\a}$
due to coupling to the holomorphic gauge field.

The Hamiltonian
$H = \minime (\det(\rambda^\text{kin}) {\mem+\mem} \det(\lambda^\text{kin}))$
gets perturbed in this canonical setting as
% \begin{align}
$
    H = H^\circ + H'
%     \,,
% \end{align}
$,
where $H^\circ = \minime (\det(\rambda) + \det(\lambda)) + m_0$ and
$H' = H'_1 + H'_2$:
\begin{align}
    % H' &= H'_1 + H'_2
    % \,,\quad
    H'_1 :=
        \frac{\hem\bDelta\hem}{2}\mem
        qA_{\smash{\b\db}\vphantom{\da}}\hnem(z)\mem (p{}^{-1})^{\db\b}
    \,,\quad
    H'_2 :=
        -\frac{1}{4\Delta}\mem q^2\mem \e^{\a\b} A_{\a\da}\hnem(z)\mem A_{\smash{\b\db}\vphantom{\da}}\hnem(z)\mem \be^{\da\db}
    \,.
\end{align}
We restrict our attention to the leading order 
and discard $H'_2$.
For the self-dual plane wave \eqref{eq:lcpw}, the $\mathcal{O}(q^1)$ interaction Hamiltonian becomes
\begin{align}
    \label{eq:H'1}
    H'_1
    = \frac{q}{\sqrt{2} \Delta}\mem
    % \i^\a p_{\a\da} \bo^\da
    \langle\i|p|\bo\rsq
    % \,\mathe^{-i\bozo}
    \,\mathe^{-i\mem \bo \mu\lambda^{-1}\hnem\o}
    \,.
\end{align}
%% [M^1] hidden!! \ref{footnote:mode-operator-dimension}

There are many kinds of eigenstates
in the free theory
that we can consider:
coherent states $\Bra{\bZ}$ and $\Ket{Z}$,
frame eigenstates $\Ket{\lambda\rambda}$,
and simultaneous eigenstates of $\hat{p}_{\a\da}$ and \smash{$\hat{\vec{W}}{}^2$}, 
which might be called ``Wigner states.''
For our scattering problem, we identify $\Ket{\lambda\rambda}$ as the relevant in-state.
Using the convention
$\{Z_\rmA{}^I,\bZ_J{}^\rmB\} = (1/\ihbar)\mem \delta_\rmA{}^\rmB \delta_J{}^I$
$\implies$
$\smash{[\hat{Z}_\rmA{}^I,\hat{\bZ}_J{}^\rmB]} = \delta_\rmA{}^\rmB \delta_J{}^I$
(see footnote \ref{footnote:hbar-restore}),
the canonically quantized operators are realized as
\begin{align}
\begin{split}
    \Bra{\lambda\rambda}\hem \hat{\lambda}_\a{}^I
    = \lambda_\a{}^I \Bra{\lambda\rambda}
    &\,,\quad
    \Bra{\lambda\rambda}\hem \hat{\rambda}_{I\da}
    = \rambda_{I\da} \Bra{\lambda\rambda}
    \,,\\
    \Bra{\lambda\rambda}\hem \hat{\mu}^{\da I}
    = -i\mem
        \frac{\partial}{\partial\rambda_{I\da}}
    \Bra{\lambda\rambda}
    &\,,\quad
    \Bra{\lambda\rambda}\hem \hat{\bmu}_I{}^\a
    = -i\mem
        \frac{\partial}{\partial\lambda_\a{}^I}
    \Bra{\lambda\rambda}
    \,.
\end{split}
\end{align}

\paragraph{AHH amplitude from first quantization}
From the interaction Hamiltonian \eqref{eq:H'1},
the matrix element 
% of $\hat{T}$ 
is given by
% reads
\begin{align}
    \label{eq:can-amp.1}
    \Bra{\lambda'\rambda{}'} \hat{T} \Ket{\lambda\rambda}
    &= 
    -\nem\int d\t\,\,
    \Bra{\lambda'\rambda{}'}
    \mem
        \mathe^{-(\t/\ihbar)\hat{H}^\circ}
        {\hat{H}'_1 \,}
        \mathe^{(\t/\ihbar)\hat{H}^\circ}
    \Ket{\lambda\rambda}
    + \mathcal{O}(q^2)
    \,,\\
    &= 
    % 2\pi\hbar\mem\hem 
    \deltabar(E'{\hhem-\mem}E)
    \bigg[\hhem\mem{
        -\frac{q}{\sqrt{2}\Delta}\mem 
        \langle\i|p|\bo\rsq\mem
        \exp\nem\nem\bigg({
            (\lambda^{-1})_I{}^\a\mem
            \three_{\a\da}\mem
            \frac{\partial}{\partial\rambda{}'_{I\da}}
        }\mem\bigg)\hem
        \BraKet{\lambda'\rambda{}'}{\lambda\rambda}
    }\mem\bigg]
    + \mathcal{O}(q^2)
    \,,
    \nonumber
\end{align}
where $E := \minime(\det(\rambda)+\det(\lambda))$,
$E' := \minime(\det(\rambda{}')+\det(\lambda'))$,
and $\deltabar(\xi) := 2\pi\hbar\mem\hem \delta(\xi)$.
The exponential operator implements a translation of the right-handed spin frame:
\begin{align}
\begin{split}
    \label{eq:can-amp.2}
    \exp\nem\nem\bigg({
        (\lambda^{-1})_I{}^\a\mem
        \three_{\a\da}\mem
        \frac{\partial}{\partial\rambda{}'_{I\da}}
    }\mem\bigg)\mem
    \BraKet{\lambda'\rambda{}'}{\lambda\rambda}
    &\,=\,
    \BraKet{\lambda', \rambda{}' {+} \lambda^{-1}\three }{\lambda\rambda}
    \,,\\
    &\,=\,
    \delta^{(4)}\hnem(\lambda' {\hem-\mem} \lambda)\,
    \delta^{(4)}\hnem(\mem{
        \rambda{}' + \lambda^{-1}\three - \rambda
    }\mem)
    \,.
\end{split}
\end{align}
When combined with $\deltabar(E'{\hhem-\mem}E)$,
this product of delta functions 
precisely describes the heavenly kinematics \eqref{eq:3kin-plus}.\footnote{
    When dealing with delta functions of the complexified spinor-helicity variables and their determinants, we are implicitly working in the $(2,2)$ signature.
}
Especially, one finds that $E' {\hhem-\mem} E \hem=\hem \three{\mem\cdot\mem}p\mem/\Delta$.
% \begin{align}
%     \deltabar(E'{\hhem-\mem}E)\,
%     \delta^{(4)}\hnem(\lambda' {\mem-\mem} \lambda)\,
%     \delta^{(4)}\hnem(\rambda{}' + \lambda^{-1}\three - \rambda)
%     \,=\,
%     % \deltabar(\minime\bDelta \three_{\a\da}(p^{-1})^{\da\a})
%     \frac{1}{\bDelta}\mem
%     \deltabar(\three{\mem\cdot\mem}p^{-1})\,
%     \delta^{(4)}\hnem(\lambda' {\mem-\mem} \lambda)\,
%     \delta^{(4)}\hnem(\mem{
%         \rambda{}' 
%         - \rambda (1 + |\bthree\rsq\lsq\bthree|/|\Delta|x)
%     }\mem)
% \end{align}
Then we obtain
\begin{align}
    \label{eq:can-amp.3}
    \Bra{\lambda'\rambda{}'} \hat{T} \Ket{\lambda\rambda}
    % \,=\,
    % -\Delta\,
    % \deltabar(\three{\mem\cdot\mem}p)\,
    % \delta^{(4)}\hnem(\lambda' {\hem-\mem} \lambda)\,
    % \delta^{(4)}\hnem\big(\mem{
    %     \lsq\rambda|
    %     + \lsq\rambda \bthree\rsq \lsq\bthree| / mx
    %     - \lsq\rambda{}'|
    % }\mem\big)
    % \,
    % \frac{q}{\sqrt{2}\Delta}\mem 
    % \langle\i|p|\bo\rsq\mem
    % + \mathcal{O}(q^2)
    =
    \sqrt{2}q\hem |\Delta|x\,\mem
    \deltabar(2\mem\three{\mem\cdot\mem}p)\,
    \delta^{(4)}\hnem(\lambda {\hem-\mem} \lambda')\,
    % \delta^{(4)}\hnem\big(\mem{
    %     \lsq\rambda|
    %     {\mem+\mem} \lsq\rambda \bthree\rsq \lsq\bthree| / |\Delta|x
    %     {\mem-\mem} \lsq\rambda{}'|
    % }\mem\big)
    \delta^{(4)}\hnem\bigg({
        \lsq\rambda|
        {\mem+\mem}
        \lsq\rambda|
        \frac{|\bthree\rsq\lsq\bthree|}{|\Delta|x}
        {\mem-\mem} \lsq\rambda{}'|
    }\bigg)
    + \mathcal{O}(q^2)
    % \,,
    \,.
\end{align}
Observe how the first-quantized massive twistor operator algebra 
in \eqref{eq:can-amp.1}-\eqref{eq:can-amp.2}
elegantly realizes Guevara, Ochirov, and Vines \cite{guevara2019scattering}'s exponential operator.
We can take \eqref{eq:can-amp.3} as a \textit{straight-up derivation}
of Arkani-Hamed, Huang, and Huang \cite{ahh2017}'s
minimally coupled
on-shell amplitude
from the first-quantized theory.

To be precise, it is actually not yet on-shell,
% even though 
albeit
the $x$-factor could be well-defined
in virtue of the delta function $\deltabar(2\mem\three{\mem\cdot\mem}p) = \deltabar(E'{\hhem-\mem}E) / \Delta$.
If we
finally
constrain the in- and out-states 
to lie on
% on
the mass shell,
% constraint surface,
% on-shell Hilbert space,
\begin{align}
    (\hat{\bDelta} {\mem-\mem} m_0) \Ket{\one\bone}
    = 0
    \,,\quad
    (\hat{\Delta} {\mem-\mem} m_0) \Ket{\one\bone}
    = 0
    % \,,\quad
    % \hat{\lambda}_\a{}^I \Ket{\one\bone}
    % = \one_\a{}^I \Ket{\one\bone}
    % \,,\quad
    % \hat{\rambda}_{I\da} \Ket{\one\bone}
    % = \bone_{I\da} \Ket{\one\bone}
    \,,
\end{align}
then
now
we can 
% finally 
\textit{identify} 
the matrix element with \cite{ahh2017}'s amplitude:
\begin{align}
    \Bra{\two\btwo} \hat{T} \Ket{\one\bone}
    =
    \sqrt{2}q\hem m_0x\,\mem
    \deltabar(2\mem\three{\mem\cdot\mem}\one)\,
    \delta^{(4)}\hnem\big(\mem{
        |\one\rangle {\mem-\mem} |\two\rangle
    }\big)\,
    \delta^{(4)}\hnem\bigg({
        \lsq\bone|
        {\mem+\mem}
        \lsq\bone|
        \frac{|\bthree\rsq\lsq\bthree|}{m_0x}
        {\mem-\mem} \lsq\btwo|
    }\bigg)
    + \mathcal{O}(q^2)
    % \,,
    \,.
\end{align}
We emphasize again that the amplitude is supported on the heavenly kinematics 
so that it is indeed tempting to introduce the 
on-shell
diagrammatic notations
\eqref{eq:3kin-plus}-\eqref{eq:3kin-minus}.
% Perhaps
% the heavenly kinematics can
% play a central role in
% a certain realization of on-shell recursion in the massive twistor space.
Perhaps a certain realization of on-shell recursion in the massive twistor space
could take the heavenly kinematics as basic building blocks.

Going further, we can try to reproduce the impulses in section \ref{sec:Flat.amplitude}
by applying the KMO'C formalism \cite{kosower2019amplitudes}
in this first-quantized setting.
Yet another direction 
is non-minimal couplings.
Even though the relation between ``kinematic'' and ``canonical'' variables becomes unclear,
we could
work with higher-spin interaction Hamiltonians
constructed from real-spacetime fields.
We leave these topics for future research.

\newpage
\bibliography{references}

\end{document}